\documentclass[11pt]{article}
\usepackage{amsmath,axodraw,cite}
\numberwithin{equation}{section}
\renewcommand{\Re}{\mathop{\rm Re}\nolimits}
\renewcommand{\Im}{\mathop{\rm Im}\nolimits}
\def\slash#1{\setbox0=\hbox{$#1$}               
   \dimen0=\wd0                                 
   \setbox1=\hbox{/} \dimen1=\wd1               
   \ifdim\dimen0>\dimen1                        
      \rlap{\hbox to \dimen0{\hfil/\hfil}}      
      #1                                        
   \else                                        
      \rlap{\hbox to \dimen1{\hfil$#1$\hfil}}   
      /                                         
   \fi}                                         %
    \advance\textheight by \topskip

\begin{document}

\begin{titlepage}
\begin{flushright}
TTP99-18\\
UM-TH-99-04\\
April 1999\\
\end{flushright}
\vskip 2cm
\begin{center}
{\Large \bf On the Precise Determination of
            the Fermi Coupling Constant\\[3mm]
            from the Muon Lifetime
            } \\[20mm]
  Timo van Ritbergen
\vskip 0.5cm {\it Institut f\"ur Theoretische Teilchenphysik,
                  Universit\"at Karlsruhe\\ D-76128 Karlsruhe, GERMANY}
\vskip 0.3cm
and
\vskip 0.3cm
  Robin G. Stuart
\vskip 0.5cm {\it Randall Physics Laboratory, University of Michigan\\
                  Ann Arbor, MI 48109-1120, USA}\\
\end{center}
\vskip 3cm
\hrule
\begin{abstract}
The determination of the Fermi coupling constant, $G_F$,
is examined in the
light of recently calculated 2-loop QED corrections and planned
experiments to measure the muon lifetime to a level below
1\,ppm. The methods used in the calculation of the QED corrections are
described in detail. Sources of the dominant theoretical and
experimental uncertainties are identified. Finally the incorporation
of $G_F$ into analyses using the full electroweak Standard Model is
discussed.
\end{abstract}
\hrule
\vspace{2mm}
\noindent {
\tiny PACS: 13.35.Bv, 12.15.Lk, 12.20.Ds, 14.60.Ef }

\end{titlepage}

\setcounter{footnote}{0} \setcounter{page}{2} \setcounter{section}{0}
\newpage

\section{Introduction}

Any renormalizable theory only becomes predictive once it has been
supplied with a sufficient number of experimental inputs so as to fix
the free parameters that appear in its Lagrangian. In order
for the subsequent theoretical predictions to be as accurate as possible,
these inputs are chosen from the experimentally best-measured quantities
available. In the case of the Standard Model of electroweak interactions
these are the electromagnetic coupling constant,
$\alpha$, the Fermi coupling constant, $G_F$, and the mass of the $Z^0$
boson, $M_Z$. Their recognized best values, along with their absolute
and relative errors that they represent are \cite{PDG,LEPEWWG}
\begin{xxalignat}{2}
{}\qquad\qquad\qquad\qquad
\alpha&=1/(137.0359895\pm0.0000061)&  (0.045\,{\rm ppm})& \\
G_F&=(1.16639\pm0.00002)\times10^{-5}\,{\rm GeV^{-2}}& (17\,{\rm ppm})& \\
M_Z&=91.1867\pm0.0021\,{\rm GeV}&           (23\,{\rm ppm})&
\end{xxalignat}
The first of these, $\alpha$, is measured at momentum $q=0$ and when used
in the analysis high energy data is afflicted with a
`hadronic uncertainty' entering already at the 1-loop level and arising
because it must be `run up' from low energy crossing, on the way,
the hadronic resonance region.

Most processes, such as $e^+e^-$ annihilation, that allow
$M_Z$ to be accurately determined, are also mediated by
the photon and thus contain a pair of characteristic energy scales.
Experiments performed at low-energy therefore do not provide useful
insight into physics at the high scale, $M_Z$.

The Fermi coupling constant is obtained from the muon lifetime,
$\tau_\mu$, via a calculation in the Fermi model, in which the weak
interactions are represented by a contact interaction.
It suffers from neither of the disadvantages described above.
There can be no hadronic effects at 1-loop in this
model and even in the full Standard Model 1-loop hadronic effects in
muon decay are suppressed by a factor $m_f^2/M_W^2$ where $m_f$ is a
light fermion mass.
Charged current processes are only mediated
by the $W$ boson whose effects are unobscured no matter at what energy
experiments are performed.
The measurement of the muon lifetime can thus
be considered a high-energy experiment that is actually performed at
  a low scale
\cite{Roberts}. This point is highlighted by the fact that the next
generation of muon lifetime measurements will be sensitive to $W$
propagator effects.

In the mid-80's,
just before the turn on of LEP, a CERN report concluded that the error
on $M_Z$ would be $\pm50$\,MeV or 550\,ppm and that ``A factor of 2--3
improvement can be reached with a determined effort'' \cite{CERN}.
It was thus generally thought that $M_Z$ would be the input parameter
limiting the accuracy with which theoretical predictions could be made.
As is often case, due to a variety of unforeseen effects,
the uncertainty turned out to have been significantly
overestimated and it now approaches that of $G_F$.
The lesson that should be learned from this is that it is extremely
difficult to predict, even in the relatively short term, the accuracy to
which fundamental parameters will be determined and it is important that
these be extracted to the limits that the current theoretical and
experimental technology allows. A great effort and expense was brought
to bear in order to reduce the uncertainty on $M_Z$ to the level given
in the table
above and whereas it is unlikely to diminish significantly in the near
future a muon collider offers the prospect of a reduction by a factor of
10 \cite{Blondel}.

The Fermi coupling constant is closely related to the $\rho$-parameter
that was introduced by Ross and Veltman \cite{RossVeltman} as a way of
interrogating the mass generation sector of the theory
\begin{equation}
\rho=\frac{M_W^2}{M_Z^2\cos^2\theta_W}=1+\delta\rho
\end{equation}
The detailed understanding of the nature of the Higgs mechanism will
come about through the scrutiny of systems and processes that are most
sensitive to the Yukawa sector of the theory. The Fermi coupling constant
obtained from the muon lifetime provides such a probe but stands
alone for the tremendous accuracy with which in can be measured.
For this reason a considerable amount of theoretical effort, reviewed in
section~\ref{sec:WeakCorr}, has been devoted to the study of higher
order weak corrections to the muon lifetime. It can be shown that
the 2-loop QED corrections to the muon lifetime affect the predicted
value of the Higgs mass at the percent level which is precisely the
level needed to probe the detailed structure of the Higgs system by its
radiative corrections.

In this paper we look closely at the theoretical and experimental
considerations necessary for the extraction of $G_F$ from $\tau_\mu$
to an accuracy of 1\,ppm or better as is anticipated for
new experiments planned at the Brookhaven National Laboratory,
the Paul Scherrer Institute and the Rutherford-Appleton Laboratory.
What is known about the relationship between the muon lifetime
and the Fermi coupling constant is reviewed in
section~\ref{sec:taumuGF}. In section~\ref{sec:Definition}
inadequacies that appear at higher orders with the usual definition
of $G_F$ are pointed out and the relative merits
of the one adopted here are discussed. In
section~\ref{sec:AlphaRenorm} it is shown how to perform the
renormalization of the electromagnetic coupling constant in such a
way as to incorporate the large logarithms, $\ln(m_e^2/m_\mu^2)$.
Section~\ref{sec:calcul} describes the methods used for the
calculation of the 2-loop quantum electrodynamic (QED)
corrections to the muon lifetime in the Fermi model.
The sources of the dominant theoretical and experimental uncertainties
are examined in sections~\ref{sec:TheoUncert} and \ref{sec:ExpUncert}
respectively. The incorporation of the value of $G_F$ obtained here
into analyses using the full electroweak Standard Model is discussed
in section~\ref{sec:WeakCorr}.

Appendix~\ref{sec:multloop} contains expressions for certain
loop integrals for which exact analytic forms are known in
dimensional regularization. Appendix~\ref{sec:diagresults} gives
the results for the individual Feynman diagrams that occur in
the calculation of the 2-loop QED corrections to muon lifetime and
Appendix~\ref{sec:elecspec} gives an expression for the electron
spectrum to ${\cal O}(\alpha)$ keeping the full electron
mass dependence. Finally Appendix~\ref{sec:BR3e} examines the
branching ratio for the process
$\mu^-\rightarrow e^-\nu_\mu\bar\nu_e e^+e^-$.

\section{The Muon Lifetime and The Fermi Coupling Constant}
\label{sec:taumuGF}

In the Minkowskian metric in which time-like momenta squared are positive,
the Lagrangian, relevant for the calculation of the
muon lifetime in the Fermi theory is
\begin{equation}
{\cal L}_F={\cal L}_{{\rm QED}}^0+{\cal L}_{{\rm QCD}}^0+{\cal L}_W
\label{eq:FullLagrangian}
\end{equation}
Here ${\cal L}_{{\rm QED}}^0$ is the usual bare Lagrangian of Quantum
Electrodynamics (QED),
\begin{equation}
{\cal L}_{{\rm QED}}^0= \sum_f\bar\psi_f^0(\slash p-m_f)\psi_f^0
-\frac{1}{4}\left(\partial_\rho A_\sigma^0
                 -\partial_\sigma A_\rho^0\right)^2
-ie^0\sum_f Q_f\bar\psi_f^0\gamma_\rho\psi_f^0 A_\rho^0.
\end{equation}
The sum is over all fermion species, $f$, with wavefunction, $\psi_f$,
mass, $m_f$, and electric charge, $Q_f$.
$A_\rho$ is the photon field. The superscript ${}^0$ indicates
bare, as opposed to renormalized, quantities.
${\cal L}_{{\rm QCD}}^0$ is the bare Quantum Chromodynamic (QCD)
Lagrangian responsible for strong interactions. The Fermi contact
interaction that mediates muon decay is
\begin{equation}
{\cal L}_W=-2\sqrt{2} G_F
    \big[\bar\psi_{\nu_\mu}^0\gamma_\lambda\gamma_L\psi_\mu^0\big].
    \big[\bar\psi_e^0\gamma_\lambda\gamma_L\psi_{\nu_e}^0\big]
\label{eq:FermiLagrange}
\end{equation}
in which $\psi_\mu$, $\psi_e$, $\psi_{\nu_\mu}$ and $\psi_{\nu_e}$
are the wavefunctions for the muon, the electron and their associated
neutrinos respectively. and $\gamma_L$ denotes the usual Dirac
left-hand projection operator. For the present purposes the Fermi
coupling constant, $G_F$, goes unrenormalized.

To leading order in $G_F$ and all orders in $\alpha$ the
formula obtained for the muon lifetime, $\tau_\mu$, by means of the
${\cal L}_{{\rm F}}$ takes the general form
\begin{equation}
\frac{1}{\tau_\mu}\equiv\Gamma_\mu=\Gamma_0(1+\Delta q).
\label{eq:QEDcorr}
\end{equation}
where
\begin{equation}
\Gamma_0=\frac{G_F^2 m_\mu^5}{192\pi^3}
\label{eq:Gamma0}
\end{equation}
and $\Delta q$ encapsulates the higher order QED corrections
and can be expressed as power series expansion in the renormalized
electromagnetic coupling constant $\alpha_r=e_r^2/(4\pi)$.
\begin{equation}
\Delta q=\sum_{i=0}^\infty\Delta q^{(i)}
\label{eq:DeltaqSeries}
\end{equation}
in which the index $i$ gives the power of $\alpha_r$ that appears in
$\Delta q^{(i)}$.

Although ${\cal L}_F$ is not renormalizable, the $\Delta q^{(i)}$ can be
shown to be finite \cite{BermSirl}. This follows from the fact that when
Fierz rearrangement is used to rewrite ${\cal L}_W$ in so-called
charge retention order,
\[
{\cal L}_W\rightarrow-2\sqrt{2} G_F
    \big[\bar\psi_e^0\gamma_\lambda\gamma_L\psi_\mu^0\big].
    \big[\bar\psi_{\nu_\mu}^0\gamma_\lambda\gamma_L\psi_{\nu_e}^0\big]
\]
the currents remain purely V$-$A in form.
By contrast, in the case of neutron decay, the analogous
transformation generates scalar and pseudo-scalar terms and the following
arguments break down. The radiative corrections in that case are not
finite. Considering the vector part
$\bar\psi_e\gamma_\mu\psi_\mu$ of this effective $\mu$-$e$ current,
one sees that after fermion mass renormalization is performed
the remaining divergences are independent of the masses and
thus cancel, as for the case of pure QED.
The QED corrections to the axial vector part may be shown to also be
finite by noting that the transformations
$\psi_e\rightarrow\gamma_5\psi_e$ and
$m_e\rightarrow -m_e$ leave ${\cal L}_{{\rm QED}}$ and
${\cal L}_{{\rm QCD}}$ invariant but exchange
$\bar\psi_e\gamma_\lambda\psi_\mu\leftrightarrow
\bar\psi_e\gamma_\lambda\gamma_5\psi_\mu$.
Thus the radiative corrections to the axial-vector contribution to
the muon decay matrix element can be obtained from those of the vector
contribution by setting $m_e\rightarrow-m_e$.

This argument was used by Roos and Sirlin \cite{RoosSirlin} to
show that terms odd in $m_e$ would cancel between vector and axial-vector
contributions in the expression for the differential decay rate.
They then went on to show, by direct examination of known analytic form
of the differential decay rate \cite{BehrFinkSirl},
that the phase-space integration could not generate
terms linear in $m_e$ and thus that the leading electron mass
corrections at 1-loop are
${\cal O}\left(\alpha(m_e^2/m_\mu^2)\ln(m_e^2/m_\mu^2)\right)$,
${\cal O}\left(\alpha(m_e^2/m_\mu^2)\right)$
and higher but they did not exclude the possibility of
${\cal O}\left(\alpha(m_e^3/m_\mu^3)\right)$ terms which, in fact,
do occur.

It is possible to show that $\Delta q$ cannot
contain any terms that are odd in $m_e$ at any order in $\alpha$. The
total decay rate may be calculated directly as the imaginary part of
muon self-energy diagrams and was done in Ref.\cite{muonprl}.
In charge retention order terms in the numerators that
are odd in the electron mass lead to a helicity flip along the internal
electron line that causes the purely left-handed V$-$A vertices to
annihilate. However the integrals themselves generate
non-analytic terms such as $m_e^2\sqrt{m_e^2}=|m_e|^3$ which may appear
to be odd in $m_e$ if the absolute value is dropped.

The above considerations are true in any regularization scheme and have
the importance consequence that in the limit $m_e\rightarrow0$
only the radiative corrections to the vector pieces in ${\cal L}_W$
need to be calculated as they are equal to those to the axial vector
part. This avoids entirely the problems associated with $\gamma_5$
when dimensional regularization is used.

Dropping the electron neutrino mass and keeping only
the leading term in that of the muon neutrino gives
\begin{equation}
\Delta q^{(0)}=-8x-12x^2\ln x+8x^3-x^4-8y+{\cal O}(xy),
\ \ \ \ \ \ x=\frac{m_e^2}{m_\mu^2},
\ \ \ \ \ \ y=\frac{m_{\nu_\mu}^2}{m_\mu^2}
\label{eq:Deltaq0}
\end{equation}
that comes from phase space integrations. Notice that the coefficient of
the leading terms in the electron and muon neutrino masses are identical
which again follows from the fact that their wavefunctions may be
interchanged by Fierz rearrangement of ${\cal L}_{{\rm W}}$.

The first order corrections to $\Delta q$ are
\begin{equation}
\Delta q^{(1)}=
\left(\frac{\alpha_r}{\pi}\right)
\left(\frac{25}{8}-3\zeta(2)
-(34+12\ln x)x+96\zeta(2)\,x^{\frac{3}{2}}
+{\cal O}(x\ln^2 x)
\right)
\label{eq:Deltaq1}
\end{equation}
where $\zeta$ is the Riemann zeta function and $\zeta(2)=\pi^2/6$.
The leading electron mass-independent term was calculated
Kinoshita and Sirlin \cite{KinoSirl} and it was the observation
that the result is well-behaved in the limit $m_e\rightarrow0$
that ultimately lead to the discovery of the Kinoshita-Lee-Nauenberg
(KLN) theorem \cite{KinoLeeNaue}. An exact expression for the
full electron mass dependence in $\Delta q^{(1)}$ has been given by
Nir \cite{Nir}. Starting from the expressions for the 1-loop QED
corrections to the electron spectrum given by
Behrends {\it et al.}\cite{BehrFinkSirl} and, after taking into account
the correction given in Appendix~C of Ref.\cite{KinoSirl},
we obtain complete agreement with Eq.(12) of Ref.\cite{Nir}.
As a spinoff we have
considerably simplified the expression for the electron spectrum of
Ref.\cite{BehrFinkSirl}. The result appears in
Appendix \ref{sec:elecspec}.
An analogous expression has been given by Czarnecki {\it et al.}
\cite{CzarJezaKuhn} for the QCD corrections to
$b\rightarrow c\tau\bar\nu_\tau$.
We have also checked the electron mass-dependent terms in
Eq.(\ref{eq:Deltaq1}) by performing a large momentum expansion of the
 imaginary part of  propagator-type diagrams.

The second order corrections to $\Delta q$ have been presented recently
\cite{muonprl,muonhad} for $m_e=0$.
The result, ignoring the effects of tau loops, is
\begin{multline}
\Delta q^{(2)}=
              \left(\frac{\alpha_r}{\pi}\right)^2
        \bigg(\frac{156815}{5184}
             -\frac{1036}{27}\zeta(2)
                         -\frac{895}{36}\zeta(3)
                         +\frac{67}{8}\zeta(4)\\
                         +53\zeta(2)\ln2
                         -(0.042\pm0.002)
                          \bigg)
\label{eq:Deltaq2}
\end{multline}
where $\zeta(3)=1.20206...$ and $\zeta(4)=\pi^4/90$.
The numerical constant is the hadronic contribution with a
conservative estimate of its error. The contribution from tau loops
has been shown to be very small \cite{muonhad} as anticipated
from the decoupling theorem
\begin{equation}
\Delta q^{(2)}_{{\rm tau}}=-\left(\frac{\alpha_r}{\pi}\right)^2\,0.00058
\label{eq:Deltaq2tau}
\end{equation}
and will be neglected unless otherwise stated.

\begin{figure}
\begin{picture}(600,280)(10,-280)


\SetScale{.25}

\SetOffset(0,0)

 \SetWidth{8}
 \Line(50,50)(75,50)
 \Line(175,50)(300,50)
 \SetWidth{2}
 \Line(75,50)(175,50)
 \Oval(125,50)(40,50)(0)
 \PhotonArc(175,73)(73,238,340){4}{11.5}
 \PhotonArc(179,58)(35,246,343){4}{5.5}

\Text(20,-5)[m]{\small A1}

\SetOffset(90,0)

 \SetWidth{8}
 \Line(50,50)(75,50)
 \Line(175,50)(300,50)
 \SetWidth{2}
 \Line(75,50)(175,50)
 \Oval(125,50)(40,50)(0)
 \PhotonArc(215,49)(22,5,175){-4}{5.5}
 \PhotonArc(190,90)(85,242,330){4}{10.5}

\Text(20,-5)[m]{\small A2}

\SetOffset(180,0)

 \SetWidth{8}
 \Line(50,50)(108,50)
 \Line(192,50)(300,50)
 \SetWidth{2}
 \Line(95,50)(195,50)
 \Oval(150,50)(30,42)(0)

 \PhotonArc(230,49)(22,5,175){-4}{5.5}


 \PhotonArc(175,110)(122,210,-30){4}{18.5}

\Text(20,-5)[m]{\small A3}

\SetOffset(270,0)

 \SetWidth{8}
 \Line(50,50)(132,50)
 \Line(218,50)(300,50)
 \SetWidth{2}
 \Line(125,50)(225,50)
 \Oval(175,50)(30,42)(0)

 \PhotonArc(175,27)(28,180,0){-4}{7.5}
 \PhotonArc(175,-2)(108,28,152){4}{17.5}

\Text(20,-5)[m]{\small A4}

\SetOffset(360,0)

 \SetWidth{8}
 \Line(50,50)(132,50)
 \Line(218,50)(300,50)
 \SetWidth{2}
 \Line(125,50)(225,50)
 \Oval(175,50)(30,42)(0)

 \PhotonArc(175,75)(80,200,-20){4}{15.5}
 \PhotonArc(175,-22)(130,35,145){4}{19.5}

\Text(20,-5)[m]{\small A5}

\SetOffset(0,-65)

 \SetWidth{8}
 \Line(50,50)(108,50)
 \Line(193,50)(300,50)
 \SetWidth{2}
 \Line(100,50)(200,50)
 \Oval(150,50)(30,42)(0)
 \PhotonArc(175,64)(60,225,345){-4}{9.5}

 \PhotonArc(165,13)(92,90,156){4}{8.5}
 \PhotonArc(165,-25)(130,34,90){4}{10}

 \Text(20,-5)[m]{\small A6}

\SetOffset(90,-65)

 \SetWidth{8}
 \Line(50,50)(100,50)
 \Line(200,50)(300,50)
 \SetWidth{2}
 \Line(100,50)(200,50)
 \Oval(150,50)(40,50)(0)
 \PhotonArc(172,12)(25,41,184){4}{5.5}


 \PhotonArc(160,48)(58,198,270){-4}{5.0}
 \PhotonArc(160,98)(108,270,332){-4}{9.0}

  \Text(20,-5)[m]{\small A7}


\SetOffset(180,-65)

 \SetWidth{8}
 \Line(50,50)(75,50)
 \Line(175,50)(300,50)
 \SetWidth{2}
 \Line(75,50)(175,50)
 \Oval(125,50)(40,50)(0)

 \PhotonArc(154,60)(60,238,350){4}{9.5}
 \PhotonArc(195,78)(67,238,335){4}{9.5}

  \Text(20,-5)[m]{\small B1}

\SetOffset(270,-65)

 \SetWidth{8}
 \Line(50,50)(75,50)
 \Line(175,50)(300,50)
 \SetWidth{2}
 \Line(75,50)(175,50)
 \Oval(125,50)(40,50)(0)
 \PhotonArc(237,45)(30,15,165){4}{6.5}
 \PhotonArc(165,90)(85,250,330){4}{10.5}

  \Text(20,-5)[m]{\small C1}

\SetOffset(360,-65)

 \SetWidth{8}
 \Line(45,50)(108,50)
 \Line(192,50)(300,50)
 \SetWidth{2}
 \Line(95,50)(195,50)
 \Oval(150,50)(30,42)(0)

 \PhotonArc(247,45)(30,15,165){4}{6.5}


 \PhotonArc(155,90)(100,205,-25){4}{16.5}

   \Text(20,-5)[m]{\small C2}

\SetOffset(0,-130)

\SetWidth{8}
 \Line(50,50)(75,50)
 \Line(175,50)(300,50)
 \SetWidth{5}
 \Oval(208,8)(16,16)(0)
 \SetWidth{2}
 \Line(75,50)(175,50)
 \Oval(125,50)(40,50)(0)
 \PhotonArc(190,90)(85,242,272){4}{4.0}
 \PhotonArc(190,90)(85,292,330){4}{4.5}

  \Text(20,-5)[m]{\small C3}

\SetOffset(90,-130)

 \SetWidth{8}
 \Line(50,50)(132,50)
 \Line(218,50)(300,50)
 \SetWidth{5}
 \Oval(175,-8)(16,16)(0)
 \SetWidth{2}
 \Line(125,50)(225,50)
 \Oval(175,50)(30,42)(0)
 \PhotonArc(160,65)(75,190,270){4}{9.5}
 \PhotonArc(190,65)(75,270,350){4}{9.5}

  \Text(20,-5)[m]{\small C4}

\SetOffset(180,-130)

 \SetWidth{8}
 \Line(50,50)(132,50)
 \Line(218,50)(300,50)
 \SetWidth{5}
 \Oval(175,-7)(16,16)(0)
 \SetWidth{2}
 \Line(125,50)(225,50)
 \Oval(175,50)(30,42)(0)
 \PhotonArc(170,20)(30,160,245){4}{4.5}
 \PhotonArc(180,20)(30,295,20){4}{4.5}

 \Text(20,-5)[m]{\small C5}

\SetOffset(270,-130)

 \SetWidth{8}
 \Line(50,50)(100,50)
 \Line(200,50)(300,50)
 \SetWidth{2}
 \Line(100,50)(200,50)
 \Oval(150,50)(40,50)(0)
 \PhotonArc(185,90)(85,250,330){4}{8.5}
 \PhotonArc(160,-5)(40,48,150){4}{5.75}

  \Text(20,-5)[m]{\small D1}

\SetOffset(360,-130)

 \SetWidth{8}
 \Line(50,50)(110,50)
 \Line(210,50)(300,50)
 \SetWidth{2}
 \Line(110,50)(220,50)
 \Oval(160,50)(40,50)(0)
 \PhotonArc(200,52)(48,239,354){-4}{7.5}

 \PhotonArc(137,25)(25,158,323){-4}{6.5}

 \Text(20,-5)[m]{\small D2}

\SetOffset(0,-195)

 \SetWidth{8}
 \Line(50,50)(115,50)
 \Line(235,50)(300,50)
 \SetWidth{2}
 \Line(110,50)(240,50)
 \Oval(175,50)(40,60)(0)
 \PhotonArc(200,-5)(40,55,155){4}{5.75}
 \PhotonArc(165,33)(40,190,330){4}{7.75}

 \Text(20,-5)[m]{\small D3}

\SetOffset(90,-195)

 \SetWidth{8}
 \Line(50,50)(115,50)
 \Line(235,50)(300,50)
 \SetWidth{2}
 \Line(110,50)(240,50)
 \Oval(175,50)(40,60)(0)

 \PhotonArc(175,-72)(105,65,115){-4}{7.5}
 \PhotonArc(175,17)(23,190,-10){-4}{5.5}

  \Text(20,-5)[m]{\small D4}

\SetOffset(180,-195)

 \SetWidth{8}
 \Line(50,50)(115,50)
 \Line(235,50)(300,50)
 \SetWidth{2}
 \Line(110,50)(240,50)
 \Oval(175,50)(40,60)(0)

 \PhotonArc(144,23)(23,160,327){-4}{5.5}
 \PhotonArc(206,23)(23,213,20){-4}{5.5}

   \Text(20,-5)[m]{\small D5}

\SetOffset(270,-195)

\SetWidth{8}
 \Line(50,50)(75,50)
 \Line(175,50)(300,50)
 \SetWidth{2}
 \Oval(208,8)(16,16)(0)
 \SetWidth{2}
 \Line(75,50)(175,50)
 \Oval(125,50)(40,50)(0)
 \PhotonArc(190,90)(85,242,272){4}{4.0}
 \PhotonArc(190,90)(85,292,330){4}{4.5}

  \Text(20,-5)[m]{\small D6}

\SetOffset(360,-195)

 \SetWidth{8}
 \Line(50,50)(132,50)
 \Line(218,50)(300,50)
 \SetWidth{2}
 \Oval(175,-8)(16,16)(0)
 \SetWidth{2}
 \Line(125,50)(225,50)
 \Oval(175,50)(30,42)(0)
 \PhotonArc(160,65)(75,190,270){4}{9.5}
 \PhotonArc(190,65)(75,270,350){4}{9.5}

  \Text(20,-5)[m]{\small D7}

\SetOffset(0,-260)

\SetWidth{8}
 \Line(50,50)(132,50)
 \Line(218,50)(300,50)
 \SetWidth{2}
 \Oval(175,-7)(16,16)(0)
 \SetWidth{2}
 \Line(125,50)(225,50)
 \Oval(175,50)(30,42)(0)
 \PhotonArc(170,20)(30,160,245){4}{4.5}
 \PhotonArc(180,20)(30,295,20){4}{4.5}

 \Text(20,-5)[m]{\small D8}

\SetOffset(90,-260)

 \SetWidth{8}
 \Line(50,50)(125,50)
 \Line(225,50)(300,50)
 \SetWidth{2}
 \Line(110,50)(240,50)
 \Oval(175,50)(40,50)(0)

 \PhotonArc(125,54)(45,188,300){4}{7.5}
 \PhotonArc(225,54)(45,240,352){4}{7.5}

  \Text(20,-5)[m]{\small E1}

\SetOffset(180,-260)

 \SetWidth{8}
 \Line(50,50)(132,50)
 \Line(217,50)(300,50)
 \SetWidth{2}
 \Line(110,50)(240,50)
 \Oval(175,50)(30,42)(0)

 \PhotonArc(152,60)(68,190,270){4}{7.0}
 \PhotonArc(152,55)(63,270,340){4}{5.5}

 \PhotonArc(198,55)(63,200,270){4}{6.0}
 \PhotonArc(198,60)(68,270,350){4}{7.5}

 \Text(20,-5)[m]{\small F1}

\SetOffset(270,-260)

 \SetWidth{8}
 \Line(50,50)(118,50)
 \Line(202,50)(300,50)
 \SetWidth{2}
 \Line(110,50)(210,50)
 \Oval(160,50)(30,42)(0)
 \PhotonArc(204,85)(77,238,332){4}{9.5}

 \PhotonArc(160,23)(80,20,160){4}{16.5}

     \Text(20,-5)[m]{\small G1}

\SetOffset(360,-260)

\SetWidth{8}
 \Line(45,50)(132,50)
 \Line(218,50)(305,50)
 \SetWidth{2}
 \Line(125,50)(225,50)
 \Oval(175,50)(30,42)(0)
 \PhotonArc(185,15)(90,90,156){4}{8.5}
 \PhotonArc(185,-25)(130,34,90){4}{10}

 \PhotonArc(165,125)(130,214,270){4}{10}
 \PhotonArc(165,85)(90,270,338){4}{8.5}

  \Text(20,-5)[m]{\small G2}

\end{picture}

\caption{
   \label{4loopdiagrams}
          4-loop diagrams  whose cuts give contributions to
       the muon decay rate.
         The diagrams are grouped according to the
        main integration topologies A--G (see Figs.
          \ref{TopoAFigs}--\ref{TopoDEFGFigs}).
     The results for the individual diagrams are given in
             Appendix \ref{sec:diagresults}
         }

\end{figure}
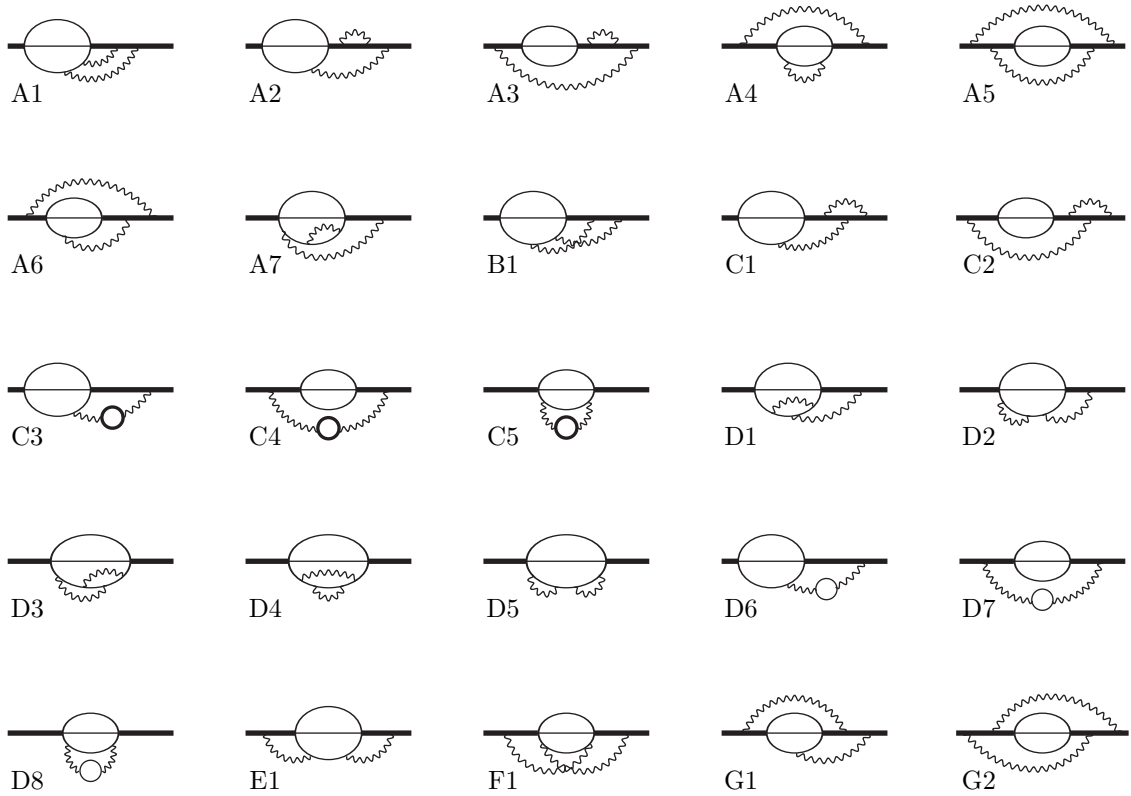

The methods used to obtain the result (\ref{eq:Deltaq2}) will be
described in some detail in section~\ref{sec:calcul}.
A key feature is that the
calculation was performed by means of cutting relations \cite{Cutkosky}
applied to the 25 1-particle irreducible (1PI) diagrams that appear
in Fig.\ref{4loopdiagrams}.
In these diagrams the thick line represents a muon with the external
legs being on-shell. The thin lines
represent the electron, electron neutrino or muon neutrino all of which
are taken to be massless. The wiggly line represents the photon.
Cuts passing only through massless internal lines give non-zero
contribution and all others vanish. In particular, since the external,
leg is on-shell, any cut through a muon line is identically zero.
The imaginary part of the diagrams of Fig.\ref{4loopdiagrams}
then generates precisely those combinations of the products
of amplitudes that appear in the calculation of the 2-loop QED
corrections to muon decay. Moreover, when the calculation is performed
in this way, the intricate cancellation of infrared (IR) divergences,
that would occur between these products of amplitudes, coming from
distinct cuts applied to a given diagram, is largely taken care
of automatically.

The result for $\Delta q^{(2)}$ in Eq.~(\ref{eq:Deltaq2})
is composed of several independent pieces. The part coming from the
purely photonic diagrams of Fig.\ref{4loopdiagrams} that contain
no closed charged-fermion loop and related 1-particle reducible (1PR)
external leg corrections is
\begin{eqnarray}
\Delta q_{\gamma\gamma}^{(2)}&=&
\left(\frac{\alpha_r}{\pi}\right)^2
\bigg(\frac{11047}{2592}-\frac{1030}{27}\zeta(2)
                         -\frac{223}{36}\zeta(3)
                         +\frac{67}{8}\zeta(4)
                         +53\zeta(2)\ln (2)\bigg)
\label{eq:Deltaq2photon}\\
                         &=&\left(\frac{\alpha_r}{\pi}\right)^2
                            3.55877.
\end{eqnarray}
This was calculated in a general $R_\xi$ gauge and was found to be
both gauge invariant and ultraviolet (UV) finite.

Feynman diagrams containing an electron loop or $e^+e^-$ pair in the
final state are obtained from cuts to diagrams D6--D8 of
Fig.\ref{4loopdiagrams}. Addition of the related 1PR external leg corrections
and diagrams containing coupling constant counterterm insertions
gives
\begin{eqnarray}
\Delta q^{(2)}_{{\rm elec}}&=&
-\left(\frac{\alpha_r}{\pi}\right)^2
\left(\frac{1009}{288}-\frac{77}{36}\zeta(2)
                      -\frac{8}{3}\zeta(3)\right)
\label{eq:Deltaq2elec}\\
                &=&\left(\frac{\alpha_r}{\pi}\right)^2
                            3.22034
\label{eq:electronnum}
\end{eqnarray}
The value obtained in Eq.(\ref{eq:electronnum}) is consistent with a
numerical study presented in Ref.\cite{LukeSavaWise} in the context of
semi-leptonic decays of heavy quarks, $Q\rightarrow X_q e\bar\nu_e$.

The contribution from diagrams containing muon loops, diagrams C3--C5 of
Fig.\ref{4loopdiagrams}
can be shown to be identical in the on-shell and
$\overline{{\rm MS}}$ renormalization schemes for a 't~Hooft mass
$\mu=m_\mu$ (see section\ref{sec:AlphaRenorm}).
These plus related 1PR external leg corrections
and diagrams containing coupling constant counterterm insertions and
give
\begin{eqnarray}
\Delta q^{(2)}_{{\rm muon}}&=&\left(\frac{\alpha_r}{\pi}\right)^2
\left(\frac{16987}{576}-\frac{85}{36}\zeta(2)
                        -\frac{64}{3}\zeta(3)\right)\\
                         &=&-\left(\frac{\alpha_r}{\pi}\right)^2
                            0.0364333.
\end{eqnarray}
This was calculated both by means of dispersion relations in the on-shell
scheme \cite{muonhad} and by the perturbative methods described in
section~\ref{sec:calcul}
in the $\overline{{\rm MS}}$ renormalization scheme. Full
agreement was found.

The hadronic contribution to (\ref{eq:Deltaq2}) is given in
Ref.\cite{muonhad}. Notice that, because it was calculated using
dispersion relations, it naturally involves a subtraction of the photon
vacuum polarization at $q=0$ which corresponds to the on-shell
scheme being adopted for electric charge renormalization.
The same is true for the stated result for tau
loops of Eq.~(\ref{eq:Deltaq2tau}).
On the other hand the photonic and electronic contributions,
(\ref{eq:Deltaq2photon}) and (\ref{eq:Deltaq2elec}), were calculated
perturbatively assuming, where
necessary, the $\overline{{\rm MS}}$ renormalization scheme. It will be
shown in section~\ref{sec:AlphaRenorm}
that the two sets of results can be combined in a
consistent manner by setting renormalized electromagnetic coupling
constant, $\alpha_r=\alpha_e(m_\mu)$=1/135.90. In doing so
all corrections up to ${\cal O}(\alpha^2)$,
${\cal O}\left(\alpha^3\ln(m_e^2/m_\mu^2)\right)$ and
${\cal O}\left(\alpha^i\ln^{i-1}(m_e^2/m_\mu^2)\right)$ for all $i\ge2$
are taken into account.

Fermion mass renormalization is always performed in the on-shell
renormalization scheme in which the renormalized mass of a stable
fermion is set equal to its pole mass.

$\Delta q^{(2)}$ of Eq.~(\ref{eq:Deltaq2}), when combined with the
tiny contribution of tau loops gives Eq.~(\ref{eq:Deltaq2tau}),
\begin{equation}
\Delta q^{(2)} =  \Gamma_0\left(\frac{\alpha_e(m_\mu)}{\pi}\right)^2
                (6.700\pm 0.002).
\end{equation}
When this is used in Eq.~(\ref{eq:QEDcorr}) along with the current best
value for the muon lifetime,
$\tau_\mu=(2.19703\pm0.00004)\,\mu$s \cite{PDG}, it gives a new value
for the Fermi coupling constant
\begin{xxalignat}{2}
{}\qquad\qquad\qquad\qquad
G_F&=(1.16637\pm0.00001)\times10^{-5}\,{\rm GeV^{-2}}&   (9\,{\rm ppm})&
\end{xxalignat}
This represents a reduction in the overall error on $G_F$ of
about a factor of 2 and a downward shift in the central value of twice
the experimental uncertainty. The error is now entirely experimental.

\section{The Definition of the Fermi Coupling Constant}
\label{sec:Definition}

At high precision it is essential to have a clear and unambiguous
definition for the Fermi coupling constant, $G_F$.
For the present purposes
$G_F$ will taken to be defined by Eq.(\ref{eq:QEDcorr})
\begin{equation}
\tau^{-1}_\mu \equiv \frac{G_F^2 m_\mu^5}{192\pi^3}
                \left( 1+\Delta q \right)
\end{equation}
with
$\Delta q$ being calculated using the Fermi Theory
Lagrangian ${\cal L}_F$ of Eq.(\ref{eq:FullLagrangian}).
This expresses $G_F$ to arbitrary accuracy in terms of
the physically observable quantities, $\tau_\mu$, $\alpha$, $m_\mu$
and $m_e$. In this way the
infrared structure, which involves both real bremsstrahlung and virtual
corrections, is entirely relegated to a single constant $\Delta q$
and does not have to be dealt with when $G_F$ is eventually used in the
calculation of other physical constants such as the $W$ boson mass, $M_W$.
Thus the, presumably well-understood, contribution of QED to the muon
lifetime is eliminated from $G_F$ which is left with an enriched
contribution from weak physics.

$G_F$ is {\bf sometimes } defined via the formula \cite{PDG}
\begin{equation}
\tau^{-1}_\mu=\frac{G_F^2 m_\mu^5}{192\pi^3}F\left(\frac{m_e^2}{m_\mu^2}\right)
          \left(1+\frac{3}{5}\frac{m_\mu^2}{M_W^2}\right)
                \left[1+\frac{\alpha(m_\mu)}{2\pi}
                \left(\frac{25}{4}-\pi^2\right)\right]
\label{eq:taumudef}
\end{equation}
where
\[
F(x)=1-8x-12x^2\ln x+8x^3-x^4
\]
and
\[
\alpha(m_\mu)^{-1}=\alpha^{-1}
                  -\frac{2}{3\pi}\ln\left(\frac{m_e}{m_\mu}\right)
                  +\frac{1}{6\pi}\sim136.
\]
In as far as this can be derived from ${\cal L}_F$ it provides an
adequate definition but it contains a number of features that become
 out of place at higher orders.

The function $F$, coming from phase space integration, does not
factorize in this way at higher orders.

The factor
$\left(1+(3m_\mu^2)/(5M_W^2)\right)$ is the effect of the $W$ boson
propagator and is not generated by the ${\cal L}_F$. It is more
naturally included along with the weak corrections in $\Delta r$ as
is described in section~\ref{sec:WeakCorr}.
Its presence is an historical artifact
of an attempt to reconcile the observed muon and neutron beta decay
rates with universality of weak interactions before the advent of the
Cabibbo angle \cite{BermSirl,Hofstadter}.
In that scenario a light $W$ boson with a mass slightly heavier than the
kaon was needed so as to forbid the unobserved decay mode,
$K^\pm\rightarrow W^\pm\gamma$.

The expression for $\alpha(m_\mu)$ used in the above definition contains
a term $1/(6\pi)$ that comes from $W$ boson loops in the photon
self-energy. As these do not come from ${\cal L}_F$ they should be omitted
so as not to risk double counting when $G_F$ is used as input
in electroweak calculations.

It is sometimes suggested that Eq.(\ref{eq:taumudef}) should be viewed
as an exact definition of $G_F$. This has significant drawbacks
and hard to justify now that the 2-loop QED corrections are available
\cite{muonprl,muonhad}.
It is already the case that these 2-loop QED corrections are larger
than the current experimental error and are roughly an order of
magnitude greater than those anticipated in the next generation of
experiments (see section~\ref{sec:ExpUncert}).
If Eq.~(\ref{eq:taumudef})
is taken to be exact then these 2-loop corrections will ultimately
have to be incorporated in formulas that relate $G_F$ to other
electroweak observables such as the mass of the $W$ boson, $M_W$.
This is extremely inconvenient as one would prefer to deal with
quantities from which QED has been eliminated as far as possible.

\section{Renormalization of the Electromagnetic Coupling Constant}
\label{sec:AlphaRenorm}

In this section it will be shown how to set up the $\overline{{\rm MS}}$
renormalization scheme so as to obtain the leading logarithmic
corrections to the muon lifetime of
${\cal O}\left(\alpha^3\ln(m_\mu^2/m_e^2)\right)$ and of
${\cal O}\left(\alpha^i\ln^{i-1}(m_\mu^2/m_e^2)\right)$ for all $i>0$.
It is also shown how to incorporate hadronic contributions \cite{muonhad},
calculated in the on-shell renormalization scheme, in a consistent manner.

In any self-consistent calculation using the Lagrangian
(\ref{eq:FullLagrangian}), the quantity $\alpha$ that appears
in the $\Delta q^{(i)}$ is the renormalized electromagnetic
coupling constant, to be denoted $\alpha_r$, in whatever
renormalization scheme has been chosen. It is generally convenient
to adopt the on-shell renormalization scheme for the fermion masses
which will be assumed throughout unless otherwise stated. No such
restriction is placed on the coupling constant renormalization.

For problems containing widely disparate scales, renormalization of
the coupling constant in the $\overline{{\rm MS}}$ scheme is to be
preferred as it automatically absorbs the dominant logarithmic
corrections into $\alpha_r$ at the outset and avoids the need for
resummation of large logarithms coming from higher-order contributions
as is required when the on-shell renormalization scheme is adopted.

In QED the numerical value of $\alpha_r$ is by obtained solving the
equation
\begin{equation}
\alpha=\frac{\alpha_r}{1-\widehat\Pi_{\gamma\gamma}^\prime(0)}.
\label{eq:alphadef}
\end{equation}
where $\alpha$ on the right-hand side of Eq.(\ref{eq:alphadef})
is the experimentally-measured quantity, $\alpha=1/137.0359895(61)$
\cite{PDG}.
$\widehat\Pi_{\gamma\gamma}^\prime(0)$ is the photon vacuum
polarization function which itself may be written as expansion in
$\alpha_r$,
\begin{equation}
\widehat\Pi_{\gamma\gamma}^\prime(0)
=\sum_{i=1}^\infty\widehat\Pi_{\gamma\gamma}^{\prime(i)}(0)
\end{equation}
and each term receives contributions from all fermion species.
The hat, \ $\widehat{}$\ , indicates that counterterm contributions in
the chosen renormalization scheme have been included. In the on-shell
renormalization scheme the counterterms are adjusted so that
$\widehat\Pi_{\gamma\gamma}^\prime(0)=0$ and it follows from
Eq.(\ref{eq:alphadef}) that the renormalized coupling constant in this
scheme satisfies,
$\alpha_r\equiv\alpha_{{\rm OS}}=\alpha$. In the $\overline{{\rm MS}}$
renormalization scheme the counterterms are chosen to contain only
divergent pieces plus certain uninteresting constants. In particular
1-loop counterterms are just proportional to
$\Delta= 1/\varepsilon -\gamma_E +\ln4\pi + O(\varepsilon)$ where
$D=4-2\varepsilon $ is the dimensionality of spacetime and
$\gamma_E=0.57721566...$ is Euler's constant. In this scheme
an appropriate choice for the 't~Hooft mass is $\mu=m_\mu$ and we write
$\alpha_r\equiv\alpha(m_\mu)$.

Defining $\widehat\Pi_{\gamma\gamma}^{\prime(i)}(0)=\alpha_r^i P^{(i)}$
and solving Eq.(\ref{eq:alphadef}) for $\alpha_r$ yields
\begin{equation}
\alpha_r=\alpha-\alpha^2 \widehat P^{(1)}
               -\alpha^3 \widehat P^{(2)}+{\cal O}(\alpha^4)
\label{eq:alphaRsoln}
\end{equation}

The contribution of leptons to
$\widehat\Pi_{\gamma\gamma}^{\prime(i)}(0)$ can be exactly
calculated in perturbation theory using dimensional regularization
\cite{dimreg,dimreg2}.
The 1-loop contribution is
\begin{eqnarray}
\widehat\Pi_{\gamma\gamma}^{\prime(1)}(0)&=&
-\frac{\alpha_r}{3\pi}(4\pi)^{2-\frac{D}{2}}
\Gamma\left(2-\frac{D}{2}\right)
\sum_l Q_l^2\left(\frac{m_l^2}{\mu^2}\right)^{\frac{D}{2}-2}
+2\frac{\delta e^{(1)}_l}{e}\\
&=&-\frac{\alpha_r}{3\pi}\sum_l Q_l^2\left(\Delta-\ln \frac{m_l^2}{\mu^2}\right)
+2\frac{\delta e^{(1)}_l}{e}+{\cal O}(D-4)
\label{eq:PiAA1}
\end{eqnarray}
where $\delta e^{(1)}_l$ is the leptonic contribution to
the 1-loop electromagnetic charge counterterm.
At this order the $\overline{{\rm MS}}$ and on-shell renormalization
schemes differ only in the the finite parts of their coupling constant
counterterms, $\delta e^{(1)}_{\overline{{\rm MS}}}$ and
$\delta e^{(1)}_{{\rm OS}}$ respectively.
Notice that the contribution from muon loops
to $\delta e^{(1)}_{\overline{{\rm MS}}}$
and $\delta e^{(1)}_{{\rm OS}}$ is identical for $\mu=m_\mu$
and thus their overall contribution to
$\widehat\Pi_{\gamma\gamma}^{\prime(1)}(0)$ vanishes in both schemes.

The 2-loop correction in a general renormalization scheme
can be obtained from Ref.\cite{MaldeStuart1} taking into account
the difference in measure used to define dimensional regularization
\begin{multline}
\widehat\Pi_{\gamma\gamma}^{\prime(2)}(0)=
\frac{\alpha_r^2}{12\pi^2}
\frac{(5D^2-33D+34)}{D(D-5)}(4\pi)^{4-D}
\Gamma\left(3-\frac{D}{2}\right)\Gamma\left(2-\frac{D}{2}\right)
\sum_l Q_l^4 \left(\frac{m_l^2}{\mu^2}\right)^{D-4}\\
+\frac{2\alpha_r}{3\pi}(4\pi)^{2-\frac{D}{2}}\Gamma\left(3-\frac{D}{2}\right)
   \sum_l Q_l^2\frac{\delta m_l^{(1)}}{m_l}
               \left(\frac{m_l^2}{\mu^2}\right)^{\frac{D}{2}-2}
+2\frac{\delta e^{(2)}_l}{e}
\label{eq:GeneralPi2}
\end{multline}
where $\delta m_l^{(1)}$ is the 1-loop lepton mass counterterm and
$\delta e^{(2)}_l$ is the leptonic part of the
2-loop electromagnetic charge counterterm.
Adopting the on-shell renormalization scheme for fermion masses gives
\begin{equation} \label{masscounterinsert}
\frac{\delta m_l^{(1)}}{m_l}=-\frac{\alpha_r}{4\pi}
     Q_l^2\left(\frac{m_l^2}{\mu^2}\right)^{\frac{D}{2}-2}
     \frac{(D-1)}{(D-3)}
     (4\pi)^{2-\frac{D}{2}}
     \Gamma\left(2-\frac{D}{2}\right)
\end{equation}
and substituting this in Eq.(\ref{eq:GeneralPi2}) yields
\begin{eqnarray}
\widehat\Pi_{\gamma\gamma}^{\prime(2)}(0)&=&
\frac{\alpha^2_r}{4\pi^2}
\frac{(D^3-12D^2+41D-34)}{D(D-3)(D-5)}(4\pi)^{4-D}
\Gamma\left(3-\frac{D}{2}\right)\Gamma\left(2-\frac{D}{2}\right)
  \nonumber\\
 & &\qquad\qquad\qquad
\times\sum_l Q_l^4 \left(\frac{m_l^2}{\mu^2}\right)^{D-4}
   +2\frac{\delta e^{(2)}_l}{e}\\
&=&-\frac{\alpha_r^2}{4\pi^2}
    \sum_lQ_l^4\left(\Delta-\ln\frac{m_l^2}{\mu^2}+\frac{15}{4}\right)
    +2\frac{\delta e^{(2)}_l}{e}+{\cal O}(D-4)
\end{eqnarray}
whose leading logarithms are in agreement with the well-known result
of Jost and Luttinger \cite{JostLuttinger}.
The leading logarithms of ${\cal O}(\alpha^3)$
were obtained by Rosner \cite{Rosner}.

The contributions to muon lifetime from hadron, muon and tau loops have
been calculated using dispersion relations \cite{muonhad}. This method
naturally generates a subtraction at momentum $q=0$ which corresponds to
on-shell renormalization of the coupling constant. The calculation of
Ref.\cite{muonhad} was performed by effectively replacing the photon
propagators
\begin{equation}
\frac{g_{\mu\nu}}{q^2+i\epsilon}\longrightarrow
\frac{g_{\mu\sigma}}{q^2+i\epsilon}
(q^2 g_{\sigma\tau}-q_\sigma q_\tau)
\widehat\Pi_{\gamma\gamma}^{\prime(1)}(q^2)
\frac{g_{\tau\nu}}{q^2+i\epsilon}.
\end{equation}
The on-shell results of Ref.\cite{muonhad} are converted to
$\overline{{\rm MS}}$ by adding a correction obtained by
replacing the photon propagator by
\begin{equation}
\frac{g_{\mu\nu}}{q^2+i\epsilon}\longrightarrow
\frac{g_{\mu\sigma}}{q^2+i\epsilon}
(q^2 g_{\sigma\tau}-q_\sigma q_\tau)
\frac{2}{e}\left(\delta e^{(1)}_{\overline{{\rm MS}}}
                -\delta e^{(1)}_{{\rm OS}}\right)
\frac{g_{\tau\nu}}{q^2+i\epsilon}.
\label{eq:MStoOS}
\end{equation}
where $\delta e^{(1)}_{\overline{{\rm MS}}}-\delta e^{(1)}_{{\rm OS}}$
is a finite constant that includes the effects of loops of all fermion
species except the electron. As usual $\epsilon$ is a positive
infinitesimal. When (\ref{eq:MStoOS}) is applied to the
calculation of the muon lifetime the terms on the right hand side
proportional to $q_\mu q_\nu$
cancel between diagrams hence its effect amounts to an overall
multiplicative factor for the photon propagator and leads to a
correction to $\Gamma_\mu$ of
\[
\Delta\Gamma_{{\rm OS}\rightarrow\overline{{\rm MS}}}
=\Gamma_0 \Delta q^{(1)}
\frac{2}{e}\left(\delta e^{(1)}_{\overline{{\rm MS}}}
                -\delta e^{(1)}_{{\rm OS}}\right)
\]
This can be accounted for by a redefinition of the renormalized
coupling constant, $\alpha_r$ of Eq.(\ref{eq:alphaRsoln}),
which has the effect of eliminating the contributions of all
fermion loops except those of the electron.
It follows that the results of Ref.\cite{muonhad}
can be used directly in the calculation of the muon lifetime provided
that the $\overline{{\rm MS}}$ value $\alpha(\mu)$ is replaced by
\begin{equation}
\alpha(\mu)\longrightarrow
\alpha_e(\mu)=\alpha
   +\alpha^2\left(\frac{1}{3\pi}+\frac{\alpha}{4\pi^2}\right)
                 \ln\frac{\mu^2}{m_e^2}+\frac{15\alpha^3}{16\pi^2}.
\label{eq:alphae}
\end{equation}
The KLN theorem \cite{KinoLeeNaue} requires
that the only electron mass singularities contributing
the muon lifetime are those that appear in (\ref{eq:alphae}) which
arise as a consequence of $\alpha$'s having been defined at the
exceptional momentum, $q^2=0$. Moreover, careful examination of
Eq.(\ref{eq:alphadef}) reveals that writing
\begin{equation}
\alpha_e(m_\mu)=\frac{\alpha}
    {1-\frac{\displaystyle \alpha}{\displaystyle 3\pi}
       \ln\frac{\displaystyle m_\mu^2}{\displaystyle m_e^2}}
                 +\frac{\alpha^3}{4\pi^2}\ln\frac{m_\mu^2}{m_e^2}
\label{eq:alphaefinal}
\end{equation}
correctly resums logarithms of the form
$\alpha^n\ln^{n-1}(m_\mu^2/m_e^2)$ for all $n>0$ and incorporates those
of ${\cal O}(\alpha^3\ln(m_\mu^2/m_e^2))$ without generating spurious
higher-order
logarithms as there would be if the second term on the right hand side
of Eq.(\ref{eq:alphaefinal}) where moved into the denominator of the
first. The constant term of ${\cal O}(\alpha^3)$ has been dropped since
terms of this order have not been accounted for in the present calculation.
Terms that are not logarithmically enhanced are correct up to
${\cal O}(\alpha^2)$.

For the extraction of $G_F$ in the Fermi theory the appropriate choice
is $\mu=m_\mu$ which gives
\[
\alpha_e(m_\mu)=1/135.90=0.0073582
\]
however when $G_F$ is used in the analysis of electroweak data obtained
at the weak scale $\mu=M_Z$ is generally a more convenient choice.

The expression for $\alpha_e(m_\mu)$ could also be obtained be adopting
a hybrid renormalization scheme in which electron loops were renormalized
in $\overline{{\rm MS}}$ and all other fermions loops in the on-shell
scheme. This is allowable in QED since the individual fermions are not
connected by any gauge symmetry and provides a simple, if unconventional,
way of deriving Eq.(\ref{eq:alphaefinal}).

\section{The Calculation of the 2-loop QED Contributions}

\label{sec:calcul}

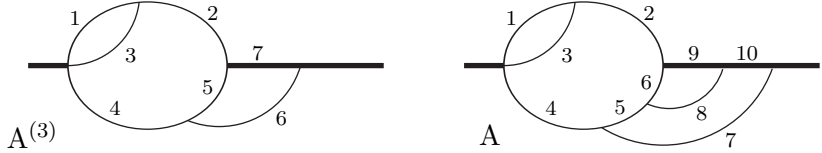
\begin{figure}
\begin{center}

\begin{picture}(350,70)(0,0)


\SetOffset(0,0)

 \SetWidth{2}
 \Line(30,30)(45,30)
 \Line(105,30)(164,30)
 \SetWidth{.5}
 \CArc(45,57)(27,270,353)
 \Oval(75,30)(24,30)(0)
 \CArc(102,39.0)(32,248,343)

 \Text(48,48)[m]{\scriptsize 1}
 \Text(100,50)[m]{\scriptsize 2}
 \Text(69,34)[m]{\scriptsize 3}
 \Text(63,14)[m]{\scriptsize 4}
 \Text(98,22)[m]{\scriptsize 5}
 \Text(126,10)[m]{\scriptsize 6}
 \Text(117,35)[m]{\scriptsize 7}

 \Text(30,4)[m]{ A$^{(3)}$}

\SetOffset(165,0)

 \SetWidth{2}
 \Line(30,30)(45,30)
 \Line(105,30)(164,30)
 \SetWidth{.5}
 \CArc(45,57)(27,270,353)
 \Oval(75,30)(24,30)(0)
 \CArc(105,43.8)(43.800,238,340)
 \CArc(107.4,34.8)(21,246,343)

 \Text(48,48)[m]{\scriptsize 1}
 \Text(100,50)[m]{\scriptsize 2}
 \Text(69,34)[m]{\scriptsize 3}
 \Text(63,14)[m]{\scriptsize 4}
 \Text(89,14.5)[m]{\scriptsize 5}
 \Text(99,24)[m]{\scriptsize 6}
 \Text(131,2)[m]{\scriptsize 7}
 \Text(120,12)[m]{\scriptsize 8}
 \Text(117,35)[m]{\scriptsize 9}
 \Text(137,35)[m]{\scriptsize 10}

 \Text(38,4)[m]{ A}

\end{picture}
\end{center}

\caption{  \label{TopoAFigs}
Integration topologies A$^{(3)}$ and A. The lines are numbered according to
  Eq.(\ref{defTopoA3})  and  Eq.~(\ref{defTopoA}) respectively }

\end{figure}

\begin{figure}
\begin{center}
\begin{picture}(350,70)(0,0)


\SetOffset(0,0)
 \SetWidth{2}
 \Line(30,30)(45,30)
 \Line(105,30)(164,30)
 \SetWidth{.5}
 \CArc(45,59)(29,270,350)
 \Oval(75,30)(24,30)(0)
 \CArc(92,36)(36,238,350)
 \CArc(117,46.8)(40.2,238,335)

\Text(48,48)[m]{\scriptsize 1}
 \Text(100,50)[m]{\scriptsize 2}
 \Text(69,34)[m]{\scriptsize 3}
 \Text(63,14)[m]{\scriptsize 4}
 \Text(86,13)[m]{\scriptsize 5}
 \Text(99,24)[m]{\scriptsize 6}
 \Text(119,19)[m]{\scriptsize 7}
 \Text(146,12)[m]{\scriptsize 8}
 \Text(117,35)[m]{\scriptsize 9}
 \Text(140,35)[m]{\scriptsize 10}

\Text(35,4)[m]{ B}

\SetOffset(165,0)

 \SetWidth{2}
 \Line(33,30)(45,30)
 \Line(105,30)(172,30)
 \SetWidth{.5}
 \CArc(45,59)(29,270,350)
 \Oval(75,30)(24,30)(0)
 \CArc(142.2,27)(18,15,165)
 \CArc(99,54)(51,250,330)

 \Text(48,48)[m]{\scriptsize 1}
 \Text(100,50)[m]{\scriptsize 2}
 \Text(69,34)[m]{\scriptsize 3}
 \Text(63,14)[m]{\scriptsize 4}
 \Text(94,18)[m]{\scriptsize 5}
 \Text(114,10)[m]{\scriptsize 6}
 \Text(142,50)[m]{\scriptsize 7}
 \Text(116,35)[m]{\scriptsize 8}
 \Text(134,35)[m]{\scriptsize 9}
 \Text(150,35)[m]{\scriptsize 10}

 \Text(39,2)[m]{ C}

\end{picture}
\end{center}

\caption{ \label{TopoBCFigs}
Integration topologies B and C.
        The lines are numbered according to Eqs.
              (\ref{defTopoB},\ref{defTopoC})   }

\end{figure}
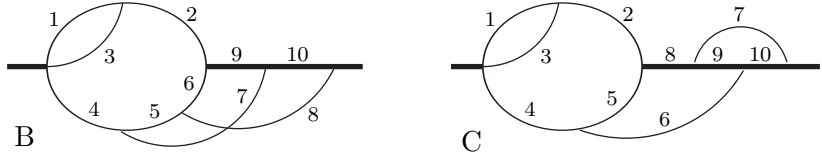

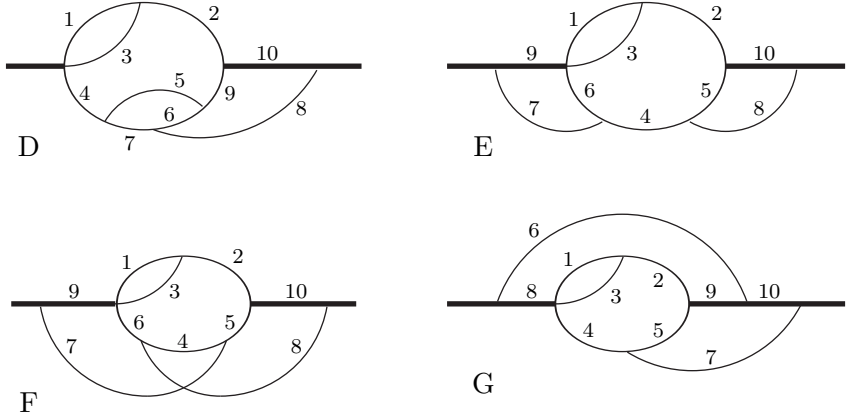
\begin{figure}
\begin{center}
\begin{picture}(350,160)(0,0)


\SetOffset(-10,100)

 \SetWidth{2}
 \Line(38,30)(60,30)
 \Line(120,30)(172,30)
 \SetWidth{.5}
 \CArc(60,59)(29,270,350)
 \Oval(90,30)(24,30)(0)
 \CArc(111,54)(51,250,330)
 \CArc(96,-3)(24,48,150)

 \Text(62,48)[m]{\scriptsize 1}
 \Text(117,50)[m]{\scriptsize 2}
 \Text(84,34)[m]{\scriptsize 3}
 \Text(68,20)[m]{\scriptsize 4}
 \Text(104,25)[m]{\scriptsize 5}
 \Text(100,12)[m]{\scriptsize 6}
 \Text(85,1)[m]{\scriptsize 7}
 \Text(150,14)[m]{\scriptsize 8}
 \Text(123,20)[m]{\scriptsize 9}
 \Text(137,35)[m]{\scriptsize 10}

 \Text(45,0)[m]{ D}

\SetOffset(165,100)

 \SetWidth{2}
 \Line(30,30)(75,30)
 \Line(135,30)(180,30)
 \SetWidth{.5}
 \CArc(75,59)(29,270,350)
 \Oval(105,30)(24,30)(0)

 \CArc(75,32.4)(27,188,300)
 \CArc(135,32.4)(27,240,352)

 \Text(78,48)[m]{\scriptsize 1}
 \Text(132,50)[m]{\scriptsize 2}
 \Text(100,35)[m]{\scriptsize 3}
 \Text(105,11)[m]{\scriptsize 4}
 \Text(128,21)[m]{\scriptsize 5}
 \Text(84,21)[m]{\scriptsize 6}
 \Text(63,14)[m]{\scriptsize 7}
 \Text(148,14)[m]{\scriptsize 8}
 \Text(62,35)[m]{\scriptsize 9}
 \Text(150,35)[m]{\scriptsize 10}

 \Text(42,0)[m]{ E}

\SetOffset(-10,10)

 \SetWidth{2}
 \Line(40,30)(79.2,30)
 \Line(130.2,30)(170,30)
 \SetWidth{.5}
  \CArc(79,57)(27,270,340)
 \Oval(105,30)(18,25.2)(0)

 \CArc(91.2,36)(40.8,190,270)
 \CArc(91.2,27)(31.8,270,340)

 \CArc(118.8,27)(31.8,200,270)
 \CArc(118.8,36)(40.8,270,350)

 \Text(84,46)[m]{\scriptsize 1}
 \Text(126,48)[m]{\scriptsize 2}
 \Text(102,35)[m]{\scriptsize 3}
 \Text(105,16)[m]{\scriptsize 4}
 \Text(123,23)[m]{\scriptsize 5}
 \Text(88.5,23)[m]{\scriptsize 6}
 \Text(63,14)[m]{\scriptsize 7}
 \Text(148,14)[m]{\scriptsize 8}
 \Text(64,35)[m]{\scriptsize 9}
 \Text(148,35)[m]{\scriptsize 10}

\Text(45,-7)[m]{ F}

\SetOffset(165,10)

 \SetWidth{2}
 \Line(30,30)(70.8,30)
 \Line(121.2,30)(180,30)
 \SetWidth{.5}
 \CArc(71,57)(27,270,340)
 \Oval(96,30)(18,25.2)(0)
 \CArc(122.4,51)(46.2,238,332)
 \CArc(96,13.8)(50,20,160)

 \Text(76,47)[m]{\scriptsize 1}
 \Text(110,39)[m]{\scriptsize 2}
 \Text(94,33)[m]{\scriptsize 3}
 \Text(83.5,20)[m]{\scriptsize 4}
 \Text(110,20)[m]{\scriptsize 5}
 \Text(63,58)[m]{\scriptsize 6}
 \Text(130,10)[m]{\scriptsize 7}
 \Text(62,35)[m]{\scriptsize 8}
 \Text(130,35)[m]{\scriptsize 9}
 \Text(152,35)[m]{\scriptsize 10}

\Text(42,0)[m]{ G}

\end{picture}
\end{center}

\caption{ \label{TopoDEFGFigs}
Integration topologies D -- G.
        The lines are numbered according to Eqs.
         (\ref{defTopoE}) and (\ref{defTopoF}) }

\end{figure}

In this section  the evaluation of Feynman
diagrams that appear in the calculation of $\Delta q^{(2)}$
 Eqs.\,(\ref{eq:DeltaqSeries}), (\ref{eq:Deltaq2}) is discussed.
For the calculation
of the 25 muon decay diagrams of Fig.\ref{4loopdiagrams}
 it is necessary to deal with only 7 basic 4-loop topologies labeled A--G
(see Figs.\ref{TopoAFigs}--\ref{TopoDEFGFigs}).
These 7 basic topologies are assumed to have scalar propagators raised
 to arbitrary (positive and negative) integer powers.

The integrals are evaluated on the mass shell $Q^2= m_\mu^2$ using
dimensional regularization \cite{dimreg,dimreg2} for both the
UV and IR divergences.  Within dimensional
regularization a few integrals are known in a simple closed form for
arbitrary powers of the propagators as a ratio of Euler $\Gamma$
functions. These are listed in Appendix~\ref{sec:multloop}.

The imaginary parts must be obtained for the 4-loop topologies.  Those
4-loop integrals that, on general grounds, have no imaginary part will  be
discarded along the way.  This includes 4-loop vacuum bubble integrals and
4-loop integrals with a through-going on-shell line, as these have no cuts
that can generate an imaginary part.

To reduce the burden of analytic integration as much as possible we use
the well-known method of integration-by-parts \cite{parialint} in
dimensional regularization to express all integrals of a given topology,
having arbitrary integer powers of the propagators, in terms of a small
set of primitive integrals.  Using integration-by-parts has the advantage
of requiring the explicit calculation of only a fixed number of primitive
integrals instead of hundreds of scalar integrals that generally appear in
a multiloop calculation.    On-shell integration-by-parts relations were
used at the two-loop level for the first time in Ref.\cite{msbarpolemass}
(but see also \cite{shell2}) and at the three-loop level in
Ref.\cite{gmin2at3loops}.

A given set of primitive integrals is not unique as one can eliminate one
set in favor of another set of integrals, as long  as they are independent
with respect to the integration-by-parts identities.  Therefore only the
total number of primitive integrals is a fixed quantity for  a given
topology, and what is used as a primitive integral is  a matter of
convenience.

Integration-by-parts relations follow from calculating the derivative in
the identity
\begin{equation}
\label{integratebyparts}
\int  {\rm d}^Dp\;  \frac{\partial}{\partial p^\mu} \;
      \left[    k^\mu f(p,l_1,\cdots, l_n) \right]=0
\end{equation}
within dimensional regularization. Here $p$ is any momentum that is integrated
over, $k\in \{ p,l_1,\cdots, l_n \} $ and $f$ is a scalar function
    that may depend in addition to $p$ on a set of momenta $l_i$.

\begin{figure}
\begin{center}

\begin{picture}(230,60)(0,0)


\SetOffset(0,10)
 \SetWidth{.5}
 \Line(0,21.65063510)(25,21.65063510)
 \Line(0,21.65063510)(12.5,0)
 \Line(25,21.65063510)(12.5,0)
 \Line(12.5,0)(12.5,-15)
  \Line(25,21.65063510)(40,34.64101616)
  \Line(0,21.65063510)(-15,34.64101616)

 \Text(12.5,27)[m]{\scriptsize 3}
 \Text(23,9)[m]{\scriptsize 1}
 \Text(2,9)[m]{\scriptsize 2}
 \Text(18,-12)[m]{\scriptsize 5}
 \Text(40,28)[m]{\scriptsize 4}
 \Text(-15,28)[m]{\scriptsize 6}
 \Text(-14,-18)[m]{$T_{0,0,0}$}

     \LongArrow(52,27)(42,19)
      \Text(53,21)[m]{\footnotesize $l$}

    \LongArrow(35,13)(29,2)
     \Text(38,6)[m]{\footnotesize $p$}

    \LongArrow(28,-8)(28,-19)
     \Text(34,-12)[m]{\footnotesize $k$}

\SetOffset(110,10)
 \SetWidth{2}
 \Line(0,21.65063510)(25,21.65063510)
 \SetWidth{.5}
 \Line(0,21.65063510)(12.5,0)
 \Line(25,21.65063510)(12.5,0)
 \Line(12.5,0)(12.5,-15)
  \SetWidth{2}
  \Line(25,21.65063510)(40,34.64101616)
  \SetWidth{.5}
  \Line(0,21.65063510)(-15,34.64101616)

 \Text(12.5,27)[m]{\scriptsize 3}
 \Text(23,9)[m]{\scriptsize 1}
 \Text(2,9)[m]{\scriptsize 2}
 \Text(18,-12)[m]{\scriptsize 5}
 \Text(40,28)[m]{\scriptsize 4}
 \Text(-15,28)[m]{\scriptsize 6}
 \Text(-14,-18)[m]{$T_{m,0,0}$}

\SetOffset(220,10)
 \SetWidth{2}
 \Line(0,21.65063510)(25,21.65063510)
 \Line(0,21.65063510)(12.5,0)
 \SetWidth{.5}
 \Line(25,21.65063510)(12.5,0)
  \SetWidth{2}
 \Line(12.5,0)(12.5,-15)
  \Line(25,21.65063510)(40,34.64101616)
  \SetWidth{.5}
  \Line(0,21.65063510)(-15,34.64101616)

 \Text(12.5,27)[m]{\scriptsize 3}
 \Text(23,9)[m]{\scriptsize 1}
 \Text(2,9)[m]{\scriptsize 2}
 \Text(18,-12)[m]{\scriptsize 5}
 \Text(40,28)[m]{\scriptsize 4}
 \Text(-15,28)[m]{\scriptsize 6}
 \Text(-14,-18)[m]{$T_{m,m,0}$}

\end{picture}
\end{center}

\caption{
\label{TriangleFigs}
Examples of triangle diagrams with different mass configurations for which
the triangle relation Eq.~(\ref{defTriangleRel}) holds.  The lines are
numbered according to Eq.~(\ref{defTriangle}).
}

\end{figure}
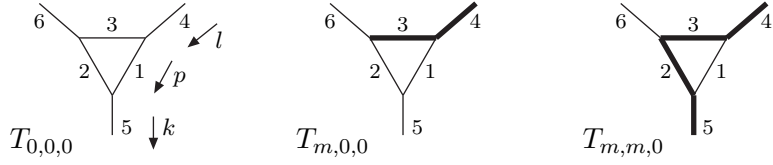

An important example of an integration-by-parts identity is the triangle
relation for one-loop 3-point integrals \cite{parialint}.  Let us define
for masses $a,b,c$
\begin{multline}
T_{a,b,c} (\alpha_1,\alpha_2,\alpha_3,\alpha_4,\alpha_5,\alpha_6
      ) =
\int {\rm d}^Dp\;
      \Biggl(  \frac{ 1}{(p^2)^{\alpha_1}
              [(p-k)^2-a^2]^{\alpha_2} [(p-l)^2-b^2]^{\alpha_3}
           }   \Biggr)\\
       \times
         \frac{ 1}{(l^2-b^2)^{\alpha_4} (k^2-a^2)^{\alpha_5}
         [(k-l)^2-c^2]^{\alpha_6} }
\label{defTriangle}
\end{multline}
Here and below $+i\epsilon$ in the propagators is not written  explicitly
to keep expressions compact (e.g. $1/p^2$ is understood to be
           $1/(p^2+i\epsilon)$).
By taking the derivative in the integral

\[   \int {\rm d}^Dp\;  \frac{\partial}{\partial p^\mu}
      \Biggl(  \frac{  p^\mu }{(p^2)^{\alpha_1}
              [(p-k)^2-a^2]^{\alpha_2} [(p-l)^2-b^2]^{\alpha_3}
           }   \Biggr)=0
        \]
one obtains the triangle relation
\begin{equation}
  \label{defTriangleRel}
      \biggl[ ( D -2\alpha_1 - \alpha_2 - \alpha_3)
       -  \alpha_2 {\bf 2^+} ( {\bf 1^-} - {\bf 5^-} )
       -  \alpha_3 {\bf 3^+} ( {\bf 1^-} - {\bf 4^-} ) \biggr]
           T_{a,b,c} (\alpha_1,\alpha_2,\alpha_3,\alpha_4,
            \alpha_5,\alpha_6) = 0
 \end{equation}
where the notation of Ref.\cite{parialint} has been adopted.  Thus the
operator ${\bf 2}^{+}$  raises  the power $\alpha_2$ in $T$ by one,
${\bf 4}^{-}$ lowers $\alpha_4$ in $T$ by one,  etc.. By repeated
use of this relation on integrals $T$ with positive integer powers
$\alpha_1,\cdots,\alpha_6$ one is able to lower at least one of the powers
$\alpha_1$, $\alpha_4$ or $\alpha_5$ to zero.  Instead of applying the
triangle relations recursively, the result can be obtained directly and
expressed as a 3-fold nested sum \cite{resolvedtriangle}.  Note that the
configuration of masses in Eq.~(\ref{defTriangle}) is not completely
general, and if all propagators carry different masses additional terms
appear in the triangle relation, leading to a less useful identity.
However, when a triangle with the mass configuration of $T$ appears as a
subgraph of a multiloop integral, one can use the triangle relation to
simplify the multiloop integrals considerably.  Some subgraphs relevant
to the present calculation are given in Fig.\ref{TriangleFigs}

To express multiloop integrals of a certain topology in terms of
primitive integrals, one generally needs to consider all
integration-by-parts identities obtained from Eq.~(\ref{integratebyparts})
and solve them.  To solve these relations there are two possible ways
to proceed\footnote{It should be mentioned that new approaches to solving
integration-by-parts identities are being developed as well
\cite{newibpsolution} }. Integration-by-parts relations may be explicitly
applied to a large set integrals with specific integer powers
$\{\cdots,-2,-1,0,1,2, \cdots \}$ and solved the resulting set of linear
equations (see, for example, Ref.\cite{gmin2at3loops}).  Alternatively the
relations may be considered as symbolic equations with unspecified powers of
the propagators and combined into a set of recurrence relations that, after
repeated application, express integrals in terms of primitives
\cite{parialint,msbarpolemass}.

Both approaches can have advantages in particular circumstances. The second
approach is more commonly used as one may encounter integrals with many
different powers in applications that involve expansions of diagrams.  For
the present work the recursive approach was used and a recurrence scheme
was implemented in a FORM \cite{form} program.  As some steps in the
recurrence relations involve hundreds of terms we limit ourselves to the
main conclusions for each of the  basic 4-loop topologies and give a list
of the primitive integrals were chosen.  The calculation of these
primitive integrals is discussed in the next section.

For the 3-loop topology A$^{(3)}$ of Fig.\ref{TopoAFigs}, it is possible
to lower the power of at least one propagator to zero through the use of
triangle identities.
It is assumed that a tensor reduction on the massless
1-loop sub-diagram formed by lines 1 and 3 is performed, such that for
A$^{(3)}$, all invariants can be expressed in terms of propagators.
Integration-by-parts recurrence relations express all integrals of this
topology in terms of one primitive integral with non-zero cuts.  Here and
below we do not count  integrals that can be calculated by repeated
application of the closed expressions
Eqs.(\ref{1loopbubble})--(\ref{bubblem00}).  For the normalization of the
primitive integrals we define
\begin{equation}
\label{defnormloop}
N_\varepsilon =  \frac{\pi^2}{ (\pi s)^{\varepsilon}}
\frac{\Gamma^2 (1-\varepsilon) \Gamma(1+\varepsilon) }
     {\Gamma(1-2\varepsilon) }
\end{equation} where $\varepsilon$ is related to
the space-time dimension as $D=4-2\varepsilon$.

\begin{multline}
I_{A^{(3)}}
     (\alpha_1,\alpha_2,\alpha_3,\alpha_4,\alpha_5,\alpha_6,\alpha_7) =
   i\; \int\int\int {\rm d}^Dp\; {\rm d}^Dk\; {\rm d}^Dl\;
      \Biggl(  \frac{ 1}{(p^2)^{\alpha_1}
              [(k+Q)^2]^{\alpha_2} }\\
              \times \frac{1}{
          [(p-k-Q)^2]^{\alpha_3}
           (k^2)^{\alpha_4} [(k-l)^2]^{\alpha_5}
         (l^2)^{\alpha_6} (l^2+2l\cdot Q)^{\alpha_{7}}
         }     \Biggr)
\label{defTopoA3}
\end{multline}

\begin{multline}
 \Im( I_{A^{(3)}}(1,0,1,1,1,0,1) ) = \pi\, s\, (N_\varepsilon)^3
     \biggl[ \frac{1}{\varepsilon} \Bigl( -\frac{1}{2} \Bigr)
        +\Bigl(  -\frac{11}{2} + \zeta_2 \Bigr)
        + \varepsilon \Bigl( -\frac{77}{2} + \frac{15}{2}\zeta_2
                    + 8 \zeta_3 \Bigr) \\
        +  \varepsilon^2 \Bigl(  - \frac{439}{2} + \frac{81}{2}\zeta_2
                + 50 \zeta_3 + \frac{69}{2}\zeta_4 \Bigr)
        + {\cal O}(\varepsilon^3)  \biggr]
\end{multline}
where for compactness the notation $\zeta_n\equiv\zeta(n)$ has been
adopted.

For 4-loop topology A of Fig.\ref{TopoAFigs} one can lower the power of at
least one propagator to zero through the use of triangle identities.  For
all 4-loop topologies A--G of Figs.\ref{TopoAFigs}--\ref{TopoDEFGFigs} we
assume that a tensor reduction on the massless 1-loop subdiagram formed by
lines 1 and 3 is performed, such that one has to deal with only one
invariant that cannot be expressed in terms of propagators.  For topology
A this invariant is chosen to be $p\cdot Q$.  Integration-by-parts
recurrence relations express all integrals of this topology in terms of 9
primitive integrals with a non-zero imaginary part.  Again we do not count
integrals that can be calculated by repeated application of the closed
expressions Eqs.(\ref{1loopbubble})--(\ref{bubblem00}).

\begin{multline}
I_A (\alpha_1,\alpha_2,\alpha_3,\alpha_4,\alpha_5,\alpha_6,\alpha_7,\alpha_8
      ,\alpha_9,\alpha_{10},\alpha_{11}) =\\
   \int\int\int\int {\rm d}^Dp\; {\rm d}^Dk\; {\rm d}^Dl\; {\rm d}^Dr \;
      \Biggl(  \frac{ (p\cdot Q)^{\alpha_{11}}}{(r^2)^{\alpha_1}
              [(k+Q)^2]^{\alpha_2} [(r-k-Q)^2]^{\alpha_3}
           (k^2)^{\alpha_4} [(k-l)^2]^{\alpha_5}  }\\
  \times \frac{1}{
         [(k-l-p)^2]^{\alpha_6} (l^2)^{\alpha_7}
             (p^2)^{\alpha_8}  [(l+p)^2 +2(l+p)\cdot Q]^{\alpha_9}
          (l^2+2l\cdot Q)^{\alpha_{10}}
         }     \Biggr)
\label{defTopoA}
\end{multline}

\begin{eqnarray*}
 \Im( I_A(1,1,1,1,0,1,1,1,1,1,0) )&=&\textstyle
 \frac{ \pi}{s} (N_\varepsilon)^4
     \left[ \frac{1}{\varepsilon} \Bigl( - 5 \zeta_4 \Bigr)
        + {\cal O}(\varepsilon^0)  \right]
\\
 \Im( I_A(1,1,1,0,1,0,1,1,1,0,0) )&=&\textstyle
 \pi\, s\, (N_\varepsilon)^4 \biggl[
      \frac{1}{\varepsilon^2}\Bigl( - \frac{1}{2}\Bigr)
                 + \frac{1}{\varepsilon} \Bigl( - \frac{31}{4} + \zeta_2
                  \Bigr)
                 + \Bigl( - \frac{591}{8} + \frac{29}{2}\zeta_2
                               + 9 \zeta_3  \Bigr)
\\ & &\textstyle
       + \varepsilon \Bigl( - \frac{8979}{16} + \frac{533}{4} \zeta_2
                 + \frac{209}{2} \zeta_3 + \frac{65}{2} \zeta_4  \Bigr)
                 + {\cal O}(\varepsilon^2) \biggr]
\\
 \Im( I_A(1,0,1,1,0,1,0,1,0,1,0) )&=&\textstyle
 \pi\, s^2\, (N_\varepsilon)^4 \biggl[
        \frac{1}{\varepsilon^2}\Bigl( -\frac{1}{4} \Bigr)
        + \frac{1}{\varepsilon} \Bigl(  -\frac{7}{3} \Bigr)
          + \Bigl( - \frac{1703}{144} - \zeta_2  \Bigr)
\\ & &\hspace{-4cm}\textstyle
              +\varepsilon \Bigl( - \frac{3697}{108}
              - \frac{119}{12} \zeta_2 - \frac{17}{2} \zeta_3  \Bigr)
  +\varepsilon^2 \Bigl(
                       \frac{159001}{5184} - \frac{2053}{36} \zeta_2
                      - \frac{511}{6} \zeta_3 - \frac{263}{4} \zeta_4
                                  \Bigr)
               + {\cal O}(\varepsilon^3) \biggr]
\\
 \Im( I_A(1,1,1,1,0,1,0,1,0,1,0) )&=&\textstyle
 \pi\, s\, (N_\varepsilon)^4 \biggl[
        \frac{1}{\varepsilon^3}\Bigl( \frac{1}{2}  \Bigr)
           + \frac{1}{\varepsilon^2} \Bigl(  4  \Bigr)
            + \frac{1}{\varepsilon} \Bigl(  \frac{165}{8}   \Bigr)
                        + \Bigl(\frac{343}{4} + \frac{3}{2} \zeta_2
                          - 3 \zeta_3  \Bigr)
\\ & &\textstyle
        + \varepsilon \Bigl( \frac{9749}{32} + \frac{25}{2} \zeta_2
                    - 9 \zeta_3 - \frac{17}{2} \zeta_4   \Bigr)
              + {\cal O}(\varepsilon^2) \biggr]\\
 \Im( I_A(1,0,1,0,1,1,1,0,1,0,0) )&=&\textstyle
 \pi\, s^2\, (N_\varepsilon)^4 \biggl[
             \frac{1}{\varepsilon} \Bigl( \frac{1}{12}  \Bigr)
                 + \Bigl( \frac{49}{18} - \zeta_2   \Bigr)
                  +  \varepsilon \Bigl( \frac{17161}{432}
                - \frac{127}{12} \zeta_2 - \frac{25}{2}\zeta_3  \Bigr)
\\ & &\textstyle
                   +  \varepsilon^2 \Bigl( \frac{260347}{648}
                                - \frac{2837}{36} \zeta_2
                             - \frac{376}{3} \zeta_3
                             - \frac{319}{4} \zeta_4 \Bigr)
                  + {\cal O}(\varepsilon^3) \biggr]
\\
  \Im( I_A(1,1,1,0,1,1,1,0,1,0,0) )&=&\textstyle
  \pi\, s\, (N_\varepsilon)^4 \biggl[
                 \frac{1}{\varepsilon^2} \Bigl( -\frac{1}{2}\Bigr)
                + \frac{1}{\varepsilon} \Bigl(  -6 \Bigr)
                 +\Bigl(   - \frac{385}{8} + \frac{11}{2}\zeta_2
                                    + \zeta_3 \Bigr)
\\ & &\textstyle
                  +  \varepsilon \Bigl(
                   - \frac{651}{2} + \frac{121}{2}\zeta_2
                      + 36 \zeta_3 + \frac{19}{2} \zeta_4 \Bigr)
                  + {\cal O}(\varepsilon^2) \biggr]
\\
  \Im( I_A(1,0,1,1,1,1,0,0,1,1,0) )&=&\textstyle
  \pi\, s\, (N_\varepsilon)^4 \biggl[
             \frac{1}{\varepsilon^2} \Bigl(  -\frac{1}{2} \Bigr)
           +  \frac{1}{\varepsilon} \Bigl( - \frac{31}{4} + 2 \zeta_2 \Bigr)
              + \Bigl( - \frac{591}{8} + 15 \zeta_2 + 22 \zeta_3 \Bigr)
\\ & &\textstyle
        +  \varepsilon \Bigl( - \frac{8979}{16} + \frac{155}{2} \zeta_2
                           + 146 \zeta_3 + 155 \zeta_4 \Bigr)
              + {\cal O}(\varepsilon^2) \biggr]
\\
\Im( I_A(1,0,1,1,0,1,1,1,1,0,0) )&=&\textstyle
\pi\, s\, (N_\varepsilon)^4 \biggl[
               \frac{1}{\varepsilon^2} \Bigl(  -\frac{1}{4} \Bigr)
           +  \frac{1}{\varepsilon} \Bigl( - \frac{33}{8} + \zeta_2  \Bigr)
              + \Bigl(  - \frac{665}{16} + \frac{13}{2} \zeta_2
                            +13 \zeta_3   \Bigr)
\\ & &\textstyle
+  \varepsilon \Bigl(  -\frac{10605}{32} + \frac{61}{4} \zeta_2
                 +\frac{179}{2}\zeta_3 + \frac{233}{2}\zeta_4  \Bigr)
              + {\cal O}(\varepsilon^2) \biggr]
\\
  \Im( I_A(1,1,1,0,0,1,1,1,1,0,0) )&=&\textstyle
  \pi\, s\, (N_\varepsilon)^4 \biggl[
           \frac{1}{\varepsilon} \Bigl( - \frac{7}{4} + \zeta_2 \Bigr)
        + \Bigl(  - \frac{119}{4} + \frac{15}{2}\zeta_2
                  + 13 \zeta_3  \Bigr)
\\ & &\textstyle
   + \varepsilon \Bigl( - \frac{4843}{16} + \frac{249}{4}\zeta_2
                      + \frac{195}{2} \zeta_3 + \frac{125}{2} \zeta_4 \Bigr)
               + {\cal O}(\varepsilon^2) \biggr]
\end{eqnarray*}

Topology B has 7 non-trivial primitive integrals with a non-zero imaginary
part, not counting the integrals that can be considered cases of Topology
A.  In choosing primitive integrals one may prefer in one case to keep a
power of $(p\cdot Q)$ in the numerator to avoid an infrared divergence
when the three momenta $p$, $k$ and $l$ go simultaneously to zero.

\begin{multline}
I_B (\alpha_1,\alpha_2,\alpha_3,\alpha_4,\alpha_5,\alpha_6,\alpha_7,\alpha_8
      ,\alpha_9,\alpha_{10},\alpha_{11}) =\\
   \int\int\int\int {\rm d}^Dp\; {\rm d}^Dk\; {\rm d}^Dl\; {\rm d}^Dr \;
      \Biggl(  \frac{ (p\cdot Q)^{\alpha_{11}}}{(r^2)^{\alpha_1}
              [(k+Q)^2]^{\alpha_2} [(r-k-Q)^2]^{\alpha_3}
           (k^2)^{\alpha_4} [(k-p)^2]^{\alpha_5}  }\\
 \times \frac{1}{
         [(k-l-p)^2]^{\alpha_6} (p^2)^{\alpha_7}
             (l^2)^{\alpha_8}  [(l+p)^2 +2(l+p)\cdot Q]^{\alpha_9}
          (l^2+2l\cdot Q)^{\alpha_{10}}
         }     \Biggr)
\label{defTopoB}
\end{multline}

\begin{eqnarray*}
 \Im( I_B(1,1,1,1,1,1,1,1,1,1,1) )&=&
\textstyle
 \frac{ \pi}{s} (N_\varepsilon)^4
     \left[ \frac{1}{\varepsilon} \Bigl( - \frac{17}{16} \zeta_4 \Bigr)
        + {\cal O}(\varepsilon^0)  \right]
\\
 \Im( I_B(1,1,1,1,1,1,0,1,1,1,0) )&=&
\textstyle
 \frac{ \pi}{s} (N_\varepsilon)^4
     \left[ \frac{1}{\varepsilon} \Bigl( - \frac{51}{8} \zeta_4 \Bigr)
        + {\cal O}(\varepsilon^0)  \right]
\\
 \Im( I_B(1,0,1,0,1,1,1,0,1,1,0) )&=&
\textstyle
 \pi\, s\, (N_\varepsilon)^4 \biggl[
            \Bigl( -\frac{7}{4}+\zeta_2 \Bigr)
   + \varepsilon \Bigl( -\frac{287}{8}+10 \zeta_2 +15\zeta_3 \Bigr)
\\ & &
\textstyle
+ \varepsilon^2 \Bigl(  -\frac{6847}{16}
      +67\zeta_2 +159 \zeta_3 +108\zeta_4 \Bigr)
                 + {\cal O}(\varepsilon^3) \biggr]
\\
 \Im( I_B(1,1,1,1,1,0,0,1,1,1,0) )&=&
\textstyle
 \pi\, (N_\varepsilon)^4 \biggl[
            \frac{1}{\varepsilon^3} \Bigl(    \frac{1}{2}  \Bigr)
           +  \frac{1}{\varepsilon^2} \Bigl( 5  \Bigr)
          +  \frac{1}{\varepsilon} \Bigl(\frac{63}{2}
                                      -2 \zeta_3 -2\zeta_2  \Bigr)
\\ & &
\textstyle
           + \Bigl( 159-19\zeta_2 -15 \zeta_3 - \frac{11}{2} \zeta_4 \Bigr)
                + {\cal O}(\varepsilon) \biggr]
\\
 \Im( I_B(1,1,1,0,1,0,1,1,1,1,0) )&=&
\textstyle
 \pi\, (N_\varepsilon)^4 \biggl[
            \frac{1}{\varepsilon}
                   +\Bigl( 20 - \zeta_2 - 4\zeta_3 \Bigr)
  + \varepsilon \Bigl( 237 - 24 \zeta_2
\\ & &
\textstyle
    - 53 \zeta_3 - \frac{67}{2}\zeta_4 \Bigr)
           + {\cal O}(\varepsilon^2) \biggr]
\\
\Im( I_B(1,1,1,1,1,1,0,0,1,2,0) )&=&
\textstyle
\frac{ \pi}{s} (N_\varepsilon)^4 \biggl[
            \frac{1}{\varepsilon} \Bigl(
                 \frac{5}{4} \zeta_3 - 3  \zeta_2 \ln(2)  \Bigr)
     + \Bigl(  5\zeta_3 - 12 \zeta_2 \ln(2)
\\ & &
\textstyle
- \frac{49}{8} \zeta_4 + 6{\rm Li_4}(1/2) + \frac{1}{4} \ln^4(2)  \Bigr)
           + {\cal O}(\varepsilon) \biggr]
\\
\Im( I_B(1,0,1,1,1,1,0,0,2,2,0) )&=&
\textstyle
\frac{ \pi}{s} (N_\varepsilon)^4 \biggl[
            \Bigl( -3 \zeta_2  +\frac{3}{4}\zeta_3 +3\zeta_2\ln(2)  \Bigr)
     + \varepsilon \Bigl(    - 21\zeta_2  - \frac{25}{2}\zeta_3
\\ & &
\textstyle
   + 6\zeta_2\ln(2) + \frac{291}{8}\zeta_4
          - 6\, {\rm Li_4}(1/2) - \frac{1}{4}\ln^4(2)  \Bigr)
                 + {\cal O}(\varepsilon^2) \biggr]
\end{eqnarray*}
Here ${\rm Li_4}(x)$ is the fourth order Polylogarithm \cite{polylog},
      ${\rm Li_4}(1/2) = 0.517479\cdots$.

For topology C one can lower the power of at least one propagator to zero
through the use of integration-by-parts recurrence relations.
 For this topology there are three
additional primitive integrals with a non-zero imaginary part, not
counting the integrals that are special cases of Topologies A and B.

\begin{multline}
I_C (\alpha_1,\alpha_2,\alpha_3,\alpha_4,\alpha_5,\alpha_6,\alpha_7,\alpha_8
      ,\alpha_9,\alpha_{10},\alpha_{11}) =\\
   \int\int\int\int {\rm d}^Dp\; {\rm d}^Dk\; {\rm d}^Dl\; {\rm d}^Dr \;
      \Biggl(  \frac{ (kl)^{\alpha_{11}}}{(r^2)^{\alpha_1}
              [(k+Q)^2]^{\alpha_2} [(r-k-Q)^2]^{\alpha_3}
           (k^2)^{\alpha_4} [(k-p)^2]^{\alpha_5}  }\\
         \times \frac{1}{
        (p^2)^{\alpha_6} (l^2)^{\alpha_7}
             (p^2+2p\cdot Q)^{\alpha_8}  [(l+p)^2 +2(l+p)\cdot Q]^{\alpha_9}
          (l^2+2l\cdot Q)^{\alpha_{10}}
         }     \Biggr)
\label{defTopoC}
\end{multline}

\begin{eqnarray*}\textstyle
\Im( I_C(1,0,1,1,1,0,0,1,1,1,0) )&=&
\textstyle
\pi\, s\, (N_\varepsilon)^4 \biggl[
            \frac{1}{\varepsilon^2} \Bigl( -\frac{1}{4}  \Bigr)
      +  \frac{1}{\varepsilon} \Bigl(   -\frac{33}{8} +\zeta_2   \Bigr)
\\ & &
\textstyle
  + \Bigl( -\frac{657}{16}  +11 \zeta_2 + 9 \zeta_3  \Bigr)
 + {\cal O}(\varepsilon) \biggr]
\\
 \Im( I_C(1,0,1,1,1,1,0,1,1,1,0) )&=&
\textstyle
  \pi\, (N_\varepsilon)^4 \biggl[
            \frac{1}{\varepsilon} \Bigl( 3 \zeta_2   \Bigr)
                + \Bigl( 11 + 26\zeta_2 + \zeta_3  \Bigr)
                + {\cal O}(\varepsilon) \biggr]
\\
 \Im( I_C(1,1,1,1,0,0,0,1,1,1,0) )&=&
\textstyle
  \pi\, s (N_\varepsilon)^4 \biggl[
       \frac{1}{\varepsilon^3} \Bigl(  \frac{3}{2} \Bigr)
       + \frac{1}{\varepsilon^2} \Bigl(  \frac{47}{4}  \Bigr)
       + \frac{1}{\varepsilon} \Bigl(  \frac{457}{8} - 3 \zeta_2  \Bigr)
      \\
    & & \textstyle
      + \Bigl(  \frac{3507}{16}
            - \frac{31}{2} \zeta_2
           - 3\zeta_3  \Bigr)
     + \varepsilon \Bigl(  \frac{22985}{32} - 48\zeta_2 \ln(2)
     \\
    & & \textstyle
            - \frac{89}{4} \zeta_2
          + \frac{9}{2} \zeta_3 - \frac{15}{2} \zeta_4 \Bigr)
       + {\cal O}(\varepsilon^2) \biggr]
\end{eqnarray*}

For topology D one can apply integration-by-parts relations in the form of the
massless triangle rule to the triangle formed by lines 5, 6 and 7.
Further use of integration-by-parts expresses integrals of this topology
into integrals that can be calculated using the closed expressions
Eqs.(\ref{1loopbubble})--(\ref{bubblem00}) or integrals with no imaginary
part, or integrals that are of topology A.  Therefore no new non-trivial
primitives with a non-zero imaginary part appear for topology D.

For topology E one can lower the power of at least one propagator
to zero through the use of integration-by-parts  relations.
There is one primitive integral with a non-zero imaginary part,
 not counting the integrals that can be considered as cases
      of topology A and B.

\begin{multline}    \label{defTopoE}
I_E (\alpha_1,\alpha_2,\alpha_3,\alpha_4,\alpha_5,\alpha_6,\alpha_7,\alpha_8
      ,\alpha_9,\alpha_{10},\alpha_{11}) =\\
   \int\int\int\int {\rm d}^Dp\; {\rm d}^Dk\; {\rm d}^Dl\; {\rm d}^Dr \;
      \Biggl(  \frac{ (p\cdot l)^{\alpha_{11}}}{(r^2)^{\alpha_1}
              [(k+Q)^2]^{\alpha_2} [(r-k-Q)^2]^{\alpha_3}
           (k^2)^{\alpha_4} [(k-l)^2]^{\alpha_5}  }\\
          \times \frac{1}{
        [(k-p)^2]^{\alpha_6} (p^2)^{\alpha_7}
             (l^2)^{\alpha_8}  (p^2 +2p\cdot Q)^{\alpha_9}
          (l^2+2l\cdot Q)^{\alpha_{10}}
         }     \Biggr)
\end{multline}

\[\textstyle
\Im( I_E(1,1,1,0,1,1,1,1,1,1,0) ) = \frac{ \pi}{s} (N_\varepsilon)^4
     \left[ \frac{1}{\varepsilon} \Bigl( -6 \zeta_4 \Bigr)
        + {\cal O}(\varepsilon^0)  \right]
\]

Non-planar topology F  has 6 non-trivial primitive integrals with a
non-zero imaginary part, not counting the integrals that can be considered
cases of topologies A--E.  In choosing primitive integrals one may prefer
in one case to keep a power of $(p\cdot l)$ in the numerator to avoid an
infrared divergence when the three momenta $p$, $k$ and $l$ go
simultaneously to zero.

\begin{multline}
I_F (\alpha_1,\alpha_2,\alpha_3,\alpha_4,\alpha_5,\alpha_6,\alpha_7,\alpha_8
      ,\alpha_9,\alpha_{10},\alpha_{11}) =\\
   \int\int\int\int {\rm d}^Dp\; {\rm d}^Dk\; {\rm d}^Dl\; {\rm d}^Dr \;
      \Biggl(  \frac{ (p\cdot l)^{\alpha_{11}}}{(r^2)^{\alpha_1}
              [(p+k+l+Q)^2]^{\alpha_2} [(r-p-k-l-Q)^2]^{\alpha_3}
           (k^2)^{\alpha_4}  }\\
       \times \frac{1}{  [(k+p)^2]^{\alpha_5}
        [(k+l)^2]^{\alpha_6} (p^2)^{\alpha_7}
             (l^2)^{\alpha_8}  (p^2 +2p\cdot Q)^{\alpha_9}
          (l^2+2l\cdot Q)^{\alpha_{10}}
         }     \Biggr)
\label{defTopoF}
\end{multline}

\begin{eqnarray*}
 \Im( I_F(1,1,1,1,1,1,1,1,1,1,1) )&=&\textstyle
 \frac{ \pi}{s} (N_\varepsilon)^4
     \biggl[  \frac{1}{\varepsilon} \Bigl( \frac{3}{4} \zeta_4 \Bigr)
        + {\cal O}(\varepsilon^0)  \bigg]\\
 \Im( I_F(1,1,1,1,1,1,1,0,1,1,0) )&=&\textstyle
 \frac{ \pi}{s} (N_\varepsilon)^4
     \biggl[  \frac{1}{\varepsilon} \Bigl( -\frac{37}{8} \zeta_4 \Bigr)
        + {\cal O}(\varepsilon^0)  \bigg]\\
 \Im( I_F(1,0,1,1,1,1,1,0,1,2,0) )&=&\textstyle
 \frac{ \pi}{s} (N_\varepsilon)^4
     \biggl[   \Bigl( -\frac{27}{8} \zeta_4 \Bigr)
        + {\cal O}(\varepsilon)  \bigg]\\
 \Im( I_F(1,1,1,1,0,1,1,0,1,1,0) )&=&\textstyle
 \pi\, (N_\varepsilon)^4
     \biggl[    \frac{1}{\varepsilon^2} \Bigl( 1 \Bigr)
              + \frac{1}{\varepsilon} \Bigl( 15 -\zeta_2 -\zeta_3 \Bigr)\\
& &\qquad\textstyle
+  \Bigl( 139  -19 \zeta_2  -17\zeta_3  - 9\zeta_4  \Bigr)
           + {\cal O}(\varepsilon)  \bigg]\\
 \Im( I_F(1,1,1,1,0,0,1,1,1,1,0) )&=&\textstyle
 \pi\, (N_\varepsilon)^4
     \biggl[    \frac{1}{\varepsilon} \Bigl( -1 + \zeta_2 \Bigr)
               + \Bigl( -20 + 11 \zeta_2 + 11 \zeta_3  \Bigr)\\
& &\qquad\textstyle
+   \varepsilon \Bigl(  -237 + 93 \zeta_2 + 121 \zeta_3
                         + 37 \zeta_4 \Bigr)
              + {\cal O}(\varepsilon^2)  \bigg]\\
\Im( I_F(1,1,1,1,1,1,0,0,1,1,0) )&=&\textstyle
\pi\, (N_\varepsilon)^4
      \biggl[   \frac{1}{\varepsilon} \Bigl(
            -2 + \zeta_2  + \frac{13}{4} \zeta_3  -3\zeta_2\ln(2) \Bigr)
        + {\cal O}(\varepsilon^0)   \bigg]
\end{eqnarray*}

For integrals of topology G one is able to remove enough propagators
through the use of integration-by-parts relations to make these integrals
 cases of topologies A--F. Therefore no new primitives appear
for topology G.

\subsection{The evaluation of primitive integrals}

In this section we will discuss the calculation of primitive integrals
on the mass shell using a large mass expansion as an intermediate
representation. The general theory of Euclidean asymptotic expansions
  that is invoked was
developed in Refs. \cite{formalexpansions1,formalexpansions2} and in
practice the techniques of Ref.\cite{largemass} were used.
In many
physical applications the large mass expansion is truncated at a finite
order, however it has been used in recent works
\cite{largemassUntrunc,largemassUntrunc2} also in an untruncated form.

Before treating the more complicated 4-loop integrals we will
illustrate some basic features of the approach using a simple massive one-loop
integral as an example
\begin{equation} \label{prototypeoneloop}
    I(\alpha_1,\alpha_2)
  =  \int {\rm d}^Dk\;
       \frac{ 1}{(k^2)^{\alpha_1} [(k+Q)^2-M^2]^{\alpha_2} }
  =  \int {\rm d}^Dp\;
       \frac{ 1}{[(p+Q)^2]^{\alpha_1} (p^2-M^2)^{\alpha_2} }
\end{equation}
Again we do not write explicitly $+i\epsilon$ in the denominators
 to keep expressions compact.
For a large mass, $M^2 > Q^2$ one may expand this integral
 by making an ordinary Taylor expansion in the integrand:
\begin{eqnarray}
 \frac{1}{[(p+Q)^2]^{\alpha_1}} & \rightarrow  & \frac{1}{(p^2)^{\alpha_1}}
           \sum_{i=0}^\infty
           (-1)^i  \left( \frac{ 2p\cdot Q+Q^2}{p^2} \right)^i
           \frac{ \Gamma(\alpha_1+i)}{ \Gamma(\alpha_1)\; i!}
      \nonumber \\
 & = & \frac{1}{(p^2)^{\alpha_1}}
           \sum_{i=0}^\infty
           \sum_{k=0}^i
          (-1)^i \frac{ (2p\cdot Q)^k\; (Q^2)^{i-k} }{(p^2)^{i}}
            \frac{ \Gamma(\alpha_1+i)}{
                    \Gamma(\alpha_1)\; k! \; (i-k)! }
\end{eqnarray}
The remaining integral over $p$ is of the vacuum bubble type and after applying
a bubble tensor reduction to simplify the powers of $(p\cdot Q)$ in the
numerator
\begin{equation}
 (p\cdot Q)^{n}\rightarrow
     \begin{cases}
         (p^2 Q^2)^{\frac{n}{2}}
       \; \frac{n!}{2^{n}\left(n/2\right)!} \;
          \frac{\Gamma\left(D/2\right)}
               {\Gamma\left(D/2+n/2\right)}
       & \text{$n$ even},\\
         0 & \text{$n>0$ odd}
     \end{cases}
 \end{equation}
the bubble integral can be performed using Eq.~(\ref{1loopbubble}).
In the resulting expression one of the two sums can be performed using
the summation identity
\begin{equation} \label{sumidexample1}
     \sum_{k=0}^n \frac{ (-1)^k\; \Gamma(a+k)}{k!\; (n-k)!\;
    \Gamma(b+k)}
    = \frac{ \Gamma(a)\; \Gamma(b-a+n)}{n!\; \Gamma(n+b)\; \Gamma(b-a)
           }
\end{equation}
and obtain the well-known result
\begin{eqnarray}
I(\alpha_1,\alpha_2)
&=& i \pi^{(D/2)} (-1)^{-\alpha_1-\alpha_2} (M^2)^{D/2-\alpha_1-\alpha_2}
\frac{ \Gamma(D/2-\alpha_1 )}
     { \Gamma( \alpha_1 ) \; \Gamma( \alpha_2 )}\nonumber\\
 & &\qquad\qquad\qquad\qquad\qquad
\times\sum_{j=0}^{\infty} \left( \frac{ Q^2}{M^2} \right)^j
      \frac{ \Gamma(\alpha_1 + j)\; \Gamma(\alpha_1 +\alpha_2 +j-D/2) }{
     j!\; \Gamma(D/2+j) }
\label{Iexaple00}\\
&=&i \pi^{(D/2)} (-1)^{-\alpha_1-\alpha_2} (M^2)^{D/2-\alpha_1-\alpha_2}
       \frac{ \Gamma(D/2-\alpha_1 )\; \Gamma(\alpha_1+\alpha_2-D/2)}{
           \Gamma(D/2)\; \Gamma(\alpha_2)}\nonumber\\
 & &\qquad\qquad\qquad\qquad\qquad
     \times {}_2F_1 \left(
          \begin{array}{c} \alpha_1\; ,\; \alpha_1+\alpha_2-D/2
             \\ D/2 \end{array} \Biggr| \frac{ Q^2}{M^2}  \right)
\label{Iexaple0}
\end{eqnarray}
where ${}_2F_1$ is the Gauss hypergeometric function. To obtain the proper
value on the mass shell $Q^2/M^2 = 1$ within dimensional regularization,
one goes on-shell in an interval of the dimension $D$ where the integral
is convergent and continues the result in the number of space-time
dimensions.  Indeed, assuming that $D$ is chosen such that the sum is
convergent one can use the Gauss summation identity
\[  {}_2F_1 \left( \begin{array}{c} a\; ,\; b \\ c \end{array} \Biggr| 1\right)
 = \frac{ \Gamma(c)\; \Gamma(c-a-b)}{\Gamma(c-a)\; \Gamma(c-b)}
    \hspace{1.5cm}    (c\neq0,-1,\cdots,\; \Re(c-a-b)>0) \]
to obtain the on-shell expression Eq.~(\ref{1looponshell}).

Depending on the values of $\alpha_1$, $\alpha_2$, integral
$I(\alpha_1,\alpha_2)$ can have  specific IR divergences on the mass-shell
since $[(k+Q)^2-M^2]$ becomes $(k^2+2Q\cdot k)$.  One may use IR
power-counting in the small $k$ region to determine whether an IR
divergence occurs on the mass-shell. Here $1/(k^2+2Q\cdot k)$ counts for
only half the power of an ordinary massless line $1/k^2$. Since such an IR
divergence appears only on-shell, it manifests itself as
non-convergence of the large mass expansion starting exactly on the mass shell.

Using integration-by-parts recurrence relations it is possible to
choose an appropriate basis set of primitive integrals for which the
large mass expansion is convergent on-shell in $D=4-2\varepsilon$.
Nevertheless it is instructive to study the behaviour of IR divergent
integrals near the mass-shell.  Let us therefore consider integral
$I(1,2)$. Using the transformation rule for the hypergeometric function
\begin{eqnarray}
{}_2F_1\left( \begin{array}{c} a\; ,\; b \\ c \end{array}  \Biggr| z  \right)
 &  = & \frac{ \Gamma(c)\; \Gamma(c-a-b)}{\Gamma(c-a)\; \Gamma(c-b)}
{}_2F_1 \left( \begin{array}{c} a\; ,\; b \\ a+b-c+1 \end{array}
    \Biggr|1- z \right)
    \nonumber \\  \label{gausstransform}
 & &  + (1-z)^{c-a-b} \frac{\Gamma(c)\; \Gamma(a+b-c)}{\Gamma(a)\; \Gamma(b)}
 {}_2F_1 \left( \begin{array}{c} c-a\; ,\; c-b \\ c-a-b+1 \end{array}
     \Biggr|1- z \right)
\end{eqnarray}
one easily obtains the behaviour near the mass shell
\begin{equation} \label{i12nearshell}
 I(1,2)= \frac{ i \pi \; \Gamma(1+\varepsilon) }{2\varepsilon
      \; (M^2)^{1+\varepsilon}  }
             \left[ 1 -  (1-x)^{-2\varepsilon}
                 \frac{ \Gamma(1+2\varepsilon)\; \Gamma(1-\varepsilon)}{
                    \Gamma(1+\varepsilon) }    \right]
              + {\cal O}(1-x)
\end{equation}
where $x= Q^2/M^2$. On shell the term $(1-x)^{-2\varepsilon}$ is nullified
within dimensional regularization and for $x<1$ the threshold term can be
expanded in $\varepsilon$
\begin{equation} \label{polesonshell}
     \frac{-1}{2\varepsilon} \left[ 1- (1-x)^{-2\varepsilon} \right] =
             - \ln(1-x) + \varepsilon \ln^2(1-x)
       - \frac{2}{3} \varepsilon^2 \ln^3(1-x)
      + {\cal O} \left( \varepsilon^3\ln^4(1-x)  \right)
   \end{equation}
The logarithms $\ln^i(1-x)$ in Eq.~(\ref{polesonshell}) are translated
on-shell into a pole. Such logarithmic divergences appear for certain
individual diagrams of the muon decay calculations but they
cancel to all orders in $\varepsilon$ when the external leg corrections
are included.  One may therefore use dimensional regularization and
evaluate all integrals at threshold.

To make the connection with methods where one expands expressions in
$\varepsilon$ at an early stage in the calculation, we also expand the
$\Gamma$ functions of Eq.~(\ref{Iexaple0}) in $\varepsilon$
\begin{equation} \label{expandGamma}
        \Gamma(n+1+\varepsilon) = n!\; \Gamma(1+\varepsilon) \left[
              1+ \varepsilon S_1(n) + \frac{\varepsilon^2}{2}(
                   S_1^2(n)-S_2(n)) + {\cal O}(\varepsilon^3)\right]
\end{equation}
where the harmonic sums $S$ are defined as
\begin{equation}
      S_k(n) = \sum_{i=1}^{n} \frac{1}{i^k}
\end{equation}
In this way one obtains for integral $I(1,2)$
\begin{eqnarray}
    I(1,2)&=&
   \frac{  -i \pi \;\Gamma(1+\varepsilon) }{ (M^2)^{1+\varepsilon} }
 \; \sum_{j=0}^{\infty}
     \frac{x^j}{1+j} \left[ 1+2\varepsilon S_1(j) +2\varepsilon^2 S^2_1(j)
          + \frac{\varepsilon}{1+j} +{\cal O}(\varepsilon^2) \right]\nonumber\\
    &=&\frac{ -i \pi \; \Gamma(1+\varepsilon) }{(M^2)^{1+\varepsilon}  }
      \Biggl[
             - \ln(1-x) + \varepsilon \ln^2(1-x)
      - \frac{2}{3} \varepsilon^2 \ln^3(1-x)
   \nonumber\\
 & &\qquad\qquad\qquad\qquad\qquad
        - 2\zeta_2 \varepsilon^2  \ln(1-x)
          + \varepsilon \zeta_2
            +{\cal O}(\varepsilon^2) + {\cal O}(1-x)  \Biggl]
\label{Iablogdivergence}
\end{eqnarray}
in agreement with Eqs.(\ref{i12nearshell}) and (\ref{polesonshell}).
However, one needs to consider the divergent logarithms to all orders in
$\varepsilon$ to obtain the proper on-shell value in dimensional
regularization.  As integration-by-parts recurrence relations are used to
reduce the number of different integrals that are required, one may
conveniently choose a set of primitive integrals that are free from
threshold singularities. These well-behaved primitive integrals can then
be calculated via a convergent large mass expansion that can be truncated
at a fixed order in $\varepsilon$. This is the general strategy followed in
the more complicated multiloop integrals.  The only caveat is that, after
applying recurrence relations, poles may appear in the coefficients
multiplying the primitive integrals and these primitive integrals then
have to be evaluated to a higher order in $\varepsilon$ than na\"\i vely
expected.

It follows from Eq. (\ref{gausstransform}) that integral
  $I(\alpha_1,\alpha_2)$ has general on-shell divergences of the form
\begin{equation} \label{oneloopthreshold}
       \lim_{x\rightarrow1}
       \left[(1-x)^{(-m-2\varepsilon)} \right]
\end{equation}
which are nullified\footnote{Studies of the various multiloop integrals
that are needed for the muon-decay calculation reveal a slightly more
general threshold behaviour
\[
 \lim_{x\rightarrow1}
 \left[(1-x)^{(-m-2n\varepsilon)} \right]
\]
where $m$ is an integer and $n$ is generally the number of independent
loop momenta that when they to zero give rise to an infrared divergence
on-shell.  The absence of unregularized singularities $(1-x)^{-m}$ for
$m>0$ is important for the validity of on-shell recurrence relations.}
in $D>4+m$ dimensions. Note that for $x>1$ the terms $+i\epsilon$ in
  the denominators of the propagators that were implicit
  in Eq. (\ref{prototypeoneloop})
   are restored by the substitution
  $x\rightarrow x +i\epsilon$ in Eq. (\ref{oneloopthreshold}) to give the
  contributions from channels that open up above threshold, but for the
  present work this is not needed.

\subsubsection{Example 2}

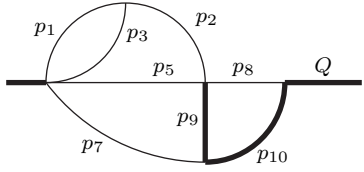
\begin{figure}
\begin{center}

\begin{picture}(200,70)(0,0)

\SetOffset(0,0)

 \SetWidth{2}
 \Line(30,30)(45,30)
 \Line(135,30)(164,30)
 \CArc(105,30)(30,270,360)
 \Line(105,0)(105,30)
 \SetWidth{.5}
  \Line(45,30)(140,30)
 \CArc(45,60)(30,270,360)
 \CArc(105,75)(75,217,270)
 \CArc(75,30)(30,0,180)

 \Text(45,50)[m]{\scriptsize $p_1$}
 \Text(106,54)[m]{\scriptsize $p_2$}
 \Text(80,48)[m]{\scriptsize $p_3$}
 \Text(90,35)[m]{\scriptsize $p_5$}
 \Text(63,6)[m]{\scriptsize $p_7$}
 \Text(99,16)[m]{\scriptsize $p_9$}
 \Text(120,35)[m]{\scriptsize $p_8$}
 \Text(131,2)[m]{\scriptsize $p_{10}$}
 \Text(150,36)[m]{\scriptsize $Q$}


\end{picture}
\end{center}

\caption{  \label{primitiveB4}
Integral $I_B(1,1,1,0,1,0,1,1,1,1,0)$. Thin lines correspond to massless
scalar propagators and thick lines  correspond to massive scalar
 propagators with mass $M$.   The momenta that flow through the propagators
 are labeled with the line numbers defined for main topology B,
  Eq.(\ref{defTopoB}).
 }
\end{figure}

Let us now consider the calculation of a primitive integral
$I_B(1,1,1,0,1,0,1,1,1,1,0)$ (see Fig.\ref{primitiveB4}) using the large
mass expansion as an intermediate representation.  The basic steps
followed will be similar to the steps in the derivation of
Eq.~(\ref{Iexaple0}). However, the sums that the Taylor expansions
introduce  cannot be reduced at the level of $\Gamma$ functions because
insufficient generalizations of Eq.~(\ref{sumidexample1}) are known.  One
is therefore forced to reduce sums after expanding the $\Gamma$ functions
in $\varepsilon$.  What makes this approach feasible is that the sums over
harmonic functions, $S$, encountered in the present calculation can be
reduced to expressions given in terms of a small number of higher order
sums. Eventually one is only interested in the value of integrals
on-shell.  For integrals with more than one massive line one may therefore
choose to make a large mass expansion in terms of the mass of only one of
the propagators, and to keep the masses of other propagators on-shell, if
this leads to less complicated intermediate expressions.

The problem of reducing harmonic sums also appears in calculations for
deep-inelastic lepton-nucleon scattering, where one deals with a light
cone expansion instead of the large mass expansion.  Large collections of
summation identities of the type that are needed for the present work can
be found in the literature  \cite{finitesums1,sumsjos}.  Although
summation reduction algebra is rather involved, relatively
compact final expressions for the coefficients of the large mass
expansion are obtained. On the mass-shell the sums can be reduced to
known mathematical constants.

We take the momentum $Q$ off-shell, $ s = Q^2<M^2$, and perform a large
mass expansion by expanding the integrand as
\begin{eqnarray}
\frac{1}%
{p_1^2\,p_2^2\, p_3^2\, p_5^2\, p_7^2\, p_8^2\,(p_9^2-M^2)(p_{10}^2-M^2)}
&\rightarrow&
   \frac{1}{p_1^2\,p_2^2\, p_3^2\, p_5^2\, p_7^2}
            \left[ \frac{1}{ p_8^2\,(p_9^2-M^2) (p_{10}^2-M^2)}
                             \right]_{\{p_2,p_5,p_7,Q\}}
      \nonumber\\
& & + \left[ \frac{1}{p_1^2\,p_2^2\, p_3^2\, p_5^2\, p_7^2\, p_8^2\,
         (p_9^2-M^2) (p_{10}^2-M^2)}
      \right]_{\{Q\}}
\label{expandB4}
\end{eqnarray}
where we use the notation that the terms between square brackets are
expanded as an infinite Taylor series around the point where the momenta
between curly brackets are zero. Here momentum $p_i$ is understood to be
the momentum that flows through line $i$ as defined for main topology B,
Eq.~(\ref{defTopoB}).  After performing the indicated Taylor expansion
in $Q$, the
last term on the rhs of (\ref{expandB4}) produces integrals of the 4-loop
massive bubble type and therefore does not contribute to the imaginary
part.  The first term on the rhs of (\ref{expandB4}) produces
integrals that are a product of a 3-loop massless propagator-type integral
and  a 1-loop bubble integral, all coupled by tensor numerators.  After
the necessary tensor reductions, one can evaluate these integrals in terms
of $\Gamma$-functions using Eqs.(\ref{1loopbubble}) and
(\ref{1loopmassless}).

A representation for the imaginary part of
integral  $I_B(1,1,1,0,$ $1,0,1,1,1,1,0)$ is obtained, in this way,
as a multiple sum over a product
of $\Gamma$ functions.  After expanding the $\Gamma$ functions in
$\varepsilon$ as in Eq.~(\ref{expandGamma}) one can reduce the sums over
the functions $S_i(k)$ to a basis of independent nested sums to obtain the
off-shell expression

\begin{multline}
\Im( I_B(1,1,1,0,1,0,1,1,1,1,0) ) =  \pi
    \left( N_\varepsilon \right)^4
       \left( \frac{s}{M^2}\right)^\varepsilon \sum_{k=1}^{\infty}
       \left(\frac{s}{M^2}\right)^k  \Biggl\{
   \frac{1}{\varepsilon } \biggl[  \frac{1}{k} -\frac{1}{k+1} \biggr]
\\
   + \biggl[  \frac{16}{k}-\frac{16}{k+1}
         -\frac{1}{k^2} +4 \frac{S_1(k)}{k}
         -4\frac{S_1(k+1)}{k+1} -2 \frac{S_1(k)}{k^2} \biggr]
 + \varepsilon \biggl[ \frac{157-8\zeta_2}{k} + \frac{-157+8\zeta_2}{k+1}
\\
      -\frac{16}{k^2}
       +\frac{1}{k^3}
   -\frac{2}{(k+1)^3}  +64\frac{S_1(k)}{k} -64\frac{S_1(k+1)}{k+1}
  -26\frac{S_1(k)}{k^2}  +2\frac{S_1(k)}{k^3}  +6\frac{S_2(k)}{k}
\\
      -6\frac{S_2(k+1)}{k+1}
     -6\frac{S_2(k)}{k^2}  +8\frac{S_1^2(k)}{k}
      -8 \frac{S_1^2(k+1)}{k+1} -6\frac{S_1^2(k)}{k^2} \biggr]
  + O( \varepsilon^2 ) \Biggr\}
\end{multline}
where $N_\varepsilon$ is defined in Eq.~(\ref{defnormloop}).  On-shell
$s/M^2 = 1$ and the sum over $k$ can be performed analytically to give the
following result
\begin{multline}
\textstyle
 \Im( I_B(1,1,1,0,1,0,1,1,1,1,0) ) = \pi\, (N_\varepsilon)^4 \biggl[
            \frac{1}{\varepsilon}
                   +\Bigl( 20 - \zeta_2 - 4\zeta_3 \Bigr)
         + \varepsilon \Bigl( 237 - 24 \zeta_2\\
       - 53 \zeta_3 - \frac{67}{2}\zeta_4 \Bigr)
           + {\cal O}(\varepsilon^2) \biggr]
\end{multline}
where the various infinite sums that appear are
\begin{eqnarray} \textstyle \label{beginsumsex3}
 \sum_{i=1}^{\infty} \frac{1}{i^k} & = & S_k(\infty)
        = \zeta_k  \hspace{1cm} (k > 1)  \\ \textstyle
 \sum_{i=1}^{\infty} \frac{S_1(i)}{i^2} & \equiv & S_{2,1}(\infty) = 2 \zeta_3
 \\ \textstyle
 \sum_{i=1}^{\infty} \frac{S_1(i)}{i^3}
    &  \equiv  & \textstyle  S_{3,1}(\infty) = \frac{5}{4}\zeta_4
\\ \textstyle
 \sum_{i=1}^{\infty} \frac{S_2(i)}{i^2} & \equiv & S_{2,2}(\infty) =
        \textstyle \frac{7}{4} \zeta_4
\\ \textstyle \label{endsumsex3}
 \sum_{i=1}^{\infty} \frac{ S^2_1(i)}{i^2}  & = &
         \textstyle     \frac{17}{4}\zeta_4
\end{eqnarray}
and are known from the literature
\cite{sumsjos,infinitesums1,infinitesums2,infinitesums3}.

\subsection{Example 3}

\begin{figure}
\begin{center}

\begin{picture}(200,70)(0,0)

\SetOffset(0,0)
 \SetWidth{2}
 \Line(30,30)(45,30)
 \Line(105,30)(170,30)
 \SetWidth{.5}
 \CArc(45,59)(29,270,350)
 \Oval(75,30)(24,30)(0)
 \CArc(92,36)(36,238,350)
 \CArc(117,46.8)(40.2,238,335)

\Text(48,48)[m]{\scriptsize $p_1$}
 \Text(100,50)[m]{\scriptsize $p_2$}
 \Text(69,34)[m]{\scriptsize $p_3$}
 \Text(63,14)[m]{\scriptsize $p_4$}
 \Text(86,13)[m]{\scriptsize $p_5$}
 \Text(99,24)[m]{\scriptsize $p_6$}
 \Text(119,19)[m]{\scriptsize $p_7$}
 \Text(146,12)[m]{\scriptsize $p_8$}
 \Text(117,35)[m]{\scriptsize $p_9$}
 \Text(140,35)[m]{\scriptsize $p_{10}$}
 \Text(165,35)[m]{\scriptsize $Q$}


\end{picture}
\end{center}

\caption{  \label{primitiveB3A}
Integral $I_B(1,1,1,1,1,1,1,1,1,1,1)$.  Thin lines correspond to massless
scalar propagators and thick lines correspond to massive scalar
propagators with mass $M$. The momenta that flow through the propagators
are labeled with the line numbers defined for main topology B,
Eq.~(\ref{defTopoB}). A factor $p_7\cdot Q$ (line 11) in the numerator  is
not shown explicitly. }
\end{figure}
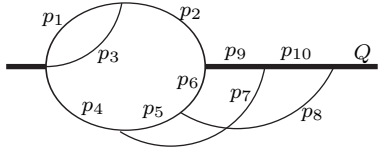

For integral $I_B(1,1,1,1,1,1,1,1,1,1,1)$ we proceed as in the previous
example and make a large mass expansion by expanding the integrand
\begin{eqnarray}
\frac{p_7\cdot Q}{p_1^2\,p_2^2\, p_3^2\, p_4^2\, p_5^2\, p_6^2\,
            p_7^2\, p_8^2\, (p_9^2-M^2) (p_{10}^2-M^2)}&\rightarrow&
\nonumber\\ & &\hspace{-3cm}\null\ \null
 \frac{p_7\cdot Q}{p_1^2\,p_2^2\, p_3^2\, p_4^2\, p_5^2\, p_6^2\,
            p_7^2\, p_8^2 }\left[ \frac{1}{(p_9^2-M^2) (p_{10}^2-M^2)}
        \right]_{\{p_9,p_{10}\}}
\nonumber\\ & &\hspace{-3cm}
+ \frac{1}{p_1^2\,p_2^2\, p_3^2\, p_4^2}
  \left[ \frac{p_7\cdot Q}{p_5^2\, p_6^2\, p_7^2\, p_8^2\,
        (p_9^2-M^2) (p_{10}^2-M^2)} \right]_{\{p_2,p_4,Q\}}
\nonumber\\ & &\hspace{-3cm}
+ \frac{p_7\cdot Q}{p_1^2\,p_2^2\, p_3^2\, p_4^2\, p_5^2\, p_7^2}
     \left[ \frac{1}{p_6^2\,p_8^2\, (p_9^2-M^2) (p_{10}^2-M^2)}
           \right]_{\{p_2,p_5,p_7,Q\}}
\nonumber\\ & &\hspace{-3cm}
 +\left[ \frac{p_7\cdot Q}{p_1^2\,p_2^2\, p_3^2\, p_4^2\, p_5^2\, p_6^2\,
            p_7^2\, p_8^2\, (p_9^2-M^2) (p_{10}^2-M^2)} \right]_{\{Q\}}
 \label{expandBprimitive}
 \end{eqnarray}
The last term is a 4-loop massive bubble and does not contribute to the
imaginary part.  After performing the indicated Taylor expansions,
integrals of the massless  propagator and massive bubble types
remain that can be evaluated in terms of $\Gamma$-functions using
Eqs.~(\ref{1loopbubble})--(\ref{bubblem00}).
The $\Gamma$-functions are then expanded in $\varepsilon$ and the
resulting sums over the functions $S_i(k)$ may be reduced
to a basis of independent nested sums.  Adding the
imaginary parts of the various contributions in
(\ref{expandBprimitive}) gives the off-shell expression
\begin{multline}
\Im(I_B(1,1,1,1,1,1,1,1,1,1,1)) =
 \left( N_\varepsilon \right)^4 \frac{\pi }{s}
     \sum_{k=1}^{\infty}
     \left(\frac{s}{M^2}
     \right)^k
     \left\{  \frac{1}{\varepsilon }
         \left[-\frac{\zeta_2}{k^2} - \frac{(-1)^k \zeta_2}{2 k^2}
               +\frac{\zeta_2 S_1(k)}{2 k}
          \right.
\right.\\
    \left.
   - \frac{(-1)^k S_{\tilde{2}}(k)}{k^2}
  +\frac{S_1(k)}{2 k^3} - \frac{S_2(k)}{4 k^2}
     +\frac{S_1(k) S_1(k)}{4 k^2}
     -\frac{S_{1,2}(k)}{k}  +\frac{S_1(k) S_2(k)}{2 k}
     -\frac{S_3(k)}{k}   +\frac{2}{k^4}   \right]\\
\left.
 +\frac{1}{\varepsilon}
  \ln^2\left(\frac{s}{M^2}\right) \left[ \frac{1}{4 k^2}
       - \frac{S_1(k)}{4 k} \right]
   +\frac{1}{\varepsilon}
      \ln\left(\frac{s}{M^2}\right) \left[ - \frac{3}{2 k^3}
      + \frac{ S_1(k)}{2k^2} +\frac{5 S_2(k)}{4 k} -\frac{S_1(k) S_1(k)}{4 k}
       \right]
 + {\cal O}(\varepsilon^0)  \right\}
\label{offshellIb}
\end{multline}
where
\begin{equation}
   S_{\tilde{k}}(i) =  \sum_{j=1}^{i} (-1)^j \frac{1}{j^k}
\end{equation}
and for higher order nested sums one defines recursively
      \begin{equation}
  S_{k,n,...,m}(i) = \sum_{j=1}^{i} \frac{ S_{n,...,m}(j)}{j^k}  ,
 \hspace{1cm}
  S_{\tilde{k},n,...,m}(i) = \sum_{j=1}^{i} (-1)^j \frac{ S_{n,...,m}(j)}{j^k}
      \end{equation}
where each of the indices $n,...,m$ can have a tilde (\ $\tilde{}$\ ) to
indicate an alternating sign component.  On-shell $s/M^2 = 1$ and
the infinite sum over $k$ can be performed to obtain
\begin{equation} \label{RESIb}
\textstyle
   \Im(I_B(1,1,1,1,1,1,1,1,1,1,1)) =
    \left( N_\varepsilon \right)^4 \frac{\pi}{s} \left\{
     -\frac{ 17 \zeta_4}{16 \varepsilon}
        + {\cal O}(\varepsilon^0)  \right\} \hspace{4cm}
  \end{equation}
To illustrate the delicate cancellation for $s/M^2 = 1$ between the
various divergent sums in Eq.(\ref{offshellIb}) let us also quote the
result for the sum over $k$ from 1 up to $y$ instead of infinity
(such that Eq.~(\ref{RESIb}) corresponds to $y \rightarrow \infty $).
\begin{multline}
\Im(I_B(1,1,1,1,1,1,1,1,1,1,1)) =
    \left( N_\varepsilon \right)^4  \frac{\pi }{s}
   \frac{1}{\varepsilon}
    \left\{
  - \zeta_2 S_2(y) -\frac{1}{2} \zeta_2 S_{\tilde{2}}(y) -
      S_{\tilde{2},\tilde{2}}(y) +3 S_{3,1}(y)
\right. \\
\left.       +\frac{1}{2}S_{2,2}(y)
  -S_{2,1,1}(y) +S_1(y) \Bigl[S_{2,1}(y)-2 S_3(y) \Bigr]
    +\frac{1}{2} S_{1,1}(y) \Bigl[ \zeta_2 - S_2(y) \Bigr]
 \right\}  + {\cal O}(\varepsilon^0)
 \end{multline}
Here one encounters in addition to the sums Eqs. (\ref{beginsumsex3})
      -- (\ref{endsumsex3}) the known infinite sums
\begin{eqnarray}
      S_{\tilde{k}}(\infty) & = & \textstyle
              - \left( 1-\frac{1}{2^{k-1}} \right)
                          \zeta_k  \hspace{1cm} (k > 1) \\
   S_{\tilde{2},\tilde{2}}(\infty) & = &\textstyle  \frac{13}{16}\zeta_4     \\
      S_{2,1,1}(\infty) & = &  3\zeta_4
\end{eqnarray}
and $S_1(y)$ and $S_{1,1}(y)$ diverge logarithmically for large $y$.

\subsubsection{Example 4}

As a last example the calculation of integral
$I_B(1,1,1,1,1,1,0,0,1,2,0)$ that produces a constant ${\rm Li}_4(1/2)$,
 where ${\rm Li}_k$ is the Polylogarithm \cite{polylog} is discussed.
 As before a large mass expansion is
performed
\begin{eqnarray}
\frac{1}{p_1^2\, p_2^2\, p_3^2\, p_4^2\, p_5^2\, p_6^2\,
             (p_9^2-M^2) (p_{10}^2-M^2)^2 }&\rightarrow&
  \frac{1}{p_1^2\,p_2^2\, p_3^2\, p_4^2}
   \left[ \frac{1}{p_5^2\, p_6^2\,
        (p_9^2-M^2) (p_{10}^2-M^2)^2} \right]_{\{p_2,p_4,Q\}}
\nonumber\\
&+& \left[  \frac{1}{p_1^2\, p_2^2\, p_3^2\, p_4^2\, p_5^2\, p_6^2\,
             (p_9^2-M^2) (p_{10}^2-M^2)^2 } \right]_{\{Q\}}
\end{eqnarray}
where the last term produces  4-loop massive bubble integrals and does
 not contribute to
the imaginary part.  After performing the indicated Taylor expansion there
remains a product of a massless two loop propagator integral and a massive
2-loop bubble integral that can be evaluated in terms of
$\Gamma$-functions using Eqs.(\ref{1loopbubble})--(\ref{bubblem00}).
After expanding the $\Gamma$-functions in $\varepsilon$ and reducing the
sums over the functions $S_i(k)$ to a basis of independent nested sums one
obtains
\begin{multline}
\Im( I_B(1,1,1,1,1,1,0,0,1,2,0) ) =  \frac{\pi }{s}
    \left( N_\varepsilon \right)^4
       \left( \frac{s}{M^2}\right)^{2 \varepsilon} \sum_{k=1}^{\infty}
       \left(\frac{s}{M^2}\right)^k  \Biggl\{
   \frac{1}{\varepsilon } \biggl[
      -\frac{2}{k^3} + \frac{ (-1)^k \zeta_2}{k}
 \\
            +2 \frac{(-1)^k S_{\tilde{2}}(k)}{k} \biggr]
     + \biggl[
         -\frac{8}{k^3}  + \frac{3}{k^4}
            + \frac{(-1)^k\, \zeta_2\,}{k^2}
        -6 \frac{S_1(k)}{k^3}   +2 \frac{(-1)^k \, S_{\tilde{2}}(k)}{k^2}
         + \zeta_2\frac{(-1)^k \, S_1(k)}{k}
 \\
               +2 \frac{(-1)^k\, S_{\tilde{2}}(k)\, S_1(k)}{k}
      +4 \zeta_2 \frac{(-1)^k}{k}  +8 \frac{(-1)^k\, S_{\tilde{2}}(k)}{k}
      - \zeta_3\frac{ (-1)^k}{k}
           + \frac{9}{2}\frac{(-1)^k\, S_2(k)\, S_{\tilde{1}}(k)}{k}\\
        +2 \frac{(-1)^k\, S_{\tilde{2},1}(k)}{k}
        + \frac{3}{2} \frac{(-1)^k\, S^3_{\tilde{1}}(k)}{k}
         -9 \frac{(-1)^k\, S_{\tilde{1},\tilde{1},\tilde{1}}(k)}{k}
       \biggr] + {\cal O}(\varepsilon^2)   \Biggr\}
\end{multline}

On-shell the infinite sum over $k$ can be performed to obtain
\begin{multline}
\textstyle
\Im( I_B(1,1,1,1,1,1,0,0,1,2,0) ) =  \frac{ \pi}{s} (N_\varepsilon)^4
\biggl[
\frac{1}{\varepsilon}
\Bigl(\frac{5}{4} \zeta_3 - 3  \zeta_2 \ln(2)  \Bigr)
 + \Bigl(  5\zeta_3 - 12 \zeta_2 \ln(2)
 \\
\textstyle
 - \frac{49}{8} \zeta_4 + 6{\rm Li_4}(1/2) + \frac{1}{4} \ln^4(2)  \Bigr)
           + {\cal O}(\varepsilon) \biggr]
\end{multline}
for which all necessary infinite sums are known
\cite{sumsjos,infinitesums1,infinitesums2,infinitesums3}
\begin{eqnarray}
       S_{\tilde{1}}(\infty) & = &  -\ln(2)
   \\
        S_{\tilde{1},1}(\infty) & = &  \textstyle
          -\frac{1}{2}\zeta_2 +\frac{1}{2}\ln^2(2)
     \\
  S_{\tilde{1},\tilde{2}}(\infty)  & = & \textstyle
             \frac{13}{8}\zeta_3 - \zeta_2 \ln(2)
      \\
      S_{\tilde{2},2}(\infty) & = & \textstyle
            - 4\, {\rm Li}_4(1/2) - \frac{1}{6}\ln^4 (2)
              +\zeta_2 \ln^2(2)  -\frac{7}{2}\zeta_3 \ln(2)
                +\frac{51}{16} \zeta_4
            \\
 S_{\tilde{1},\tilde{3}}(\infty) & = & \textstyle
                  2\, {\rm Li}_4(1/2)
            + \frac{1}{12} \ln^4 (2) -\frac{1}{2} \zeta_2 \ln^2(2)
                + \frac{3}{4} \zeta_3 \ln(2) -\frac{1}{2} \zeta_4
      \\
 S_{\tilde{1},1,\tilde{2}}(\infty) & = & \textstyle
     2 \, {\rm Li}_4(1/2) +\frac{1}{12} \ln^4 (2) + \frac{1}{8} \zeta_3 \ln(2)
            -\frac{1}{2} \zeta_4
           \\
 S_{\tilde{1},\tilde{2},1}(\infty) & = & \textstyle
           3\, {\rm Li}_4(1/2) +\frac{1}{8} \ln^4 (2)
                        -\frac{3}{4}\zeta_2 \ln^2 (2)
                         +\frac{5}{8} \zeta_3 \ln(2)
                         -\frac{9}{16} \zeta_4
            \\
  S_{\tilde{1},\tilde{1},\tilde{1},\tilde{1}}(\infty)
       & = & \textstyle
                      \frac{1}{24} \ln^4(2)
                  +\frac{1}{4} \zeta_2 \ln^2 (2)
                  +\frac{1}{4}\zeta_3 \ln(2)
                  +\frac{9}{16} \zeta_4
\end{eqnarray}

\section{The Theoretical Uncertainty}
\label{sec:TheoUncert}

With the incorporation of the 2-loop QED corrections in $\Delta q$ the
largest missing theoretical contributions come from three possible
 sources. The first is the
hadronic uncertainty \cite{muonhad} conservatively estimated to introduce
an error of $2\times10^{-8}$ in $\Delta q$ which introduces a relative
error of $10^{-8}$ in the extracted value of $G_F$.

By examining the logarithms that appear at tree- and 1-loop
levels one would expect the leading unknown 2-loop QED corrections
to be proportional to\\
$(\alpha/\pi)^2(m_e^2/m_\mu^2)\ln^2(m_e^2/m_\mu^2)=1.4\times10^{-8}$
which would need a coefficient of roughly 140 to introduce a 1\,ppm
error in $G_F$. At tree- and 1-loop level the coefficient of the leading
logarithm is 12. Allowing for a coefficient as large as 24 gives an
an estimate of the theoretical error from this source of
$1.7\times10^{-7}$ in the value of $\delta G_F/G_F$.

The 3-loop QED corrections may be estimated in the same way as
the 2-loop corrections were, before they had been actually
calculated, by assuming them to the equal to the known leading
logarithm of next order. In this case that translates a relative error
of $1.4\times10^{-7}$ in $G_F$.

Overall the theoretical uncertainty arising from missing higher-order
corrections should not exceed a few parts in $10^{7}$.

\section{Experimental Uncertainties}
\label{sec:ExpUncert}

On the experimental side the accuracy of the known values for the
muon lifetime, $\tau_\mu$, the muon mass $m_\mu$ and the muon neutrino
mass can each exert a significant effect on the extracted value of
$G_F$. From Eqs.(\ref{eq:Gamma0}) and (\ref{eq:Deltaq0}), the change in
$G_F$ in response to changes in these quantities is given by
\begin{equation}
\frac{\delta G_F}{G_F}=-\frac{1}{2}\frac{\delta\tau_\mu}{\tau_\mu}
                       -\frac{5}{2}\frac{\delta m_\mu}{m_\mu}
                       +4\frac{m_{\nu_\mu}^2}{m_\mu^2}.
\end{equation}

The measured value of the muon lifetime,
$\tau_\mu=(2.19703\pm0.00004)\,\mu$s \cite{PDG}
is currently the source of the dominant error.
Experiments are planned to reduce this error to $\pm4$\,ps
Brookhaven National Laboratory \cite{BNL} and
$\pm2$\,ps at the Paul Scherrer Institute \cite{PSI}.
A new measurement is also expected at the
Rutherford-Appleton Laboratory \cite{RAL}.
It is therefore likely that uncertainty on $G_F$ coming from this
source will be reduced to somewhere in the range 0.5--1\,ppm.

The measured value of $m_\mu$ expressed in unified atomic mass
units is \cite{PDG}
\[
m_\mu=(0.113428913\pm0.000000017)\,{\rm u}
\]
corresponding to
an accuracy of 0.15\,ppm.
A new measurement of Planck's constant, $h$, \cite{Planck}
means that this can now be converted to units of MeV without introducing
additional errors. The current error on $m_\mu$ leads to a 0.38\,ppm
uncertainty on the extracted value of $G_F$. Some reduction can be
expected in the error on the mass whose effect would then be
insignificant.

If $m_{\nu_\mu}$ is assumed to be non-zero
then setting it to the current upper bound of $m_{\nu_\mu}\le170$\,keV
shifts the extracted value of $G_F$ by 10\,ppm. This upper bound is
expected to be reduced to some where below 30\,keV by studying
the decays of pions in in flight at the Brookhaven muon storage ring
\cite{Prisca}. This would affect the extracted value of $\delta G_F/G_F$
at the level of 0.3\,ppm.

\section{Weak Corrections to the Muon lifetime}
\label{sec:WeakCorr}

The weak corrections to the muon lifetime may be encapsulated in a
quantity $\Delta r$ defined by the relation
\begin{equation}
\frac{G_F}{\sqrt{2}}=\frac{g^2}{8 M_W^2}\left(1+\Delta r\right)
\label{eq:GFdef}
\end{equation}
As with Eq.(\ref{eq:alphadef}) the parameters on the left-hand side of
Eq.(\ref{eq:GFdef}) represent experimentally-measurable quantities
and those on the right-hand side are the renormalized
parameters in whatever renormalization scheme has been chosen. Thus
$g$ and $M_W^2$ are the renormalized $SU(2)_L$ coupling constant
and square of the renormalized $W$ boson mass respectively.

$\Delta r$ will be defined through Eq.(\ref{eq:GFdef}) in such a way that
that Eq.(\ref{eq:QEDcorr}) is true exactly.

In an analogous way to $\Delta q$, $\Delta r$ can be expressed as a
power series in $\alpha_r$
\begin{equation}
\Delta r=\sum_{i=0}^\infty\Delta r^{(i)}
\end{equation}
where, once again, the index $i$ indicates the power of $\alpha_r$ that
appears in $\Delta r^{(i)}$.

Sirlin \cite{Sirlin78,Sirlin80} has described a strategy
that, starting from the full electroweak theory, makes the separation
of contributions to $\Delta q$ and $\Delta r$ automatic at least up
to ${\cal O}(\alpha m_\mu^2/M_W^2)$. In diagrams exhibiting
IR divergences, the photon propagator is replaced by
\begin{equation}
\frac{1}{k^2}\longrightarrow\left\{\frac{1}{k^2}-\frac{1}{k^2
                                  -\Lambda^2}\right\}
                           +\frac{1}{k^2-\Lambda^2}.
\label{eq:photonsplit}
\end{equation}
where it generally convenient to take $\Lambda=M_W$.
The term in curly brackets is simply the original photon propagator with
a Pauli-Villars regulator. It has the same IR behaviour
and gives contributions that are identical to those of Fermi
theory up to ${\cal O}(\alpha m_\mu^2/\Lambda^2)$ and thus are contained
in $\Delta q$. The second term in (\ref{eq:photonsplit}) gives
contributions that retain the original UV behaviour but are free from
IR singularities and therefore belong in $\Delta r$.
It should  be
noted that, in contrast to neutral currents, it is not generally the
case that charged current processes can be separated into QED and weak
parts in a gauge-invariant manner.
Sirlin \cite{Sirlin84} has also discussed in general how the strategy
may be applied at the 2-loop level.

Note  that the photons in the second term on the right-hand side of
(\ref{eq:photonsplit})
have a mass larger than the muon mass such that in contributions to
$\Delta r$ no ``photons" of this type can appear in the final state.
This means that one does not need to perform many body phase space
integrals for the calculation of $\Delta r$ and in practice the matrix
element can be evaluated even at zero external momenta.
In contrast the QED corrections to muon decay in the Fermi theory do
involve many body phase space integrals, but these corrections contain
only one mass scale, ignoring the electron and neutrino masses, and
the relatively simple Fermi contact interaction.

It is known that
\begin{equation}
\Delta r^{(0)}=\frac{3m_\mu^2}{10M_W^2}
              +{\cal O}\left(\frac{m_e^2}{m_\mu^2}\right)
\end{equation}
that is due to $W$ propagator effects and shifts the extracted value
of $G_F$ by 0.52\,ppm.
A factor $(1+\Delta r^{(0)})$ is traditionally included
in Eq.(\ref{eq:QEDcorr}) that is used to define $G_F$ and up to now
this has not been of any significance. It is, however, clearly
inappropriate in the definition of $G_F$ as it does not arise from the
Fermi theory Lagrangian (\ref{eq:FullLagrangian}) and is most
properly included with the weak corrections. It is striking that at the
1\,ppm level experiments performed at low energy are sensitive to the
finite range effects of the $W$ boson.

\subsection{1-loop Electroweak Corrections to $\Delta r$}

The 1-loop corrections to the muon lifetime were first calculated in
the Standard Model of electroweak interactions by Sirlin\cite{Sirlin80}.
In a general renormalization scheme $\Delta r^{(1)}$ can be written
\cite{ZMass4}
\begin{eqnarray}
\Delta r^{(1)}&=&-\frac{\Pi_{WW}^{(1)}(0)}{M_W^2}
  +\frac{g^2}{16\pi^2}\left\{ 4\left(\Delta-\ln\frac{M_Z^2}{\mu^2}\right)
  +6+\left(4+c_\theta^2-\frac{6c_\theta^2}{s_\theta^2}\right)
     \ln\frac{1}{c_\theta^2}\right\}\nonumber\\
      & &\qquad\qquad\qquad\qquad+\frac{g^2}{16\pi^2}
        \frac{5c_\theta^4-3s_\theta^4}{2s_\theta^2}\ln\frac{1}{c_\theta^2}
       -\frac{\delta M_W^{2(1)}}{M_W^2}+2\frac{\delta g^{(1)}}{g}
       \label{eq:Deltar1}\\
&=&-\frac{\Pi_{WW}^{(1)}(0)}{M_W^2}
  +\frac{g^2}{16\pi^2}\left\{ 4\left(\Delta-\ln\frac{M_Z^2}{\mu^2}\right)
  +\frac{12s_\theta^2-7}{2s_\theta^2}\ln\frac{1}{c_\theta^2}+6\right\}
\nonumber\\ & &\qquad\qquad\qquad\qquad
       -\frac{\delta M_W^{2(1)}}{M_W^2}+2\frac{\delta g^{(1)}}{g}
\end{eqnarray}
where all parameters are renormalized parameters in whatever
renormalization scheme has been chosen. $\Pi_{WW}^{(1)}(0)$ is
the 1-loop $W$ boson self-energy and $s_\theta$ and $c_\theta$ are,
respectively the sine and cosine of the weak mixing angle, $\theta_W$,
defined as in Ref.\cite{MaldeStuart1} so as to diagonalize the mass
matrix of renormalized $Z$ and photon fields.
In Eq.(\ref{eq:Deltar1}) the term in curly brackets comes from
vertex corrections and the following one from box diagram.
Ultraviolet divergences cancel in the overall expression.

\subsection{2-loop Electroweak Corrections to $\Delta r$}

Certain classes of contributions to $\Delta r$ are simply
related to the corresponding ones in $\delta\rho$ because they
only enter through the self-energies of the $Z$ and $W$ bosons.
In such cases
\begin{equation}
\Delta r=-\frac{c_\theta^2}{s_\theta^2}\delta\rho.
\label{eq:DelrDelrho}
\end{equation}
Contributions to $\delta\rho^{(2)}$ in the limit of large Higgs mass,
$M_H$, were obtained by van~der~Bij and Veltman \cite{BijVelt}.
Subsequently van~der~Bij and Hoogeveen \cite{Frank} calculated
corrections to $\delta\rho^{(2)}$ arising from a heavy fermion doublet.
Their results can be used to obtain the well-known asymptotic
expression
\begin{equation}
\delta\rho^{(2)}=3(2\pi^2-19)x_f^2
\label{eq:Frank}
\end{equation}
for large top quark mass, $m_t$, and for which
\[
x_t=\frac{\alpha}{16\pi s_\theta^2}\frac{m_t^2}{M_W^2}.
\]
It was this quadratic $m_t$-dependence in $\delta\rho$ that
provided some of the strongest constraints on the mass of the
top quark before it was directly observed. Thus it was the measurement
of $G_F$ was that allowed the top mass to be successfully predicted
from precision electroweak data.

Consoli, Hollik and Jegerlehner \cite{ConsHollJege} showed how to
combine (\ref{eq:Frank}) with 1PR corrections to
obtain the asymptotic dependence of $M_W$ on $m_t$. The analogous
result for $\Delta r^{(2)}$ as defined via Eqs.(\ref{eq:QEDcorr})
and (\ref{eq:GFdef}) is
\begin{equation}
\Delta r^{(2)}=\frac{9c_\theta^4}{s_\theta^4}x_f^2
              -\frac{3c_\theta^2}{s_\theta^2}(2\pi^2-19)x_f^2.
\end{equation}

At 2-loop order the contributions that are quartic in $m_t$ are
generated via the Yukawa couplings to fermions and therefore come
entirely from the scalar sector of the theory. These corrections are
given, keeping the full $M_H$ dependence, in
Refs.\cite{Barbieri1,Barbieri2,Fleischer1,Fleischer2,DegrFancGamb}.

Since it turns out that $m_t$ is roughly the same order as $M_Z$ it is to
be expected that the ${\cal O}(\alpha^2 m_t^2 M_Z^2)$ contributions to
$\Delta r^{(2)}$ could be similar in size to those that are quartic in
$m_t$ and has been borne out by direct calculation \cite{DegrFancGamb}.
These subleading corrections do not now just come from $W$ and
$Z$ self-energy diagrams alone and so are not simply related to
$\delta\rho$. The exact sensitivity of $\Delta r^{(2)}$ to $M_H$, without
making an expansion in $m_t$ has been studied in Ref.\cite{BaubWeig}.

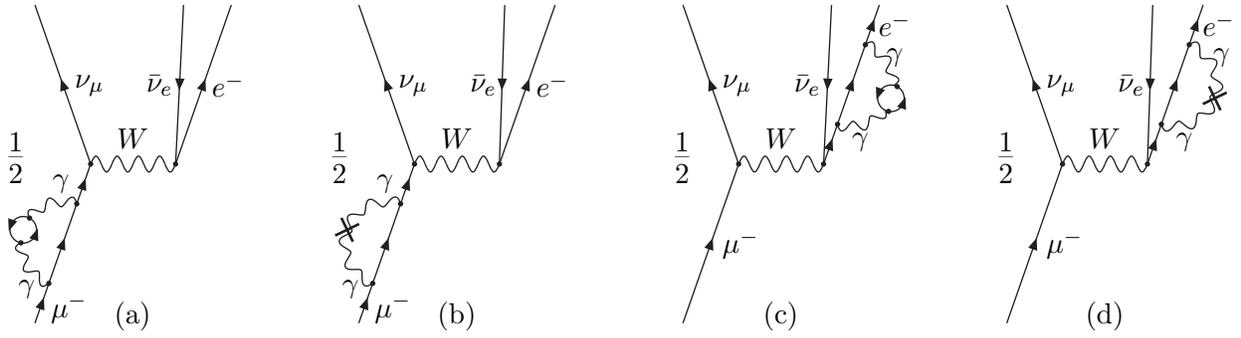
\begin{figure}
\begin{center}
\begin{picture}(75,120)(0,0)
\ArrowLine(0,0)(5.25,15)       \Vertex(5.25,15){1}
\Text(6,0)[bl]{$\mu^-$}
\ArrowLine(5.25,15)(15.75,45)  \Vertex(15.75,45){1}
\ArrowLine(15.75,45)(21,60)    \Vertex(21,60){1}
\ArrowLine(21,60)(0,120)
\Text(15,90)[l]{$\nu_\mu$}
\PhotonArc(10.5,30)(15.89,178.60,250.71){2}{2.5}
\Text(0,13)[r]{$\gamma$}
\ArrowArc(-4.39,35.21)(5,-100.35,61.77)  \Vertex(-5.29,30.29){1}
\ArrowArc(-4.39,35.21)(5,61.77,-100.35)  \Vertex(-2.02,39.62){1}
\PhotonArc(10.5,30)(15.89,70.71,142.83){2}{2.5}
\Text(13,53)[r]{$\gamma$}
\Photon(21,60)(53,60){3}{4}
\Text(37,66)[b]{$W$}
\ArrowLine(56,120)(53,60) \Vertex(53,60){1}
\Text(52,90)[r]{$\bar\nu_e$}
\ArrowLine(53,60)(74,120)
\Text(67,90)[l]{$e^-$}
\Text(-10,65)[bl]{1} \Text(-10,60)[bl]{--} \Text(-10,53)[bl]{2}
\Text(37,-3)[b]{(a)}
\end{picture}
\hfill
\begin{picture}(75,120)(0,0)
\ArrowLine(0,0)(5.25,15)       \Vertex(5.25,15){1}
\Text(6,0)[bl]{$\mu^-$}
\ArrowLine(5.25,15)(15.75,45)  \Vertex(15.75,45){1}
\ArrowLine(15.75,45)(21,60)    \Vertex(21,60){1}
\ArrowLine(21,60)(0,120)
\Text(15,90)[l]{$\nu_\mu$}
\PhotonArc(10.5,30)(15.89,70.71,250.71){2}{5.5}
\Text(0,13)[r]{$\gamma$}
\Text(13,53)[r]{$\gamma$}
\Photon(21,60)(53,60){3}{4}
\Text(37,66)[b]{$W$}
\ArrowLine(55,120)(53,60) \Vertex(53,60){1}
\Text(52,90)[r]{$\bar\nu_e$}
\ArrowLine(53,60)(74,120)
\Text(67,90)[l]{$e^-$}
\SetWidth{1}
\Line(-6.56,39.72)(-2.22,30.71)
\Line(-8.90,33.04)(0.11,37.38)
\Text(-10,65)[bl]{1} \Text(-10,60)[bl]{--} \Text(-10,53)[bl]{2}
\Text(37,-3)[b]{(b)}
\end{picture}
\hfill
\begin{picture}(75,120)(0,0)
\ArrowLine(0,0)(21,60)    \Vertex(21,60){1}
\ArrowLine(21,60)(0,120)
\Text(15,90)[l]{$\nu_\mu$}
\Photon(21,60)(53,60){3}{4}
\Text(37,66)[b]{$W$}
\ArrowLine(56,120)(53,60) \Vertex(53,60){1}
\Text(52,90)[r]{$\bar\nu_e$}
\ArrowLine(53,60)(58.25,75)        \Vertex(58.25,75){1}
\Text(15,30)[l]{$\mu^-$}
\ArrowLine(58.25,75)(68.75,105)    \Vertex(68.75,105){1}
\ArrowLine(68.75,105)(74,120)
\PhotonArc(63.5,90)(15.89,250.71,322.82){2}{2.5}
\Text(77,97)[bl]{$\gamma$}
\ArrowArc(78.39,84.79)(5,-100.35,61.77)  \Vertex(77.49,79.87){1}
\ArrowArc(78.39,84.79)(5,61.77,-100.35)  \Vertex(80.76,89.20){1}
\PhotonArc(63.5,90)(15.89,-1.4,70.71){2}{2.5}
\Text(64,72)[tl]{$\gamma$}
\Text(74,113)[l]{$e^-$}
\Text(-3,65)[bl]{1} \Text(-3,60)[bl]{--} \Text(-3,53)[bl]{2}
\Text(37,-3)[b]{(c)}
\end{picture}
\hfill
\begin{picture}(75,120)(0,0)
\ArrowLine(0,0)(21,60)    \Vertex(21,60){1}
\ArrowLine(21,60)(0,120)
\Text(15,90)[l]{$\nu_\mu$}
\Text(78,97)[bl]{$\gamma$}
\PhotonArc(63.5,90)(15.89,250.71,70.71){2}{5.5}
\Text(64,72)[tl]{$\gamma$}
\Photon(21,60)(53,60){3}{4}
\Text(37,66)[b]{$W$}
\ArrowLine(55,120)(53,60) \Vertex(53,60){1}
\Text(52,90)[r]{$\bar\nu_e$}
\ArrowLine(53,60)(58.25,75)        \Vertex(58.25,75){1}
\Text(15,30)[l]{$\mu^-$}
\ArrowLine(58.25,75)(68.75,105)    \Vertex(68.75,105){1}
\ArrowLine(68.75,105)(74,120)
\Text(74,113)[l]{$e^-$}
\SetWidth{1}
\Line(76.22,89.30)(80.56,80.29)
\Line(73.88,82.62)(82.89,86.96)
\Text(-3,65)[bl]{1} \Text(-3,60)[bl]{--} \Text(-3,53)[bl]{2}
\Text(37,-3)[b]{(d)}
\end{picture}
\caption{Infrared divergent external leg corrections contributing to
         muon decay at ${\cal O}(N_f\alpha^2)$.
\label{fig:VertexDiags}
        }
\end{center}
\end{figure}

\begin{figure}
\begin{center}
\null\hfill
\begin{picture}(80,120)(0,0)
\ArrowLine(0,0)(10,30)       \Vertex(10,30){1}
\ArrowLine(10,30)(10,90)     \Vertex(10,90){1}
\ArrowLine(10,90)(0,120)
\Text(7,15)[l]{$\mu^-$}     \Text(8,104)[l]{$\nu_\mu$}
\ArrowLine(80,120)(70,90)    \Vertex(70,90){1}
\ArrowLine(70,90)(70,30)     \Vertex(70,30){1}
\ArrowLine(70,30)(80,60)
\Text(78,106)[l]{$\bar\nu_e$} \Text(78,46)[l]{$e^-$}
\Photon(10,30)(33,30){-4}{2}    \Vertex(33,30){1}
\Photon(47,30)(70,30){4}{2}     \Vertex(47,30){1}
\ArrowArc(40,30)(7,180,0)
\ArrowArc(40,30)(7,0,180)
\Text(43,82)[t]{$W$}
\Photon(10,90)(70,90){4}{5.5}
\Text(21.5,37)[b]{$\gamma$} \Text(58.5,37)[b]{$\gamma$}
\Text(40,0)[b]{(a)}
\end{picture}
\hfill
\begin{picture}(80,120)(0,0)
\ArrowLine(0,0)(10,30)       \Vertex(10,30){1}
\ArrowLine(10,30)(10,90)     \Vertex(10,90){1}
\ArrowLine(10,90)(0,120)
\Text(7,15)[l]{$\mu^-$}     \Text(8,104)[l]{$\nu_\mu$}
\ArrowLine(80,120)(70,90)    \Vertex(70,90){1}
\ArrowLine(70,90)(70,30)     \Vertex(70,30){1}
\ArrowLine(70,30)(80,60)
\Text(78,106)[l]{$\bar\nu_e$} \Text(78,46)[l]{$e^-$}
\Photon(10,30)(70,30){-4}{5.5}
\Text(21.5,37)[b]{$\gamma$} \Text(58.5,37)[b]{$\gamma$}
\Text(43,82)[t]{$W$}
\Photon(10,90)(70,90){4}{5.5}
\SetWidth{1}
\Line(35,35)(45,25)
\Line(35,25)(45,35)
\Text(40,0)[b]{(b)}
\end{picture}
\hfill\null
\caption{The infrared divergent ${\cal O}(N_f\alpha^2)$ box diagrams
         occurring in muon decay at ${\cal O}(N_f\alpha^2)$.
\label{fig:BoxDiagrams}
        }
\end{center}
\end{figure}
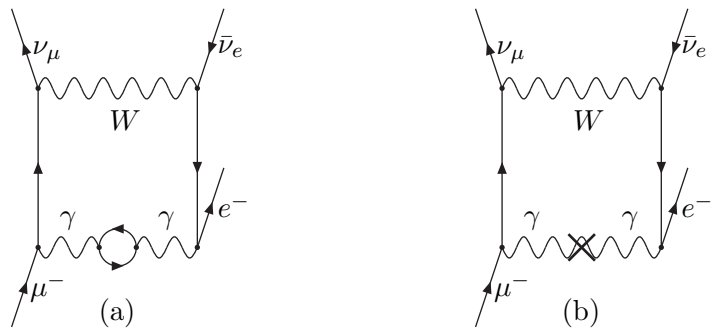

None of the corrections mentioned above encounters IR problems. All
contribute solely to $\Delta r$ and do not require the invocation
of (\ref{eq:photonsplit}) in order to separate their QED and weak
corrections.
Recently the ${\cal O}(N_f\alpha^2)$ corrections to muon decay have
been calculated \cite{MaldeStuart2}. These are all 2-loop corrections
containing a massless fermion loop and (\ref{eq:photonsplit}) must be
applied. In this case, however, the separation is particularly
natural. IR divergent contributions come from the Feynman diagrams
shown in Figs.~\ref{fig:VertexDiags} and \ref{fig:BoxDiagrams}.

IR divergent external leg corrections are shown in
Fig. \ref{fig:VertexDiags}(a)--(d) with
the `$\times$' in Figs.\ref{fig:VertexDiags}(b) and (d)
representing the fermionic contribution
to the 1-loop photon 2-point counterterm,
$(q^2 g_{\mu\nu}-q_\mu q_\nu)\;2\delta e^{(1f)}/e$.
These can be split into QED and weak
contributions using (\ref{eq:photonsplit}) in the manner explicitly
described in the appendix of Ref.\cite{Sirlin84}.
The QED part is already
included in Eq.(\ref{eq:Deltaq2}) for $\Delta q^{(2)}$.
Remarkably for the weak part, the dependence on the separation mass,
$\Lambda$, cancels in any renormalization scheme between the
pairs of diagrams Fig.\ref{fig:VertexDiags}(a)
with Fig.\ref{fig:VertexDiags}(c)
and Fig.\ref{fig:VertexDiags}(b) with Fig.\ref{fig:VertexDiags}(d).
The combined weak parts of the Feynman diagrams contains a simple pole
at $D=4$ that eventually cancels with other divergences from the weak
sector \cite{MaldeStuart2}.

The box diagrams, Fig.\ref{fig:BoxDiagrams},
both vanish identically due to a conspiracy in
the $\gamma$-matrix algebra of the fermion currents. The general result
is proportional to products of left-handed couplings with right-handed
couplings; the latter being zero for the $W$ boson.

In a general renormalization scheme the ${\cal O}(N_f\alpha^2)$ box
diagrams containing a virtual photon and a counterterm insertion on
the $W$ propagator produce contributions proportional
to the IR divergent 1-loop box diagram,
\begin{equation}
\begin{picture}(40,60)(0,0)
\ArrowLine(0,0)(5,15)
\ArrowLine(5,15)(5,45)
\ArrowLine(5,45)(0,60)
\ArrowLine(40,60)(35,45)
\ArrowLine(35,45)(35,15)     \Vertex(35,15){1}
\ArrowLine(35,15)(40,30)
\Photon(5,15)(35,15){-2}{5.5}
\Text(21.5,41)[t]{${\scriptstyle W}$}
\Photon(5,45)(35,45){2}{5.5}
\Text(21.5,11)[t]{${\scriptstyle \gamma}$}
\SetWidth{1}
\Line(1.5,41.5)(8.5,48.5)
\Line(1.5,48.5)(8.5,41.5)
\end{picture}
\quad\raisebox{30pt}{+}\quad
\begin{picture}(40,60)(0,0)
\ArrowLine(0,0)(5,15)        \Vertex(5,15){1}
\ArrowLine(5,15)(5,45)       \Vertex(5,45){1}
\ArrowLine(5,45)(0,60)
\ArrowLine(40,60)(35,45)     \Vertex(35,45){1}
\ArrowLine(35,45)(35,15)     \Vertex(35,15){1}
\ArrowLine(35,15)(40,30)
\Photon(5,15)(35,15){-2}{5.5}
\Text(11.5,41)[t]{${\scriptstyle W}$} \Text(28.5,41)[t]{${\scriptstyle W}$}
\Photon(5,45)(35,45){2}{5.5}
\Text(21.5,11)[t]{${\scriptstyle \gamma}$}
\SetWidth{1}
\Line(16.5,41.5)(23.5,48.5)
\Line(16.5,48.5)(23.5,41.5)
\end{picture}
\quad\raisebox{30pt}{+}\quad
\begin{picture}(40,60)(0,0)
\ArrowLine(0,0)(5,15)        \Vertex(5,15){1}
\ArrowLine(5,15)(5,45)       \Vertex(5,45){1}
\ArrowLine(5,45)(0,60)
\ArrowLine(40,60)(35,45)     \Vertex(35,45){1}
\ArrowLine(35,45)(35,15)     \Vertex(35,15){1}
\ArrowLine(35,15)(40,30)
\Photon(5,15)(35,15){-2}{5.5}
\Text(21.5,41)[t]{${\scriptstyle W}$}
\Photon(5,45)(35,45){2}{5.5}
\Text(21.5,11)[t]{${\scriptstyle \gamma}$}
\SetWidth{1}
\Line(31.5,41.5)(38.5,48.5)
\Line(31.5,48.5)(38.5,41.5)
\end{picture}
\quad\raisebox{30pt}%
{$=\left(2\frac{{\displaystyle\delta g^{(1f)}}}{{\displaystyle g}}
         -\frac{{\displaystyle\delta M_W^{2(1f)}}}{{\displaystyle M_W^2}}
   \right)$}\quad
\begin{picture}(40,60)(0,0)
\ArrowLine(0,0)(5,15)        \Vertex(5,15){1}
\ArrowLine(5,15)(5,45)       \Vertex(5,45){1}
\ArrowLine(5,45)(0,60)
\ArrowLine(40,60)(35,45)     \Vertex(35,45){1}
\ArrowLine(35,45)(35,15)     \Vertex(35,15){1}
\ArrowLine(35,15)(40,30)
\Photon(5,15)(35,15){-2}{5.5}
\Text(21.5,41)[t]{${\scriptstyle W}$}
\Photon(5,45)(35,45){2}{5.5}
\Text(21.5,11)[t]{${\scriptstyle \gamma}$}
\end{picture}
\label{eq:boxCTs}
\end{equation}
and diagrams would need to be treated by applying (\ref{eq:photonsplit}).
However in the $\overline{{\rm MS}}$ renormalization scheme adopted here
the sum of the counterterms that appear on the right hand side of
Eq.(\ref{eq:boxCTs}) vanishes. The superscript ${}^{(1f)}$ denotes
the 1-loop contribution to the counterterm from a light fermion species
\cite{MaldeStuart1,MaldeStuart2}.

\subsection{$G_F$ in the Analysis of Electroweak Data}

There is an important proviso that must be remembered when using
(\ref{eq:photonsplit}) to split up the radiative corrections.
The resulting sets of QED and weak corrections must be calculated
in a consistent renormalization scheme.
In the present paper, the $\overline{{\rm MS}}$ scheme with
$\mu=m_\mu$ was used for the QED corrections but in other circumstances
this need not be the appropriate choice.

In the analysis of electroweak data the numerical values of the
renormalized parameters of the theory are initially obtained from
a set of simultaneous equations for the physical observables
as calculated from the full electroweak Lagrangian.
To 1-loop \cite{ZMass4,Z0Pole} in the Standard Model
\begin{eqnarray}
\sqrt{4\pi\alpha}&=&e\left(
         1+\frac{1}{2}\Pi_{\gamma\gamma}^{(1)\prime}(0)
          +\frac{s_\theta}{c_\theta}\frac{\Pi_{Z\gamma}^{(1)}(0)}{M_Z^2}
               +s_\theta^2\frac{\delta g^{(1)}}{g}
               +c_\theta^2\frac{\delta g^{\prime(1)}}{g^\prime}\right),
\label{eq:alphaexp}\\
\tau_\mu^{-1}&=&\frac{m_\mu^5}{192\pi^3}
                \left\{\frac{\sqrt{2}g^2}{8M_W^2}
               (1+\Delta r^{(0)}+\Delta r^{(1)})\right\}^2
               (1+\Delta q^{(0)}+\Delta q^{(1)}),
\label{eq:taumuexp}\\
s_p&=&M_Z^2+\delta M_Z^2+\Pi_{ZZ}^{(1)}(M_Z^2).
\label{eq:spexp}
\end{eqnarray}
where $\Pi_{Z\gamma}^{(1)}(0)$ is the transverse part of the
1-loop $Z$-$\gamma$ mixing and $\delta g^{\prime(1)}$ is the
1-loop $U(1)$ coupling constant counterterm.
$s_p$ is the position of the pole on the complex plane associated with
the unstable $Z^0$ boson. A form similar to Eq.(\ref{eq:alphaexp})
can be found in Ref.\cite{Sirlin84}, Eq.(11a).

The quantities on the left hand side of
Eqs.(\ref{eq:alphaexp})--(\ref{eq:spexp}) are the physical observables
whose values are obtained by experiment and those on the right hand side
are the renormalized parameters and counterterms in the particular
renormalization scheme that has been chosen which, for the purposes of
this discussion, will be assumed to be the $\overline{{\rm MS}}$ scheme.
The subscript ${}_r$ has been dropped for the renormalized parameters.
For the analysis of electroweak data obtained near the $Z^0$ resonance
it is convenient and natural to take for the 't~Hooft mass, $\mu=M_Z$.
Upon solving the equations Eqs.(\ref{eq:alphaexp})--(\ref{eq:spexp})
one finds $\alpha_r\sim1/128$ and thus the running of $\alpha$ is
automatically incorporated in a self-consistent manner at
tree-level which avoids the need to resum large logarithms as is
required when the on-shell renormalization scheme is used. It also has
the consequence that the quantity in curly brackets in
Eq.(\ref{eq:taumuexp}) must also be evaluated at $\mu=M_Z$. This can
be implemented by substituting $\alpha_r=\alpha_e(M_Z^2)$ in the
formulas (\ref{eq:Deltaq1}) and (\ref{eq:Deltaq2})
for $\Delta q^{(1)}$ and $\Delta q^{(2)}$. The net result is that the
quantity in curly brackets effectively defines a running Fermi coupling
constant, $G_F(\mu)$ for which
$G_F(M_Z)=1.16639\times10^{-5}\,{\rm GeV}^{-2}$
very close to quoted value obtained for $\mu=m_\mu$ but including only
1-loop QED corrections.

\section*{Acknowledgments}

RGS wishes to thank the Max-Planck-Institute f\"ur Physik, Munich,
for hospitality while part of this work was carried out.
Helpful and informative discussions with K. Melnikov and A. Sirlin
 are gratefully acknowledged.
This work was supported in part by the US Department of Energy.
 The work of TR was also supported in part by BMBF under contract
  No. 057KA92P and DFG Forschergruppe under contract KU 502/8-1.

\newpage

\appendix

\section{Elementary Multiloop Integrals}
\label{sec:multloop}
In this appendix expressions are given for a number of integrals in
dimensional regularization for which exact analytic results are known.

For a 1-loop massive bubble integral one has \cite{dimreg2}
\begin{multline}
\label{1loopbubble} \int
  \frac{ {\rm d}^Dp\; }{(p^2-m^2+i\epsilon)^{\alpha_1}
       (p^2+i\epsilon)^{\alpha_2}} = \\
            i\; \pi^{D/2} (-1)^{-\alpha_1-\alpha_2}
                    (m^2)^{D/2-\alpha_1-\alpha_2}
        \frac{\Gamma(\alpha_1+\alpha_2-D/2)
          \Gamma(D/2-\alpha_2)}{\Gamma(\alpha_1)\Gamma(D/2)}.
\end{multline}
A 1-loop massless propagator-type integral has the simple form
\begin{multline}
  \label{1loopmassless} \int
 \frac{{\rm d}^Dp\; }{(p^2+i\epsilon)^{\alpha_1}
         [(p+Q)^2+i\epsilon]^{\alpha_2}} = \\
     i^{1-D}\; \pi^{D/2}\; (Q^2)^{D/2-\alpha_1-\alpha_2}
       \frac{\Gamma(D/2-\alpha_1)\Gamma(D/2-\alpha_2)
         \Gamma(\alpha_1+\alpha_2-D/2)}{
        \Gamma(\alpha_1)\Gamma(\alpha_2)\Gamma(D-\alpha_1-\alpha_2)}
\end{multline}
and a compact expression for this integral with a general tensor numerator
can be found in Ref. \cite{1loopclosed}.
For a 1-loop on-shell propagator-type integral one can obtain
\begin{multline}
       \label{1looponshell} \int
 \frac{{\rm d}^Dp\; }{(p^2+i\epsilon)^{\alpha_1}
      (p^2+2p\cdot Q+i\epsilon)^{\alpha_2}} = \\
   i\; \pi^{D/2}(Q^2)^{D/2-\alpha_1-\alpha_2} (-1)^{-\alpha_1-\alpha_2}
       \frac{\Gamma(\alpha_1+\alpha_2-D/2)\Gamma(D-2\alpha_1-\alpha_2)}{
        \Gamma(\alpha_2)\Gamma(D-\alpha_1-\alpha_2)}
  \end{multline}
and for a 2-loop bubble integral with one massless and two massive lines
one finds \cite{BijVelt,gegenbauer}
\begin{multline}
\int\int \frac{{\rm d}^Dp\; {\rm d}^Dk\;}{(p^2-m^2+i\epsilon)^{\alpha_1}\;
     [(p+k)^2+i\epsilon]^{\alpha_2}\;(k^2-m^2+i\epsilon)^{\alpha_3}} = \\
    \pi^D  (m^2)^{D-\alpha_1-\alpha_2-\alpha_3} \;
                (-1)^{1 -\alpha_1-\alpha_2-\alpha_3}
           \frac{\Gamma(-D+\alpha_1+\alpha_2+\alpha_3)}{
               \Gamma(\alpha_1)\Gamma(\alpha_3)}
                   \\
    \times \frac{ \Gamma(-D/2+\alpha_1+\alpha_2)\Gamma(-D/2+\alpha_2+\alpha_3)
       \Gamma(D/2-\alpha_2)}{\Gamma(D/2)
       \Gamma(\alpha_1+2\alpha_2+\alpha_3-D)}  \hspace{2cm}
\label{bubblem0m}
\end{multline}
also for this integral with a general tensor numerator
a compact expression is known \cite{2loopbubclosed}.
Several more simple cases follow by using these expressions recursively
as the powers $\alpha_1$,$\alpha_2$ and  $\alpha_3$, are allowed to be
non-integer, possibly containing $D$.
In this way one can obtain for instance
\begin{multline}
\int\int
 \frac{{\rm d}^Dp\; {\rm d}^Dk }{(p^2+i\epsilon)^{\alpha_1}\;
      (k^2-m^2+i\epsilon)^{\alpha_2}\;
       (k^2+i\epsilon)^{\alpha_4}\; [(p+k)^2+i\epsilon]^{\alpha_3}} = \\
     \pi^D  (m^2)^{D-\alpha_1-\alpha_2-\alpha_3-\alpha_4} \;
                (-1)^{1 -\alpha_1-\alpha_2-\alpha_3-\alpha_4}
      \frac{\Gamma(\alpha_1+\alpha_2+\alpha_3+\alpha_4 -D)}{
          \Gamma(\alpha_1)\Gamma(\alpha_2)\Gamma(\alpha_3)} \\
       \times  \frac{
     \Gamma(\alpha_1+\alpha_3-D/2)\Gamma(D/2-\alpha_1)
       \Gamma(D/2-\alpha_3)\Gamma(D-\alpha_1-\alpha_3-\alpha_4)}{
 \Gamma(D/2) \Gamma(D-\alpha_1-\alpha_3)}
\label{bubblem00}
\end{multline}

\section{Results for Individual Diagrams}
\label{sec:diagresults}
\begin{figure}
\begin{center}

\begin{picture}(400,70)(0,0)

\SetScale{.30}

\SetOffset(0,0)

 \SetWidth{8}
 \Line(50,50)(125,50)
 \Line(225,50)(300,50)
 \SetWidth{2}
 \Line(110,50)(240,50)
 \Oval(175,50)(40,50)(0)

  \Text(20,-5)[m]{\small A1$^{(2)}$}

\SetOffset(100,0)

 \SetWidth{8}
 \Line(50,50)(100,50)
 \Line(200,50)(300,50)
 \SetWidth{2}
 \Line(100,50)(200,50)
 \Oval(150,50)(40,50)(0)
 \PhotonArc(185,90)(85,250,330){4}{8.5}

  \Text(20,-5)[m]{\small A1$^{(3)}$}

\SetOffset(200,0)

 \SetWidth{8}
 \Line(50,50)(132,50)
 \Line(218,50)(300,50)
 \SetWidth{2}
 \Line(125,50)(225,50)
 \Oval(175,50)(30,42)(0)

 \PhotonArc(175,75)(80,200,-20){4}{15.5}

\Text(20,-5)[m]{\small A2$^{(3)}$}

\SetOffset(300,0)

 \SetWidth{8}
 \Line(50,50)(132,50)
 \Line(218,50)(300,50)
 \SetWidth{2}
 \Line(125,50)(225,50)
 \Oval(175,50)(30,42)(0)

 \PhotonArc(175,27)(28,180,0){-4}{7.5}

\Text(20,-5)[m]{\small A3$^{(3)}$}

\end{picture}
\end{center}

\caption{  \label{twoand3loopdiagrams}
    Two and 3-loop diagrams whose cuts give contributions to
       the muon decay rate. }
\end{figure}
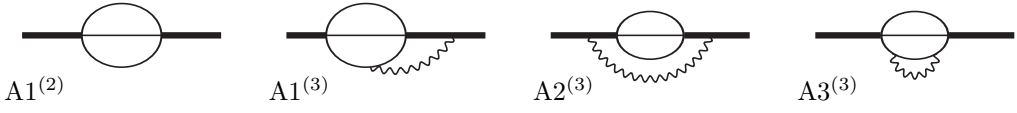

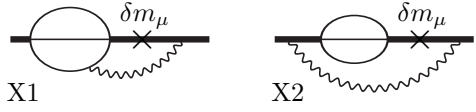
\begin{figure}
\begin{center}

\begin{picture}(200,70)(0,0)

\SetScale{.30}

\SetOffset(0,0)

 \SetWidth{8}
 \Line(50,50)(75,50)
 \Line(175,50)(300,50)
 \SetWidth{2}
 \Line(75,50)(175,50)
 \Oval(125,50)(40,50)(0)
 \PhotonArc(190,90)(85,242,330){4}{10.5}
    \Text(65,15)[m]{\Large $\times$}
    \Text(66,24)[m]{\small $\delta m_\mu$}

\Text(20,-5)[m]{\small X1}

\SetOffset(100,0)

 \SetWidth{8}
 \Line(50,50)(108,50)
 \Line(192,50)(300,50)
 \SetWidth{2}
 \Line(95,50)(195,50)
 \Oval(150,50)(30,42)(0)

 \PhotonArc(175,110)(122,210,-30){4}{18.5}

   \Text(70,15)[m]{\Large $\times$}
    \Text(71,24)[m]{\small $\delta m_\mu$}

\Text(20,-5)[m]{\small X2}

\end{picture}
\end{center}

\caption{  \label{masscounterdiagrams}
 Diagrams X1, X2 that have a muon mass counterterm insertion $\delta m_\mu$
  given in Eq. (\ref{masscounterinsert}). }
\end{figure}

The results for the 4-loop diagrams A1 -- G2
(see Fig.\ref{4loopdiagrams}), the 2- and 3-loop diagrams
A1$^{(2)}$ -- A3$^{(3)}$ (see Fig. \ref{twoand3loopdiagrams}) and
the mass counterterm diagrams X1, X2 (see Fig. \ref{masscounterdiagrams})
are listed below.
The diagrams are calculated in a general covariant gauge for the photon
field, i.e. the photon propagator reads
$ - i\; \left[ g^{\mu\nu} - \xi q^\mu q^\nu
        /(q^2+i\epsilon)\right]/(q^2+i\epsilon) $.
The average over muon spins is taken in the standard way by closing
the $\gamma$ string of the external muon line with
 $(\slash q + m_\mu )/2 $.
Diagrams are calculated in the Fierzed form (charge retention order)
 with only the vector part of the V$-$A interaction;
restoring the axial vector contributions
yields an overall factor 4 that is not included in the diagrams below.
Furthermore, diagrams A1$^{(3)}$,A1,A2,A3,A6,A7,B1,C1,C2,D1,D2,G1,X1 and X2
still have to be multiplied by a symmetry factor 2.
The regularization parameter $\varepsilon$ is defined through
$D=4-2\varepsilon$ with $D$ the space-time dimension.

For the imaginary part of the 2 and 3-loop diagrams
in Fig. \ref{twoand3loopdiagrams} one has
\begin{eqnarray}
    \Im ( {\rm A1^{(2)}}) /{\rm N}^{(2)} & = & \textstyle
        \frac{4}{3}  +  \varepsilon \bigg( \frac{23}{3} \bigg)
         +  \varepsilon^2 \bigg( \frac{127}{4}    - \frac{16}{3} \zeta_2
            \bigg) +{\cal O}(\varepsilon^3) \nonumber \\
    \Im ( {\rm A1^{(3)}}) /{\rm N}^{(3)} & = &  \textstyle
            \frac{1}{\varepsilon} \bigg(
           \frac{4}{3}
          - \frac{4}{3} \xi
          \bigg)
       + \Bigg(
           \frac{121}{9}
          - 8 \zeta_2
          - \frac{112}{9} \xi
          \bigg) \nonumber \\
   & &  \textstyle
      + \varepsilon \Bigg(
           \frac{10855}{108}
          - \frac{218}{3} \zeta_2
          - 24 \zeta_3
          - \frac{2122}{27} \xi
          + 12 \zeta_2 \xi
          \bigg)   + {\cal O}(\varepsilon^2)   \nonumber \\
  \Im ( {\rm A2^{(3)}}) /{\rm N}^{(3)} & = & \textstyle
         \frac{1}{\varepsilon} \bigg(
           \frac{8}{3}
          + \frac{4}{3} \xi
          \bigg)
       +  \bigg(
           \frac{275}{9}
          + \frac{112}{9} \xi
          \bigg) \nonumber \\
    &  & \textstyle
      + \varepsilon \bigg(
           \frac{23915}{108}
            - 24 \zeta_2
            + \frac{2122}{27} \xi
            - 12 \zeta_2 \xi
          \bigg)  + {\cal O}(\varepsilon^2)  \nonumber \\
  \Im ( {\rm A3^{(3)}}) /{\rm N}^{(3)} & = & \textstyle
         \frac{1}{\varepsilon} \bigg(
          - \frac{4}{3}
          + \frac{4}{3} \xi
          \bigg)
      +  \bigg(
          - \frac{112}{9}
          + \frac{112}{9} \xi
          \bigg) \nonumber \\
   & & \textstyle
      +  \varepsilon \bigg(
          - \frac{2122}{27}
          + 12 \zeta_2
          + \frac{2122}{27} \xi
          - 12 \zeta_2 \xi
              \bigg) +{\cal O}(\varepsilon^2)    \nonumber
  \end{eqnarray}
 here and below the normalization factor for $k$-loop diagrams is
 \begin{equation}
  {\rm N}^{(k)} = \frac{ G_F^2 s^3 }{ 1024 \pi^3}
                 \left( \frac{ \alpha_r}{4\pi}\right)^{k-2}
           \left(
 \frac{ (4\pi )^\varepsilon \; \mu^{2\varepsilon} \;
       \Gamma^2 (1-\varepsilon) \Gamma(1+\varepsilon) }{
           s^{\varepsilon} \; \Gamma(1-2\varepsilon) } \right)^k
             \end{equation}
with $s = m_\mu^2$. For the 4-loop diagrams in Fig \ref{4loopdiagrams}
one obtains up to ${\cal O}(\varepsilon)$
\begin{eqnarray}
    \Im ( {\rm A1}) /{\rm N}^{(4)} & = & \textstyle
       \frac{1}{\varepsilon^2} \biggl(
          \frac{2}{3}
          - \frac{4}{3}\xi
          + \frac{2}{3}\xi^2  \biggr)
     + \frac{1}{\varepsilon} \biggl(
          \frac{203}{18}
         - 8 \zeta_2
         - \frac{176}{9}\xi
          + 8\zeta_2 \xi
          + \frac{149}{18}\xi^2      \biggr)
     +  \biggl(
           \frac{3161}{24}
        - \frac{1894}{27} \zeta_2  \nonumber \\
 & & \textstyle
          - \frac{1936}{27} \zeta_3
          + 32 \zeta_4
        - \frac{10109}{54} \xi
          + \frac{994}{9} \zeta_2 \xi
          + \frac{56}{3} \zeta_3 \xi
          + \frac{14815}{216} \xi^2
          - \frac{32}{3} \zeta_2 \xi^2
          \biggr) \nonumber \\
 \Im ( {\rm A2}) /{\rm N}^{(4)} & = & \textstyle
       \frac{1}{\varepsilon^2} \biggl(
          - \frac{122}{3}
          - \frac{56}{3} \xi
          - \frac{2}{3} \xi^2
           \biggr)
     + \frac{1}{\varepsilon} \biggl(
          - \frac{5435}{18}
           - 88 \zeta_2
          - \frac{1288}{9}\xi
           - 8 \zeta_2 \xi
          - \frac{155}{18} \xi^2
        \biggr) \nonumber \\
   & &  \textstyle   +  \biggl(
          - \frac{382951}{216}
          - \frac{1124}{3} \zeta_2
          - \frac{728}{3} \zeta_3
          - \frac{98471}{108} \xi
          + \frac{256}{3} \zeta_2 \xi
          - \frac{88}{3} \zeta_3 \xi
          - \frac{15745}{216} \xi^2
         + \frac{20}{3} \zeta_2 \xi^2
           \biggr) \nonumber \\
  \Im ( {\rm A3}) /{\rm N}^{(4)} & = & \textstyle
         \frac{1}{\varepsilon^2} \biggl(
           \frac{140}{3}
          + \frac{74}{3} \xi
          + \frac{2}{3} \xi^2
          \biggr)
       +  \frac{1}{\varepsilon} \biggl(
           \frac{4006}{9}
          + \frac{3569}{18} \xi
          + \frac{155}{18} \xi^2
          \biggr)   \nonumber  \\
 & &  \textstyle   +  \biggl(
           \frac{340409}{108}
          - \frac{1136}{3} \zeta_2
          + \frac{290629}{216} \xi
          - \frac{608}{3} \zeta_2 \xi
          + \frac{15745}{216} \xi^2
          - \frac{20}{3} \zeta_2 \xi^2
           \biggr) \nonumber  \\
 \Im ( {\rm A4}) /{\rm N}^{(4)} & = & \textstyle
         \frac{1}{\varepsilon^2} \biggl(
          - \frac{8}{3}
          + \frac{4}{3} \xi
          + \frac{4}{3} \xi^2
          \bigg)
       + \frac{1}{\varepsilon} \bigg(
          - \frac{361}{9}
          + \frac{206}{9} \xi
          + \frac{155}{9} \xi^2
         \biggr)   \nonumber  \\
 & &  \textstyle   +  \biggl(
          - \frac{40715}{108}
          + \frac{128}{3} \zeta_2
          + \frac{12467}{54} \xi
          - \frac{64}{3} \zeta_2 \xi
          + \frac{15781}{108} \xi^2
          - \frac{64}{3} \zeta_2 \xi^2
        \biggr) \nonumber
  \end{eqnarray}

\begin{eqnarray}
    \Im ( {\rm A5}) /{\rm N}^{(4)} & = & \textstyle
       \frac{1}{\varepsilon^2} \biggl(
           \frac{8}{3}
          + \frac{8}{3} \xi
          + \frac{2}{3} \xi^2
          \bigg)
   + \frac{1}{\varepsilon} \biggl(
           \frac{460}{9}
          + \frac{409}{9} \xi
          + \frac{179}{18} \xi^2
          \bigg) \nonumber \\
     & & \textstyle + \biggl(
           \frac{59759}{108}
          - \frac{68}{3} \zeta_2
           + \frac{40295}{108} \xi
          + \frac{64}{3} \zeta_2 \xi
           + \frac{20689}{216} \xi^2
          - \frac{32}{3} \zeta_2 \xi^2
         \bigg) \nonumber \\
    \Im ( {\rm A6}) /{\rm N}^{(4)} & = & \textstyle
       \frac{1}{\varepsilon^2} \bigg(
           \frac{8}{3}
          - \frac{4}{3} \xi
          - \frac{4}{3} \xi^2
          \bigg)
     + \frac{1}{\varepsilon} \bigg(
           \frac{379}{9}
          - 16 \zeta_2
          - \frac{197}{9} \xi
          - 8 \zeta_2 \xi
          - \frac{155}{9} \xi^2
          \bigg) \nonumber  \\
     & & \textstyle
      + \bigg(
           \frac{23293}{54}
          - 226 \zeta_2
          - \frac{304}{3} \zeta_3
          - \frac{10499}{54} \xi
          - 68 \zeta_2 \xi
          - \frac{152}{3} \zeta_3 \xi
          - \frac{15763}{108} \xi^2
          + \frac{52}{3} \zeta_2 \xi^2
       \bigg) \nonumber \\
   \Im ( {\rm A7}) /{\rm N}^{(4)} & = & \textstyle
       \frac{1}{\varepsilon^2} \bigg(
          - \frac{2}{3}
          + \frac{4}{3} \xi
          - \frac{2}{3} \xi^2
          \bigg)
       + \frac{1}{\varepsilon} \bigg(
          - \frac{191}{18}
          + 8 \zeta_2
          + \frac{173}{9} \xi
          - 8 \zeta_2 \xi
          - \frac{155}{18} \xi^2
          \bigg) \nonumber \\
    & & \textstyle   + \bigg(
          - \frac{25177}{216}
          + 90 \zeta_2
          + \frac{88}{3} \zeta_3
          + \frac{20479}{108} \xi
          - \frac{302}{3} \zeta_2 \xi
          - \frac{88}{3} \zeta_3 \xi
          - \frac{15781}{216} \xi^2
          + \frac{32}{3} \zeta_2 \xi^2
           \bigg) \nonumber \\
 \Im ( {\rm B1}) /{\rm N}^{(4)} & = & \textstyle
       \frac{1}{\varepsilon} \bigg(
          - \frac{8}{3}
          + \frac{4}{3} \xi
          + \frac{1}{3} \xi^2
          \bigg)
       + \bigg(
          - \frac{73841}{486}
           - \frac{4808}{27} \zeta_2
          + \frac{1024}{3} \zeta_2 \ln (2)
          + \frac{1700}{27} \zeta_3
          - \frac{56}{3} \zeta_4     \nonumber \\
       & & \textstyle
          + \frac{148}{9} \xi
          + \frac{28}{9}\zeta_2 \xi
          + \frac{161}{36} \xi^2
             \bigg) \nonumber \\
\Im ( {\rm C1}) /{\rm N}^{(4)} & = & \textstyle
       \frac{1}{\varepsilon^2} \bigg(
           \frac{2}{3}
          - \frac{4}{3} \xi
          + \frac{2}{3} \xi^2
          \bigg)
       + \frac{1}{\varepsilon} \bigg(
           \frac{191}{18}
          + 8 \zeta_2 \xi
          - \frac{173}{9} \xi
          + \frac{155}{18} \xi^2
          - 8 \zeta_2
          \bigg)
       + \bigg(
          - \frac{13003}{216}
          - \frac{77}{9} \zeta_2   \nonumber \\
       & & \textstyle
          - \frac{32}{3} \zeta_2 \ln(2)
          + 60 \zeta_3
          - \frac{8339}{54} \xi
          + \frac{332}{3} \zeta_2 \xi
          + \frac{88}{3} \zeta_3 \xi
          + \frac{15745}{216} \xi^2
          - \frac{20}{3} \zeta_2 \xi^2
          \bigg) \nonumber \\
\Im ( {\rm C2}) /{\rm N}^{(4)} & = & \textstyle
       \frac{1}{\varepsilon^2} \bigg(
           \frac{16}{3}
          + \frac{4}{3} \xi
          - \frac{2}{3} \xi^2
          \bigg)
      + \frac{1}{\varepsilon} \bigg(
           \frac{509}{9}
          + \frac{23}{9} \xi
          - \frac{155}{18} \xi^2
          \bigg)  \nonumber \\
   & & \textstyle + \bigg(
           \frac{159511}{324}
          - \frac{112}{3} \zeta_2
          - 128 \zeta_3
          - \frac{8507}{108} \xi
          - \frac{16}{3} \zeta_2 \xi
          - \frac{15745}{216} \xi^2
          + \frac{20}{3} \zeta_2 \xi^2
    \bigg) \nonumber \\
\Im ( {\rm C3}) /{\rm N}^{(4)} & = & \textstyle
       \frac{1}{\varepsilon} \bigg(
          - \frac{8}{3}
          + \frac{32}{3} \zeta_2
          \bigg)
      + \bigg(
           \frac{328307}{1350}
          + \frac{2080}{27} \zeta_2
          - \frac{1760}{9} \zeta_3
          \bigg) \nonumber \\
\Im ( {\rm C4}) /{\rm N}^{(4)} & = & \textstyle
      \frac{1}{\varepsilon^2} \bigg(
          - \frac{16}{3}
            \bigg)
       + \frac{1}{\varepsilon} \bigg(
          - \frac{172}{3}
          \bigg)
          + \bigg(
          - \frac{270041}{675}
          + \frac{128}{3} \zeta_2
          \bigg)   \nonumber \\
\Im ( {\rm C5}) /{\rm N}^{(4)} & = & \textstyle
      \frac{1}{\varepsilon} \bigg(
           \frac{4}{3}
          \bigg)
        +\bigg(
           \frac{4409}{675}
          \bigg)  \nonumber \\
\Im ( {\rm D1}) /{\rm N}^{(4)} & = & \textstyle
      \frac{1}{\varepsilon^2} \bigg(
           \frac{2}{3}
          - \frac{4}{3} \xi
          + \frac{2}{3} \xi^2
          \bigg)
     + \frac{1}{\varepsilon} \bigg(
           \frac{191}{18}
           - 8 \zeta_2
          - \frac{173}{9} \xi
          + 8 \zeta_2 \xi
          + \frac{155}{18} \xi^2
          \bigg)   \nonumber \\
       & & \textstyle  + \bigg(
           \frac{22537}{216}
          - \frac{1054}{9} \zeta_2
          - \frac{40}{3} \zeta_3
          + \frac{160}{3} \zeta_4
          - \frac{20671}{108} \xi
          + \frac{790}{9} \zeta_2 \xi
          + \frac{136}{3} \zeta_3 \xi
          + \frac{15493}{216} \xi^2
          - \frac{32}{3} \zeta_2 \xi^2
          \bigg)  \nonumber \\
\Im ( {\rm D2}) /{\rm N}^{(4)} & = & \textstyle
      \frac{1}{\varepsilon^2} \bigg(
          - \frac{4}{3}
          + \frac{8}{3} \xi
          - \frac{4}{3} \xi^2
          \bigg)
      + \frac{1}{\varepsilon} \bigg(
          - \frac{164}{9}
          + 8 \zeta_2
          + \frac{319}{9} \xi
          - 8 \zeta_2 \xi
          - \frac{155}{9} \xi^2
          \bigg)  \nonumber \\
      &  & \textstyle  + \bigg(
          - \frac{9389}{54}
          + \frac{302}{3} \zeta_2
          + \frac{104}{3} \zeta_3
          + \frac{34415}{108} \xi
          - 122 \zeta_2 \xi
          - \frac{104}{3} \zeta_3 \xi
          - \frac{15637}{108} \xi^2
          + \frac{64}{3} \zeta_2 \xi^2
       \bigg)  \nonumber
    \end{eqnarray}

\begin{eqnarray}
    \Im ( {\rm D3}) /{\rm N}^{(4)} & = & \textstyle
           \frac{1}{\varepsilon^2} \bigg(
          - \frac{4}{3}
          + \frac{8}{3} \xi
          - \frac{4}{3} \xi^2
          \bigg)
        + \frac{1}{\varepsilon} \bigg(
          - \frac{149}{9}
          + \frac{316}{9} \xi
          - \frac{158}{9} \xi^2
          \bigg) \nonumber \\
      & & \textstyle + \bigg(
          - \frac{14887}{108}
          + \frac{64}{3} \zeta_2
          + \frac{8078}{27} \xi
          - \frac{128}{3} \zeta_2 \xi
          - \frac{4039}{27} \xi^2
          + \frac{64}{3} \zeta_2 \xi^2
         \bigg)  \nonumber  \\
  \Im ( {\rm D4}) /{\rm N}^{(4)} & = & \textstyle
           \frac{1}{\varepsilon^2} \bigg(
           \frac{2}{3}
          - \frac{4}{3} \xi
          + \frac{2}{3} \xi^2
          \bigg)
      + \frac{1}{\varepsilon} \bigg(
           \frac{161}{18}
          - \frac{161}{9} \xi
          + \frac{161}{18} \xi^2
          \bigg)  \nonumber \\
      & & \textstyle + \bigg(
           \frac{16819}{216}
          - \frac{32}{3} \zeta_2
           - \frac{16819}{108} \xi
          + \frac{64}{3} \zeta_2 \xi
          + \frac{16819}{216} \xi^2
          - \frac{32}{3} \zeta_2 \xi^2
            \bigg)  \nonumber \\
 \Im ( {\rm D5}) /{\rm N}^{(4)} & = & \textstyle
           \frac{1}{\varepsilon^2} \bigg(
           \frac{4}{3}
          - \frac{8}{3} \xi
          + \frac{4}{3} \xi^2
          \bigg)
        +  \frac{1}{\varepsilon} \bigg(
           \frac{155}{9}
          - \frac{310}{9} \xi
          + \frac{155}{9} \xi^2
          \bigg) \nonumber \\
    & & \textstyle + \bigg(
           \frac{15637}{108}
          - \frac{64}{3} \zeta_2
          - \frac{15637}{54} \xi
          + \frac{128}{3} \zeta_2 \xi
          + \frac{15637}{108} \xi^2
          - \frac{64}{3} \zeta_2 \xi^2
          \bigg)  \nonumber \\
 \Im ( {\rm D6}) /{\rm N}^{(4)} & = & \textstyle
        \frac{1}{\varepsilon} \bigg(
          - \frac{8}{3}
          + \frac{32}{3} \zeta_2
          \bigg)
       + \bigg(
          - \frac{1585}{27}
          + \frac{3088}{27} \zeta_2
          + \frac{544}{9} \zeta_3
          \bigg)  \nonumber \\
   \Im ( {\rm D7}) /{\rm N}^{(4)} & = & \textstyle
        \frac{1}{\varepsilon^2} \bigg(
          - \frac{8}{3}
          \bigg)
        +  \frac{1}{\varepsilon} \bigg(
          - \frac{410}{9}
          \bigg)
       + \bigg(
          - \frac{22937}{54}
          + \frac{16}{3} \zeta_2
          \bigg)  \nonumber \\
    \Im ( {\rm D8}) /{\rm N}^{(4)} & = & \textstyle
            \frac{1}{\varepsilon} \bigg(
           \frac{4}{3}
          \bigg)
       + \bigg(
           \frac{173}{9}
          \bigg)  \nonumber \\
 \Im ( {\rm E1}) /{\rm N}^{(4)} & = & \textstyle
            \frac{1}{\varepsilon^2} \bigg(
           \frac{4}{3}
          - \frac{8}{3} \xi
          + \frac{4}{3} \xi^2
          \bigg)
      + \frac{1}{\varepsilon} \bigg(
          + \frac{173}{9}
          - 16 \zeta_2
          - \frac{328}{9} \xi
          + 16 \zeta_2 \xi
          + \frac{155}{9} \xi^2
          \bigg) \nonumber \\
      & & \textstyle + \bigg(
           \frac{68945}{324}
          - \frac{16700}{81} \zeta_2
          - \frac{928}{9} \zeta_3
          + \frac{448}{3} \zeta_4
           - \frac{9389}{27} \xi
          + \frac{604}{3} \zeta_2 \xi
          + \frac{208}{3} \zeta_3 \xi
          + \frac{15637}{108} \xi^2
          - \frac{64}{3} \zeta_2 \xi^2
          \bigg)  \nonumber \\
 \Im ( {\rm F1}) /{\rm N}^{(4)} & = & \textstyle
            \bigg(
           \frac{38200}{243}
          - \frac{16142}{81} \zeta_2
          + \frac{1024}{3} \zeta_2 \ln(2)
          - \frac{680}{9} \zeta_3
          - 104 \zeta_4
          + \frac{56}{9} \xi
          + \frac{232}{9} \zeta_2 \xi
          - 32 \zeta_3 \xi
          + \frac{4}{3} \xi^2
         \bigg)   \nonumber \\
  \Im ( {\rm G1}) /{\rm N}^{(4)} & = & \textstyle
          \bigg(
          - \frac{53}{12}
          - \frac{1238}{27} \zeta_2
          + \frac{244}{3} \zeta_3
          - \frac{425}{36} \xi
          - \frac{26}{9} \zeta_2 \xi
          + 32 \zeta_3 \xi
          - \frac{1}{6} \xi^2
          + 4 \zeta_2 \xi^2
          \bigg) \nonumber \\
   \Im ( {\rm G2}) /{\rm N}^{(4)} & = & \textstyle
      \frac{1}{\varepsilon} \bigg(
          - \frac{16}{3}
          - \frac{16}{3} \xi
          - \frac{4}{3} \xi^2
          \bigg)
      + \bigg(
          - \frac{3103}{81}
          + \frac{16}{9} \zeta_2
          - 64 \zeta_3
         + \frac{35}{9} \xi
          - 64 \zeta_2 \xi
          - \frac{409}{18} \xi^2
          \bigg)   \nonumber
    \end{eqnarray}

For the mass counterterm diagrams X1,X2 of
    Fig. \ref{masscounterdiagrams} we obtain
    up to ${\cal O}(\varepsilon)$

\begin{eqnarray}
    \Im ( {\rm X1}) /{\rm N}^{(4)} & = & \textstyle
           \frac{1}{\varepsilon^2} \bigg(
           40
          + 20 \xi
          \Bigg)
      + \frac{1}{\varepsilon} \bigg(
           \frac{874}{3}
          + 96 \zeta_2
          + \frac{487}{3} \xi
          \Bigg) \nonumber \\
     & & \textstyle
      + \bigg(
           \frac{3345}{2}
          - 196 \zeta_2 \xi
          + \frac{38383}{36} \xi
          + 416 \zeta_2
          + 288 \zeta_3
         \bigg)   \nonumber \\
   \Im ( {\rm X2}) /{\rm N}^{(4)} & = & \textstyle
        \frac{1}{\varepsilon^2} \bigg(
          - 52
          - 26 \xi
          \bigg)
       + \frac{1}{\varepsilon} \bigg(
          - \frac{1505}{3}
          - \frac{1205}{6} \xi
          \bigg)
        + \bigg(
          - \frac{125633}{36}
          + 416 \zeta_2
          - \frac{91205}{72} \xi
          + 208 \zeta_2 \xi
           \bigg)   \nonumber
    \end{eqnarray}

Finally, for the external muon-leg renormalization we need to add 1PR
diagrams that are generated by multiplying the 1PI diagrams of Fig.
\ref{twoand3loopdiagrams}  by the gauge independent factor
 \begin{equation}
      Z(\alpha,\varepsilon) = \frac{ 1}{1-\tilde{Z}(\alpha,\varepsilon)}
        =  1 + \tilde{Z}(\alpha,\varepsilon)
         + \tilde{Z}^2(\alpha,\varepsilon)  + \cdots
       \end{equation}
Here  $\tilde{Z}(\alpha,\varepsilon)$ follows from considering the proper
muon self-energy
 \begin{equation}
         \Sigma(m_\mu,q) = ( \slash q - m_\mu)\Sigma_1(m_\mu,q^2)
                     + m_\mu \Sigma_2(m_\mu,q^2)
 \end{equation}
  and calculating
\begin{equation}
      \tilde{Z} =  \Sigma_1(m_\mu,m_\mu^2)
          + 2 m_{\mu}^2 \frac{\partial}{\partial q^2}
                \Sigma_2(m_\mu,q^2) \Bigg|_{q^2 = m_\mu^2}
 \end{equation}
After evaluating the necessary diagrams contributing to the self-energy and
the derivative one obtains
\begin{eqnarray}
    \tilde{Z}(\alpha,\varepsilon)  & = &
          -a \bigg[    \frac{3 }{\varepsilon}  +4
                + \varepsilon (8 +3 \zeta_2 )
               + O (\varepsilon^2)  \bigg]
     - a^2 \bigg[
         \frac{1}{\varepsilon^2} \bigg(
          \frac{9}{2} - 2 n_f - 4 n_H
          \bigg)  \nonumber \\
  & &
       - \frac{1}{\varepsilon}  \bigg(
           \frac{45}{4} - 9 n_f - \frac{19}{3} n_H
          \bigg)
     - \bigg(
            \frac{79}{8}
          + 87 \zeta_2
          - 96 \zeta_2 \ln(2)
          + 24 \zeta_3  \nonumber \\
    & &
          - \frac{59}{2} n_f
          - 12 \zeta_2 n_f
         - \frac{1139}{18} n_H
            + 24 \zeta_2 n_H
             \bigg)
      + O (\varepsilon) \bigg]  + O(a^3)
 \end{eqnarray}
where
  \begin{equation}
       a =   \left( \frac{ \alpha_r}{4\pi}\right)
                 \frac{ (4\pi )^\varepsilon \;
               \mu^{2\varepsilon} \;
           \Gamma^2 (1-\varepsilon) \Gamma(1+\varepsilon) }{
            (m_\mu)^{2 \varepsilon} \; \Gamma(1-2\varepsilon) }
    \end{equation}
and $n_f$ labels the number of light fermions (in the present case only
the electron, $n_f = 1$) and $n_H$ labels the number of fermions with a
mass equal to the muon-mass (in the present case only the muon itself,
$n_H = 1$).

\section{The $O(\alpha)$ Electron Spectrum}
\label{sec:elecspec}
The complete ${\cal O}(\alpha)$ corrections to the
electron energy spectrum in muon decay were first calculated by
Behrends {\it et al.}\cite{BehrFinkSirl} who gave the results in terms of
hyperbolic functions. As was pointed out by Berman \cite{Berman}
this calculation did not take into account
longitudinal degrees of freedom for the photon that must be included
when infrared regularization is done using a photon mass. The necessary
correction term was provided in Appendix~C of Ref.\cite{KinoSirl}.
The calculation was repeated and confirmed by Grotch \cite{Grotch}
but dropping terms of ${\cal O}(\alpha m_e^2/m_\mu^2)$ in well-defined
and reconstructible places. A related calculation, in the context of
QCD corrections to heavy quark decays,
has been performed in Ref.\cite{HokimPham}.
The expressions given were used by Nir \cite{Nir} to obtain the
mass-dependent corrections given in Eq.(\ref{eq:Deltaq1}).

We have taken the results of Ref.\cite{BehrFinkSirl} and rewritten them
as function of $z=\sqrt{x} e^\theta$ where $x\equiv m_e^2/m_\mu^2$ and
$\theta$ is
the variable, introduced in Ref.\cite{BehrFinkSirl}, defined by
$\cosh\theta=p_e\cdot p_\mu/(m_e m_\mu)$.
In the rest frame of the muon, $\cosh\theta=E_e/m_e$ and hence
$e^\theta=(E_e+p_e)/m_e$ where $E_e$ and $p_e$ are the energy and
3-momentum of the electron respectively. The electron energy spectrum
is then
\begin{equation}
P(z)\,dz=\frac{G_F^2 m_\mu^5}{192\pi^3}
         \left(P_0(z)+\frac{\alpha}{\pi} P_1(z)\right)\,dz
\end{equation}
in which
\begin{eqnarray}
P_0(z)&=&-\frac{2(z^2-x)^2}{z^5}
         \left(2x^2-3x(1+x)z+8x z^2
                      -3(1+x)z^3+2z^4\right),\\
P_1(z)&=&-\frac{2(z^2-x)^2}{3z^5}
          \left(5x^2-10x(1+x)z+(11+8x+11x^2)z^2
               -10(1+x)z^3+5z^4\right)\nonumber\\
    & &-\frac{(z^2-x)^2}{2z^5}
          \left(4x^2-3x(1+3x)z+16x z^2
               -3(1+3x)z^3+4z^4\right)\ln x\nonumber\\
    & &+\frac{z^2-x}{6z^5}
          \big(4x^3-3x^2(5-3x)z
               +12x(1+4x-2x^2)z^2\nonumber\\
    & &\qquad\qquad
               +(5-87x+9x^2+5x^3)z^3\nonumber\\
    & &\qquad\qquad
               +12(1+4x-2x^2)z^4-3(5-3x)z^5
               +4z^6\big)\ln\frac{z^2}{x}\\
    & &-\frac{2(z^4-x^2)}{z^5}
           \left(2x^2-3x(1+x)z+8x z^2
                        -3(1+x)z^3+2z^4\right)
           \nonumber\\
    & &\qquad\ \
       \times\bigg\{
                   \frac{(1-z)(z+x)}{z^2+x}
                        \ln\left(1-\frac{x}{z}\right)
                  -\frac{(1+z)(z-x)}{z^2+x}\ln(1-z)\nonumber\\
    & &\qquad\qquad
            +\ln\frac{1-z}{z}\ln\frac{z}{x}
            +\ln z\ln\frac{z-x}{z^2}\nonumber\\
    & &\qquad\qquad
                  -2{\rm Li}_2\left(1-\frac{x}{z^2}\right)
                  +2{\rm Li}_2\left(1-\frac{x}{z}\right)
                  -2{\rm Li}_2(1-z)
            \bigg\}.\nonumber
\end{eqnarray}
When integrated in the interval $z\in(\sqrt{x},1)$, $P_0(z)$
yields the expression for $\Delta q^{(0)}$,
with $y\equiv m_{\nu_\mu}^2/m_\mu^2=0$, as given in Eq.(\ref{eq:Deltaq0})
and $\left(\alpha/\pi\right)P_1(z)$ gives $\Delta q^{(1)}$ of
Eq.(\ref{eq:Deltaq1}).

The 1-loop QED corrections to the differential decay rate of radiative
muon decay, $\mu^-\rightarrow e^-\nu_\mu\bar\nu_e\gamma$, are given in
Ref.\cite{FiscKuroSava}.

\section{The Branching Ratio for
         $\mu^-\rightarrow e^-\nu_\mu\bar\nu_e e^+e^-$}
\label{sec:BR3e}

The contribution $\Delta q^{(2)}_{{\rm elec}}$ to the muon inverse
lifetime, given in Eq.(\ref{eq:Deltaq2elec}), includes pieces that
come from the processes
$\mu^-\rightarrow e^-\nu_\mu\bar\nu_e$, $e^-\nu_\mu\bar\nu_e\gamma$
and $e^-\nu_\mu\bar\nu_e e^+e^-$. In order to obtain the branching
ratio for the last of these processes alone,
the contributions from the other two
need to be subtracted. Contributions from the process
$\mu^-\rightarrow e^-\nu_\mu\bar\nu_e\gamma$
can be eliminated by adopting the on-shell
renormalization scheme since fermion loops on real photon lines,
with $q^2=0$, are forced to vanish. By the same argument used in
section~\ref{sec:AlphaRenorm}, the conversion from the
$\overline{{\rm MS}}$ renormalization scheme ($\mu=m_\mu$)
to on-shell scheme can be performed
for the contribution of electron loops (\ref{eq:Deltaq2elec})
by adding a term
\begin{equation}
\Delta q^{(1)}
\frac{2}{e}\left(\delta e^{(1)}_{{\rm OS}}
                -\delta e^{(1)}_{\overline{{\rm MS}}}
           \right)=
-\left(\frac{\alpha}{\pi}\right)^2
 \left(\frac{25}{24}-\zeta(2)\right)\ln\frac{m_e^2}{m_\mu^2}.
\end{equation}

The contribution in Eq.(\ref{eq:Deltaq2elec}) coming from the process
$\mu^-\rightarrow e^-\nu_\mu\bar\nu_e$, in which there is virtual photon
line containing an electron loop, may be calculated using the dispersion
relation methods of Ref.\cite{muonhad}. Note that the use of dispersion
relations naturally invokes a subtraction at $q^2=0$ and is therefore
consistent with the on-shell renormalization scheme.
>From Eq.(9) of Ref.\cite{muonhad} the contribution from Feynman
diagrams with virtual photons containing electron loops is given by
\begin{equation}
\Delta\Gamma_{\rm had}=
\frac{\alpha}{3\pi}\int_{4\rho}^\infty\frac{dz}{z}R(m_\mu^2 z)
                       \,\Delta\Gamma(z)
\end{equation}
where $\Delta\Gamma(z)$ is given in Eq.(7) of that work and in the
present case $\rho=m_e^2/m_\mu^2$ and
\begin{equation}
R(m_\mu^2z)=\left(1+\frac{2\rho}{z}\right)\sqrt{1-\frac{4\rho}{z}}.
\end{equation}
The integration may be conveniently divided into two regions,
$z\in[4\rho,4]$ and $z\in[4,\infty]$. In the latter interval the
integral contains no singularities in the limit $m_e\rightarrow0$
and $\rho$ can be safely set to zero.
Making the substitution $u=\sqrt{1-4/z}$, the integral in this interval is
\begin{eqnarray}
\Delta\Gamma_{[4,\infty]}&=&\Gamma_0\left(\frac{\alpha}{\pi}\right)^2
                 \int_0^1 K(u) du\\
              &=&\Gamma_0\left(\frac{\alpha}{\pi}\right)^2
                 \left\{\frac{7199}{5184}-\frac{23}{36}\zeta(2)
                       -\ln2\left(\frac{37}{216}
                       -\frac{2}{3}\zeta(2)\right)
                       +\frac{2}{9}\ln^2 2(1+\ln 2)
                       -\zeta(3)\right\}\nonumber\\
              & & \label{eq:ConstInt}\\
              &=&-\Gamma_0\left(\frac{\alpha}{\pi}\right)^2 0.0421308
\end{eqnarray}
where $K(u)$ is given in Eq.(11) of Ref.\cite{muonhad}.
On the interval $z\in[4\rho,4]$, the integral does exhibit singularities
as $m_e\rightarrow0$. Setting $u=iv\sqrt{\rho^{-1}-1}$ in the integral
over the first interval gives
\begin{eqnarray}
\Delta\Gamma_{[4\rho,4]}&=&-\Gamma_0\left(\frac{\alpha}{\pi}\right)^2
                 \int_0^1 \frac{i}{2}\frac{(1-\rho)}{\sqrt{\rho}}
                 (2+\rho+(1-\rho)v^2)\sqrt{1-v^2}
                 K\left(i\sqrt{\frac{1-\rho}{\rho}}v\right)\,dv\ \ {}
\label{eq:Integral3e}\\
              &=&-\Gamma_0\left(\frac{\alpha}{\pi}\right)^2 9.47056
\label{eq:NIntegral3e}
\end{eqnarray}
where the integration has been performed numerically.
As mentioned above, the integral (\ref{eq:Integral3e}) contains
terms that behave like $\ln^n\rho$ for non-negative integer $n\le3$ that
are singular in the limit $m_e\rightarrow0$. It will also contain terms
that vanish in that limit but such terms were discarded in the original
calculation of the kernel $K(u)$. Thus the above result is correct only
in the terms that do not vanish as $m_e\rightarrow0$.

Adding Eq.(\ref{eq:ConstInt}) to Eq.(\ref{eq:Deltaq2elec}) to
convert to the on-shell scheme and subtracting (\ref{eq:NIntegral3e})
to leave only the contribution of
$\mu^-\rightarrow e^-\nu_\mu\bar\nu_e e^+e^-$ gives the branching ratio
\begin{eqnarray}
\frac{\Gamma_{3e}}{\Gamma_0}&=&-\left(\frac{\alpha}{\pi}\right)^2
                 \left\{\frac{25361}{5184}-\frac{25}{9}\zeta(2)
                       -\ln2\left(\frac{37}{216}-\frac{2}{3}\zeta(2)\right)
                       +\frac{2}{9}\ln^2 2(1+\ln 2)\right.\nonumber\\
              & &\qquad\qquad\left.-\frac{11}{3}\zeta(3)
                 +\left(\frac{25}{24}-\zeta(2)\right)
                        \ln\frac{m_e^2}{m_\mu^2}
              -9.47056\right\}\\
 &=&\left(\frac{\alpha}{\pi}\right)^2 6.30028\\
 &=&3.40\times10^{-5}
\end{eqnarray}
in good agreement with the experimentally measured value of
$(3.4\pm0.4)\times10^{-5}$ \cite{PDG}.


\begin{thebibliography}{99}

\bibitem{PDG} C. Caso {\it et al.},
        {\sl European Physical Journal}\ {\bf C 3} (1998) 1.

\bibitem{LEPEWWG} LEP Electroweak Working Group, CERN-PPE/97-154.

\bibitem{Roberts} B. L. Roberts, {\it private communication}.

\bibitem{CERN} CERN report 86-02 vol.\ 1 (1986)  p.\ 10,
               ed.s J. Ellis and R. Peccei.

\bibitem{Blondel} A. Blondel, {\it to appear in\/}
          Proceedings of the IVth International
          Symposium on Radiative Corrections (RADCOR98), Barcelona,
          Spain, 8--12 September, 1998, edited by J. Sola.

\bibitem{RossVeltman} D. A. Ross and M. Veltman,
        {\sl Nucl.\ Phys.}\ {\bf B 95} (1977) 135.

\bibitem{BermSirl} S. M. Berman and A. Sirlin,
    {\sl Ann.\ Phys.}\ {\bf 20} (1962) 20.

\bibitem{RoosSirlin} M. Roos and A. Sirlin,
                    {\sl Nucl. Phys.}\ {\bf B 29} (1971) 296.

\bibitem{BehrFinkSirl} R. E. Behrends, R. J. Finkelstein and A. Sirlin,
    {\sl Phys.\ Rev.}\ {\bf 101} (1956) 866.

\bibitem{muonprl} T. van Ritbergen and R. G. Stuart,
    {\sl Phys.\ Rev.\ Lett.}\ {\bf 82} (1999) 488.

\bibitem{KinoSirl} T. Kinoshita and A. Sirlin,
    {\sl Phys.\ Rev.}\ {\bf 113} (1959) 1652.

\bibitem{KinoLeeNaue} T. Kinoshita,
      {\sl J.\ Math.\ Phys.}\ {\bf 3} (1962) 650;\\
   T. D. Lee and M. Nauenberg, {\sl J.\ Math.\ Phys.}\ {\bf 3} (1962) 650.

\bibitem{Nir} Y. Nir, {\sl Phys.\ Lett.}\ {\bf B 221} (1989) 184.

\bibitem{CzarJezaKuhn} A. Czarnecki, M. Je\.zabek and J. H. K\"uhn,
    {\sl Phys.\ Lett.}\ {\bf B 346} (1995) 335.

\bibitem{muonhad} T. van Ritbergen and R. G. Stuart,
        {\sl Phys.\ Lett.}\ {\bf B 437} (1998) 201.

\bibitem{Cutkosky} R. E. Cutkosky, {\sl J. Math. Phys.} {\bf1} (1960) 429.

\bibitem{LukeSavaWise} M. Luke, M. J. Savage and M. B. Wise,
        {\sl Phys.\ Lett.}\ {\bf B 343} (1995) 327.

\bibitem{Hofstadter} R. Hofstadter,
    {\sl Ann.\ Revs.\ Nuclear Sci.}\ {\bf 7} (1957) 231.

\bibitem{dimreg} C. G. Bollini and J. J. Giambiagi,
        {\sl Phys.\ Lett.}\ {\bf 40 B} (1972) 566.

\bibitem{dimreg2}
     G. 't\ Hooft and M. Veltman, {\sl Nucl.\ Phys.}\ {\bf B44} (1972) 189.

\bibitem{MaldeStuart1} P. J. Malde and R. G. Stuart, hep-ph/9805364.

\bibitem{JostLuttinger} R. Jost and J. M. Luttinger,
     {\sl Helv. Phys. Acta} {\bf 23} (1950) 201.

\bibitem{Rosner} J. Rosner, {\sl Annals of Physics}\ {\bf 44} (1967) 11.

\bibitem{parialint} K. G. Chetyrkin and F. V. Tkachov,
        {\sl Nucl.\ Phys.}\ {\bf B 192} (1981) 159.

 \bibitem{msbarpolemass} N. Gray, D. J. Broadhurst, W. Grafe and
         K. Schilcher, {\sl Z.\ Phys.}\ {\bf C48} (1990) 673.

 \bibitem{shell2} J. Fleischer and O.V. Tarasov,
         {\sl Comput.\ Phys.\ Comm.}\ {\bf 71} (1992) 193.

\bibitem{gmin2at3loops} S. Laporta and E. Remiddi,
        {\sl  Phys.\ Lett.}\ {\bf B 379} (1996) 283;\\
        S. Laporta, unpublished, (private communications E. Remiddi).

\bibitem{resolvedtriangle} F. V. Tkachov,
        {\sl Theor.\ Math.\ Fiz.}\ {\bf 56} (1983) 350.

  \bibitem{newibpsolution}
       P. A. Baikov, {\sl Phys.\ Lett.}\ {\bf B 385} (1996) 404;
                    {\sl Nucl.\ Instrum.\ Meth.}\ {\bf A 389} (1997) 347.

 \bibitem{form} J. A. M. Vermaseren, Symbolic Manipulation with Form,
         Computer Algebra Nederland, Amsterdam (1991).

 \bibitem{polylog} L. Lewin, {\sl Polylogarithms and associated functions}
         (North Holland, 1981).

 \bibitem{formalexpansions1}
      F. V. Tkachov, preprint INR P-358 (Moscow, 1984);
      {\sl Int.\ J.\ Mod.\ Phys.} {\bf A 8} (1993) 2047;
      G. B. Pivovarov, F.V. Tkachov,
      {\sl Int.\ J.\ Mod.\ Phys.} {\bf A 8} (1993) 2241.

 \bibitem{formalexpansions2}
       K. G. Chetyrkin, V. A. Smirnov, preprint INR P-0518;\\
       K. G. Chetyrkin, {\sl Theor.\ Math.\ Fiz.}\ {\bf 76} (1988) 207;\\
       S. G. Gorishny, {\sl Nucl.\ Phys.}\ {\bf B 319} (1989) 633;\\
       V. A. Smirnov, {\sl Commun.\ Math.\ Phys.}\ {\bf 134} (1990) 109.

 \bibitem{largemass}
        S. A. Larin, T. van Ritbergen and
        J. A. M. Vermaseren, {\sl Nucl.\ Phys.}\ {\bf B 438} (1995) 278.

 \bibitem{largemassUntrunc}  A. Czarnecki and J. H. K\"uhn,
         {\sl Phys.\ Rev.\ Lett.}\ {\bf 77} (1996) 3955 ;\\
   A. Czarnecki and K. Melnikov, preprint TTP 98-23 (Karlsruhe 1998),
        hep-ph/9806258

\bibitem{largemassUntrunc2} J. Fleischer, A. V. Kotikov, O. L. Veretin,
        preprint BI-TP-98-20, hep-ph/9808242

\bibitem{finitesums1} A. Gonz\'alez-Arroyo, C. L\'opez, F. J. Yndurain,
         {\sl Nucl.\ Phys.}\ {\bf B 153} (1979) 161;\\
        A. Devoto and D. W. Duke,
        {\sl Rivista del Nuovo Cimento}\/ {\bf 7} (1984) 1.

\bibitem{sumsjos} J. A. M. Vermaseren,
        preprint FTUAM-98-7, hep-ph/9806280

\bibitem{infinitesums1}
        L. Euler,
        {\sl Novi Comm.\ Acad.\ Sci.\ Petropol.}\ {\bf 20} (1775) 140;\\
        D. Zagier, in proceedings of {\sl First European Congress of
               Mathematics}, vol. II, Birkh\"auser, Boston (1994) p. 497.

\bibitem{infinitesums2}
       D. Borwein, J. M. Borwein, R. Girgensohn,
        {\sl Proc. Edinb. Math. Soc.}\ {\bf 38} (1995) 277;\\
       D. H. Bailey, J. M. Borwein, R. Girgensohn,
         RNR technical report  RNR-93-014;\\
       J. M. Borwein, D. M. Bradley, D.J. Broadhurst,
         preprint CECM-96-067, hep-th/9611004.

\bibitem{infinitesums3}
      D. J. Broadhurst, preprint OUT-4102-62, hep-th/9604128;\\
      D. J. Broadhurst,  preprint OUT-4102-72, hep-th/9803091;\\
      O. M. Ogreid and P. Osland,  preprint DESY-97-245 (Bergen 1997)
                  hep-th/9801168.

\bibitem{QuarkMass} J. Fleischer, O. V. Tarasov, F. Jegerlehner and
        O. L. Veretin, preprint DESY~98-026, hep-ph/9803493.

\bibitem{BNL} R. M. Carey {\it et al.}, BNL Letter of intent (1996).

\bibitem{PSI} F. Navarria {\it et al.}, preprint ETHZ-IPP PR-98-04.

\bibitem{RAL} S. N. Nakamura {\it et al}, RIKEN-RAL research proposal
              R14 (1996); R77 (1997).

\bibitem{Planck} E. R. Williams {\it et al.},
     {\sl Phys.\ Rev. Lett.}\ {\bf 81} (1998) 2404.

\bibitem{Prisca} P. Cushman {\it et al.},
        {\sl AGS2000 White Paper}\/ (1996).

\bibitem{Sirlin78} A. Sirlin,
        {\sl Rev.\ Mod.\ Phys.}\ {\bf 50} (1978) 573.

\bibitem{Sirlin80} A. Sirlin, {\sl Phys.\ Rev.}\ {\bf D 22} (1980) 971.

\bibitem{Sirlin84} A. Sirlin, {\sl Phys.\ Rev.}\ {\bf D 29} (1984) 89.

\bibitem{ZMass4} R. G. Stuart,
            {\sl Phys.\ Lett.}\ {\bf B 272} (1991) 353.

\bibitem{BijVelt} J. J. van der Bij and M. Veltman,
        {\sl Nucl.\ Phys.}\ {\bf B 231} (1984) 205.

\bibitem{Frank} J. J. van der Bij and F. Hoogeveen,
        {\sl Nucl.\ Phys.}\ {\bf B 283} (1987) 477.

\bibitem{ConsHollJege} M. Consoli, W. Hollik and F. Jegerlehner,
        {\sl Phys.\ Lett.}\ {\bf B 227} (1989) 167.

\bibitem{Barbieri1} R. Barbieri {\it et al.},
        {\sl Phys.\ Lett.}\ {\bf B 288} (1992) 95;
        errata {\it ibid}\/ {\bf B 312} (1993) 511.

\bibitem{Barbieri2} R. Barbieri {\it et al.},
        {\sl Nucl.\ Phys.}\ {\bf B 409} (1993) 105.

\bibitem{Fleischer1} J. Fleischer, O. V. Tarasov and F. Jegerlehner,
        {\sl Phys.\ Lett.}\ {\bf B 319} (1993) 249.

\bibitem{Fleischer2} J. Fleischer, O. V. Tarasov and F. Jegerlehner,
        {\sl Phys.\ Rev.}\ {\bf D 51} (1995) 3820.

\bibitem{DegrFancGamb} G. Degrassi, S. Fanchiotti and P. Gambino,
        {\sl Int.\ J.\ Mod.\ Phys.}\ {\bf A 10} (1995) 1337.

\bibitem{BaubWeig} S. Bauberger and G. Weiglein,
        {\sl Phys.\ Lett.}\ {\bf B 419} (1998) 333.

\bibitem{MaldeStuart2} P. J. Malde and R. G. Stuart, hep-ph/9903403.

\bibitem{Z0Pole} B. A. Kniehl and R. G. Stuart,
        {\sl Comput.\ Phys.\ Commun.} {\bf 72} (1992) 175.

\bibitem{1loopclosed} F. V. Tkachov,
        {\sl Phys.\ Lett.}\ {\bf 100B} (1981) 65.

\bibitem{gegenbauer} K.G.  Chetyrkin, A.L. Kataev, F.V. Tkachov,
            {\sl Nucl. Phys.}\ {\bf B174} (1980) 345.

\bibitem{2loopbubclosed} K. G. Chetyrkin, in {\sl New computing
        techniques in Physics research III},
        ed.s K.-H. Becks, D. Perret-Gallix, World Scientific, Singapore
        (1994) p 559.

\bibitem{Berman} S. M. Berman, {\sl Phys.\ Rev.}\ {\bf 112} (1958) 267.

\bibitem{Grotch} H. Grotch, {\sl Phys.\ Rev.}\ {\bf 168} (1968) 1872.

\bibitem{HokimPham} Q. Hokim and X.-Y. Pham,
    {\sl Ann.\ Phys.}\ {\bf 155} (1984) 202.

\bibitem{FiscKuroSava} A. Fischer, T. Kurosu and F. Savatier,
    {\sl Phys.\ Rev.}\ {\bf D 49} (1994) 3426.


\end{thebibliography}
\end{document}